\documentclass[aoas,preprint]{imsart}

\RequirePackage{amsthm,amsmath,amsfonts,amssymb}
\RequirePackage[authoryear]{natbib}
\RequirePackage[colorlinks,citecolor=blue,urlcolor=blue]{hyperref}
\RequirePackage{graphicx}

\startlocaldefs
\numberwithin{equation}{section}
\theoremstyle{plain}
\newtheorem{proposition}{Proposition}
\newtheorem{lemma}{Lemma}
\theoremstyle{remark}
\newtheorem{mydef}{Definition}
\newtheorem{remark}{Remark}
\usepackage[utf8]{inputenc}
\usepackage{amsbsy}
\usepackage{amscd}
\usepackage{array}

\usepackage{xspace} 
\usepackage{tikz}
\usepackage{booktabs} 
\usepackage{placeins} 
\usepackage{nameref} 

\usepackage{etoolbox}
\makeatletter
\newcommand\bibstyle@comma{\bibpunct(),a,,}
\newcommand\bibstyle@semicolon{\bibpunct();a,,}
\makeatother
\pretocmd\citet{\citestyle{comma}}\relax\relax
\pretocmd\Citet{\citestyle{comma}}\relax\relax
\pretocmd\citep{\citestyle{semicolon}}\relax\relax
\pretocmd\Citep{\citestyle{semicolon}}\relax\relax

\def\tree{\mathcal{T}}
\def\treeHeight{h}
\def\ntaxa{n}
\def\nObs{N}
\def\nObsOne{n}
\def\nNodes{m}

\def\branchLength{\ell}
\def\nodeHeight{t}
\newcommand{\sharedHeight}[2]{t_{{#1}{#2}}}

\def\iNode{j}
\def\iObs{i}
\def\iAny{k}
\def\iAnyPrime{k'}
\def\iTot{tot}
\def\iAnyBis{l}
\def\iOne{i}
\def\iTwo{j}
\def\iPar{p}
\def\iRoot{r}

\DeclareMathOperator{\pa}{pa}

\DeclareMathOperator{\children}{\mathcal{C}}
\DeclareMathOperator{\siblings}{\mathcal{S}}

\def\dataOne{Y}

\def\dataMatr{\matr{\dataOne}}

\newcommand{\datam}[1]{\matr{\dataOne}^{#1}}
\newcommand{\dataObsm}[1]{\datam{#1}_{\text{o}}}
\newcommand{\dataMissm}[1]{\datam{#1}_{\text{m}}}
\newcommand{\permObs}[1]{\matr{\Pi}_{#1,\text{o}}}
\newcommand{\permMiss}[1]{\matr{\Pi}_{#1,\text{m}}}
\def\dimTrait{p}
\newcommand{\dimTraitObs}[1]{\dimTrait^{#1}_{\text{o}}}
\newcommand{\dimTraitMiss}[1]{\dimTrait^{#1}_{\text{m}}}

\def\allParams{\vvect{\theta}}
\def\dimParam{d}

\def\unconstrParams{\vvect{\eta}}
\newcommand{\transform}{f}
\newcommand{\invTransform}{\inverse{f}}

\def\momentum{\vect{p}}

\DeclareMathOperator{\totalEnergy}{H}
\DeclareMathOperator{\potentialEnergy}{U}
\DeclareMathOperator{\kineticEnergy}{K}

\def\latentOne{Z}
\def\latentVect{\vect{\latentOne}}
\def\latentMatr{\matr{\latentOne}}
\def\latentMatrTip{\matr{\latentOne}^{\text{tip}}}
\newcommand{\latent}[1]{\latentOne^{#1}}
\newcommand{\latentm}[1]{\matr{\latentOne}^{#1}}

\def\dimLatent{q}

\def\completedOne{X}

\def\completedMatr{\matr{\completedOne}}

\newcommand{\completedm}[1]{\matr{\completedOne}^{#1}}

\newcommand{\below}[1]{\lfloor #1 \rfloor}
\renewcommand{\above}[1]{\lceil #1 \rceil}
\newcommand{\dataAbovem}[1]{\matr{\dataOne}_{\above{#1}}}
\newcommand{\dataBelowm}[1]{\matr{\dataOne}_{\below{#1}}}

\newcommand{\processm}[1]{\latentVect_{#1}}

\newcommand{\brownianm}[1]{\vect{B}_{#1}}

\def\der{\text{d}}
\newcommand{\gradient}[1]{\nabla_{#1}}
\newcommand{\partialDer}[2]{\frac{\partial \transpose{#2}}{\partial #1}}

\def\rootMean{\mu}
\def\shift{\delta}
\def\rootMeanm{\vvect{\mu}}
\def\variance{\sigma^2}
\def\variancemOne{R}
\def\variancem{\matr{\variancemOne}}

\def\sdm{\matr{R}^{1/2}}

\def\rootVariancemOne{\Gamma}
\def\rootVariancem{\matr{\rootVariancemOne}}

\newcommand{\identityMatrix}[1]{\matr{I}_{#1}}
\def\driftm{\vvect{\nu}}

\def\optimm{\vvect{\beta}}
\def\strength{\alpha}
\def\strengthm{\matr{A}}
\def\halfLife{t_{1/2}}

\newcommand{\actum}[1]{e^{-\strengthm #1}}
\newcommand{\actumt}[1]{e^{-\transpose{\strengthm} #1}}

\def\stationaryVariancem{\matr{V}} 
\def\passage{\matr{P}}
\def\strengthEigen{\matr{\Lambda}}
\def\strengthEigenValue{\lambda}

\def\factorNoise{f}

\newcommand{\branchVariance}[1]{\mathbf{\Sigma}_{#1}}
\newcommand{\branchActualization}[1]{\mathbf{q}_{#1}}
\newcommand{\branchDisplacement}[1]{\mathbf{d}_{#1}}

\newcommand{\obsVariance}[1]{\mathbf{S}_{#1}}
\newcommand{\obsVarianceUni}[1]{s^2_{#1}}
\newcommand{\obsPrecision}[1]{\mathbf{R}_{#1}}

\newcommand{\genBranchVariance}[1]{\mathbf{c}_{#1}}
\newcommand{\genBranchActualization}[1]{\mathbf{a}_{#1}}
\newcommand{\genBranchDisplacement}[1]{\mathbf{b}_{#1}}

\newcommand{\nodeMean}[1]{\mathbf{m}_{#1}}
\newcommand{\nodePrecision}[1]{\mathbf{P}_{#1}}
\newcommand{\nodeRemainder}[1]{\mathbf{r}_{#1}}
\newcommand{\deflatedNodePrecision}[1]{\nodePrecision{#1}^{\star}}

\newcommand{\sumChildren}[2]{\sum_{#1 \in \children(#2)}}

\newcommand{\nodePreMean}[1]{\mathbf{n}_{#1}}
\newcommand{\nodePrePrecision}[1]{\mathbf{Q}_{#1}}
\newcommand{\nodePreVariance}[1]{\inverse{\nodePrePrecision{#1}}}
\newcommand{\deflatedNodePrePrecision}[1]{\nodePrePrecision{#1}^{\star}}
\newcommand{\deflatedNodePreVariance}[1]{\nodePrePrecision{#1}^{\star-}}
\newcommand{\deflatedNodePreMean}[1]{\nodePreMean{#1}^{\star}}
\newcommand{\deflatedNodePrecisionSibling}[1]{\nodePrecision{-#1}^{\star}}
\newcommand{\deflatedNodeMeanSibling}[1]{\nodeMean{-#1}^{\star}}

\newcommand{\fullConditionalMean}[1]{\mathbf{M}_{#1}}
\newcommand{\fullConditionalVariance}[1]{\mathbf{V}_{#1}}
\newcommand{\fullConditionalMeanMiss}[1]{\mathbf{M}_{#1,\text{m}}}
\newcommand{\fullConditionalVarianceMiss}[1]{\mathbf{V}_{#1,\text{m}}}

\def\seqData{\matr{S}}
\def\seqParams{\vvect{\phi}}

\newcommand{\altPhi}[3]{\hat{\phi}\left(#1; #2, #3\right)}
\newcommand{\specialInverse}[1]{#1^{-}}
\newcommand{\lowDimInverse}[1]{#1^{\sim}}
\newcommand{\altDet}[1]{\hat{\text{det}}\left( #1 \right)}
\newcommand{\indMissing}[1]{\mmatr{\Delta}_{#1}}

\def\anyMatrix{\matr{M}}
\def\anySymMatrix{\matr{S}}
\def\anyVector{\vect{X}}
\def\setIndices{\mathcal{I}}
\newcommand{\subMatrix}[2]{#1_{#2}}

\newcommand{\sss}[2]{\transpose{\left(#1\right)}#2\left(#1\right)} 

\def\derivativeVarianceVariance{\matr{F}}
\def\derivativeVarianceAttenuation{\matr{G}}

\def\model{\mathcal{M}}
\def\iModel{m}
\def\nModel{K}

\newcommand{\workingPrior}[2]{p_0\!\left(\sachant{#1}{#2}\right)}
\newcommand{\pathLikelihood}[2]{q_{#1}\!\left(#2\right)}
\def\pathParameter{\beta}

\def\correlation{\matr{C}}
\def\diagVariance{\matr{D}_{\sigma}}
\def\dimMat{p}
\def\dimAny{k}
\def\dimOne{k}
\def\dimTwo{l}
\def\cholCorrelation{\matr{W}}
\def\cholCorrelationCoef{W}
\def\diffeo{\simeq}
\newcommand{\corSpaceDim}[1]{\mathcal{C}_{#1}}
\newcommand{\corSpace}{\corSpaceDim{\dimMat}}
\newcommand{\cholCorSpaceDim}[1]{\mathcal{C}^{\text{ch}}_{#1}}
\newcommand{\cholCorSpace}{\cholCorSpaceDim{\dimMat}}
\newcommand{\halfSphereDim}[1]{\mathcal{S}_{#1}^{\text{pos}}}

\newcommand{\halfSphereAny}{\halfSphereDim{\dimAny}}
\newcommand{\ballDim}[1]{\mathcal{B}_{#1}}

\newcommand{\ballAny}{\ballDim{\dimAny-1}}
\newcommand{\ballInfDim}[1]{\mathcal{B}_{#1}^{\infty}}

\newcommand{\ballInfAny}{\ballInfDim{\dimAny-1}}
\def\ballInfToBallLKJ{\vect{L}}
\DeclareMathOperator{\LKJ}{LKJ}
\DeclareMathOperator{\LKJChol}{LKJChol}
\def\paramLKJ{\eta}
\def\constantLKJ{c_{\dimMat}(\paramLKJ)}
\DeclareMathOperator{\SBeta}{SBeta}
\newcommand{\sphericalBeta}[3]{\SBeta\left(\sachant{\vect{#3}}{#1, #2}\right)}
\def\paramSB{\beta}
\newcommand{\constantSB}[2]{C\left(#1, #2\right)}

\def\weasels{\taxon{Musteloidea}\xspace}

\newcommand{\expProcess}{\Esp_{\allParams}}
\newcommand{\varProcess}{\Var_{\allParams}}
\newcommand{\popVarOne}{V}
\newcommand{\popVar}{\matr{\popVarOne}}
\newcommand{\heritability}{H}
\newcommand{\heritabilityMatrix}{\matr{\heritability}}

\newcommand{\taxon}[1]{\textit{#1}} 
\newcommand{\enquote}[1]{``#1''}

\newcommand{\code}[1]{\textsf{#1}}

\newcommand{\OU}{Ornstein-Uhlenbeck\xspace}
\newcommand{\OUa}{OU\xspace}
\newcommand{\BM}{Brownian motion\xspace}
\newcommand{\BMa}{BM\xspace}
\newcommand{\OUBMa}{OU-BM\xspace}

\newcommand{\PCMa}{PCM\xspace}
\newcommand{\PCMs}{PCMs\xspace}

\newcommand{\beast}{BEAST\xspace}

\newcommand{\GLInv}{$\mathcal{G}_{\text{\tiny LInv}}$\xspace}
\newcommand{\wrt}{with respect to\xspace}

\newcommand{\eg}{e.g.\xspace}
\newcommand{\ie}{i.e.\xspace}

\newcommand{\field}[1]{\mathbb{#1}}
\newcommand{\R}{\field{R}}

\newcommand{\Or}{\mathcal{O}}

\newcommand{\Normal}[2]{\mathcal{N}\hspace*{-0.2em}\left(#1,~#2\right)}

\newcommand{\sachant}[2]{\left.#1\mathrel{}\middle|\mathrel{}#2\right.} 

\newcommand{\Esp}{\field{E}}
\newcommand{\Espe}[1]{\field{E}\left[ #1 \right]}
\newcommand{\Espesq}[2]{\field{E}\left[\sachant{#1}{#2}\right]}

\newcommand{\Var}{\field{V}\text{ar}}

\newcommand{\Covas}[2]{\field{C}\text{ov}[ #1; #2 ]}

\newcommand{\cDensity}[2]{p\left(\sachant{#1}{#2}\right)}
\newcommand{\density}[1]{p\left(#1\right)}

\newcommand{\vect}[1]{\mathbf{#1}}
\newcommand{\vvect}[1]{\boldsymbol{#1}}

\newcommand{\matr}[1]{\mathbf{#1}}
\newcommand{\mmatr}[1]{\boldsymbol{#1}}
\newcommand{\muv}{\vvect{\mu}}

\newcommand{\thetav}{\vvect{\theta}}

\newcommand{\transpose}[1]{#1^T}

\DeclareMathOperator{\rank}{rank}

\newcommand{\inverse}[1]{#1^{-1}}
\def\bigcdot{\boldsymbol{\cdot}}

\DeclareMathOperator{\mtov}{vec} 
\DeclareMathOperator{\mtovh}{vech} 
\DeclareMathOperator{\diag}{diag} 

\DeclareMathOperator{\mtovhm}{vechm} 
\newcommand{\mtovhmn}[1]{\mtovhm_{#1}}
\newcommand{\duplicationMatrixMissing}[1]{\matr{D}_{h,#1}}

\newcommand{\hadamard}{\odot}

\usepackage{stmaryrd}

\newcommand{\set}[1]{\mathopen{\{}#1\mathclose{\}}}

\newcommand{\Det}[1]{\left\lvert#1\right\rvert}
\newcommand{\norm}[1]{\left\lVert#1\right\rVert}

\usepackage{mathtools}

\theoremstyle{plain} 

\makeatletter
\def\maxwidth{ %
  \ifdim\Gin@nat@width>\linewidth
    \linewidth
  \else
    \Gin@nat@width
  \fi
}
\makeatother

\definecolor{fgcolor}{rgb}{0.345, 0.345, 0.345}

\usepackage{framed}
\makeatletter
 {\par\unskip\endMakeFramed%
 \at@end@of@kframe}
\makeatother

\definecolor{shadecolor}{rgb}{.97, .97, .97}
\definecolor{messagecolor}{rgb}{0, 0, 0}
\definecolor{warningcolor}{rgb}{1, 0, 1}
\definecolor{errorcolor}{rgb}{1, 0, 0}
\newenvironment{knitrout}{}{} 

\usepackage{alltt}

\endlocaldefs

\begin{document}
\begin{frontmatter}
    \title{
        Efficient Bayesian Inference of
        General Gaussian Models
        on Large Phylogenetic Trees
    }
\runtitle{Efficient HMC for OU Evolutionary Models}

\begin{aug}
\author[A,B]{\fnms{Paul} \snm{Bastide}\ead[label=e1]{paul.bastide@umontpellier.fr}},
\author[C]{\fnms{Lam Si Tung} \snm{Ho}\ead[label=e5]{lam.ho@dal.ca}},
\author[B]{\fnms{Guy} \snm{Baele}\ead[label=e2]{guy.baele@kuleuven.be}},
\author[B]{\fnms{Philippe} \snm{Lemey}\ead[label=e4]{philippe.lemey@kuleuven.be}}
\and
\author[D]{\fnms{Marc A.} \snm{Suchard}\ead[label=e3]{msuchard@ucla.edu}}
\address[A]{
    IMAG,
    CNRS,
    Universit\'e de Montpellier,
    \printead{e1}
}
\address[B]{
    Department of Microbiology, Immunology and Transplantation,
    Rega Institute,
    KU Leuven,
    \printead{e2,e4}
}
\address[C]{
    Department of Mathematics and Statistics,
    Dalhousie University,
    \printead{e5}
}
\address[D]{
    Departments of Biostatistics, Biomathematics, and Human Genetics,
    University of California, Los Angeles,
    \printead{e3}
}

\end{aug}

\begin{abstract}
    Phylogenetic comparative methods correct
    for shared evolutionary history among a set of non-independent organisms 
    by modeling sample traits as arising from
   a diffusion process 
   along
    the branches of a possibly unknown history.
    To incorporate such uncertainty, we present a scalable Bayesian inference
    framework under a general
    Gaussian
    trait evolution model
    that exploits Hamiltonian Monte Carlo (HMC).
    HMC enables efficient sampling of the constrained model parameters and takes advantage of
    the tree structure for fast likelihood and gradient computations, yielding
    algorithmic complexity linear in the number of observations. 
    This approach
    encompasses a wide
    family of stochastic processes, including the general \OU (\OUa) process,
    with possible missing data and measurement errors.
    We implement inference tools for a biologically relevant subset of all these models
    into the BEAST phylogenetic software package and develop model comparison through marginal likelihood estimation.
    We apply our approach to study
    the morphological evolution in the superfamilly of \weasels
    (including weasels and allies) as well as
    the heritability of HIV virulence.
    This second problem furnishes a new measure of evolutionary heritability
    that demonstrates its utility through a targeted simulation
    study.
\end{abstract}



\end{frontmatter}

\section{Introduction}\label{sec:intro}

\subsection{Motivation}\label{sec:intro:motivations}
The evolutionary history of organisms shapes the distribution of their
observed characteristics \citep{Felsenstein1985}.
To account for correlation induced by this shared history, phylogenetic
comparative methods (\PCMs) have been developed for the analysis of quantitative
traits \citep[see e.g.][for a review]{Pennell2013}.
These methods can be applied to a wide range of organisms and traits, to answer a large
spectrum of biological questions on various evolutionary time frames,
ranging from decades or even years in virology \citep{Dudas2017}
to millions of years in evolutionary biology \citep{Aristide2016}.
Modern studies in PCMs routinely include a growing number of taxa
(\eg more than fifteen hundred in \citealt{Blanquart2017}),
with possibly a large number of missing data for the
multivariate measurements of continuous traits
(\eg almost $25\%$ in \citealt{Schnitzler2017}),
and with intricate phylogenetic and temporal structure.
The underlying biological processes at play often have complex
dynamics, and are only measured imperfectly, with a variable amount of noise.
\subsection{Model}\label{sec:intro:model}
\PCMs posit a continuous-valued stochastic process running on the branches of a
phylogenetic tree that gives rise to trait values, possibly measured with noise, at the tree tips for observed
organisms.

\paragraph*{Phylogenetic Tree}
The phylogenetic tree represents the evolutionary relationship among the
organisms studied. We assume, without loss of generality, that the tree is calibrated
in time, so that branch lengths represent actual time. For organisms evolving
rapidly, such as viruses, the observations at the $\ntaxa$ tips of the
tree are not necessarily contemporaneous (see Figure~\ref{fig:BM_on_tree}, left).
We denote by $\nNodes$ the total number of internal and external nodes in the tree
($\nNodes = 2 \ntaxa - 1$ if the tree is binary).

\paragraph*{Stochastic Process on the Tree}
We assume that a continuous trait evolves over time according to a stochastic process.
When a speciation event occurs in the tree, the process is split into two conditionally independent
processes with the same distribution (see Figure~\ref{fig:BM_on_tree}).
Only the values of the process at the tips of the tree are potentially observed.
This model enforces a phylogenetic correlation structure on the observations, as shown below.


\begin{figure}[!ht]
    \begin{center}
        \input{figure/BM_on_tree.tex}
        \caption{
            Realization of a univariate \BMa process (right)
            with variance $\variance = 0.05$
            and mean root value $\rootMean = 0$
            on a timed phylogenetic tree (left).
            Observations span 3 years (2006 -- 2008, tips $\latentm{1}$ to $\latentm{5}$).
            $\sharedHeight{1}{2}$ is the time of shared evolution between tips $1$ and $2$.
            $\branchLength_{1}$ is the length of the branch ending at tip $1$. 
        \label{fig:BM_on_tree}}
    \end{center}
\end{figure}

\paragraph*{Brownian Motion (\BMa)}
The simplest stochastic process \citep{Cavalli-Sforza1967,Felsenstein1985} assumes that a
multivariate trait $\processm{t}$ of dimension $\dimTrait$
evolves in time $t$ following a \BM $\brownianm{t}$
with variance $\variancem$, 
such that
\(
\der \processm{t} = \sdm \der \brownianm{t}
\)
for all $t \geq 0$ and at the tree root
$\latentm{\iRoot} \sim \Normal{\rootMeanm}{\rootVariancem}$.
Under this process, the covariance between
trait $\dimOne$ at node $\iOne$
and trait $\dimTwo$ at node $\iTwo$
($1 \leq \dimOne, \dimTwo \leq \dimTrait$ and $1 \leq \iOne, \iTwo \leq \nNodes$)
is the product of
(i) the covariance $\variancemOne_{\dimOne\dimTwo}$ between traits $\dimOne$ and $\dimTwo$
and (ii) the shared evolutionary time $\sharedHeight{\iOne}{\iTwo}$
between species $\iOne$ and $\iTwo$, \ie the time from the root to
their most recent common ancestor
(see Figure~\ref{fig:BM_on_tree}),
plus the contribution of the root itself:
\(
\Covas{\latent{\iOne}_{\dimOne}}{\latent{\iTwo}_{\dimTwo}}
=
\sharedHeight{\iOne}{\iTwo}
\variancemOne_{\dimOne\dimTwo}
+
\rootVariancemOne_{\dimOne\dimTwo}
\)
(see \eg \citealp{Clavel2015}).
A constant drift $\driftm$ can be added to the mean value of the trait
\citep{Gill2016},
in which case the expectation of trait $\dimOne$ at species $\iOne$ at time
$\nodeHeight_{\iOne}$ is
$\Espe{\latent{\iOne}_{\dimOne}} = \nodeHeight_{\iOne} \driftm$.

\paragraph*{\OU (\OUa)}
The OU process was proposed as a model for traits evolving under stabilizing
selection \citep{Hansen1997} and has become widely used across evolutionary biology
\citep[see \eg][and references therein]{Cooper2016}.
The OU process generalizes BM by adding a deterministic call-back term to a given
value $\optimm$ that is interpreted as the optimal value
of the trait of a species in a given environment:
\(
    \der \processm{t}
    =
    - \strengthm (\processm{t} - \optimm) \der t
    + \sdm \der \brownianm{t}
\)
for all $t \geq 0$.
Matrix $\strengthm$ is the \enquote{selection strength} that is constrained
to have positive real parts of its eigenvalues and controls the dynamics of the pull toward the optimum.
The covariance between two multivariate traits at two nodes
can be explicitly formulated \citep{Bartoszek2012,Clavel2015},
and, compared to the \BMa where the variance increases linearly in time,
is bounded by the stationary variance $\stationaryVariancem$ of the process.

\paragraph*{Observation Model and Individual Variation}
The processes described above are meant to capture the evolutionary dynamics of traits across organisms.
However, various other sources of variation may contribute to the observed data,
such as measurement error, independent environmental variation, or
intra-specific variation \citep[see \eg][for a review]{Hadfield2010}.
As outlined in the next section, we include these biological phenomena in our
model as an extra layer, that links the realization of the process at the tips
$\latentm{i}$ to the actual measurements $\datam{i}$ through a Gaussian
observation model.

\subsection{Scope of the article}\label{sec:intro:scope}
In this work, we propose a general and efficient Bayesian framework to rigorously
analyze this broad class of evolutionary models.
%

\paragraph*{State of the Art}
Since their introduction in the seminal article of \citet{Felsenstein1985},
\PCMs have undergone extensive development, resulting in increasingly realistic models.
We limit references here to those that specifically relate to the inference
problem in a general setting and refer to \eg \citet{Harmon2019} for a recent and
more comprehensive overview of these models.
\citet{Clavel2015} provide a comprehensive maximum-likelihood framework to fit
a wide range of models, using explicit estimators that can in some cases be
computationally prohibitive.
\citet{Pybus2012,Freckleton2012,Mitov2018PCM} describe and implement likelihood computation algorithms
that are linear in the number of observations in a framework similar to the one
described here.
This algorithm is exploited in \citet{Mitov2019} to conduct maximum-likelihood
inference.
From a Bayesian perspective, \citet{Pybus2012} and \citet{Cybis2015} describe
a standard Markov chain Monte Carlo (MCMC) inference framework for \BMa, while
\citet{Hassler2019} extend it to
include measurement errors with possible missing values, and
\citet{Fisher2019} to use an efficient Hamiltonian Monte Carlo (HMC) sampler on the rates of a relaxed random
walk \citep{Lemey2010}.
%
Other more general models have been developed, including:
non Gaussian models (using \eg Levy processes, see \citealt{Landis2013,Duchen2017},
or general Fokker - Plank equations, see \citealt{Boucher2018});
models where the trait explicitly impacts the tree (with BiSSE and related methods,
see \citealt{Maddison2007,Fitzjohn2009,Fitzjohn2010,Goldberg2011,Fitzjohn2012});
models where species interact with each other (through mutualism or competition,
see \citealt{Nuismer2015,Manceau2016,Drury2016,Bartoszek2017Int,Drury2018,Aristide2019},
or through migration \citealt{Bartoszek2017,Duchen2020migration});
models with varying and heritable intraspecific variance \citep{Kostikova2016};
or models in high-dimensional trait settings (using pseudo or penalized
likelihood, see \citealt{Goolsby2016,Clavel2018}).
These models are however outside of the scope of the present work, that focuses
on the general Gaussian model as presented in the next section
(see Definition~\ref{def:general_model}).


\paragraph*{Outline}
We complement recent advances in computing the likelihood under a general class of Gaussian models with an algorithm to analytically evaluate its gradient \wrt any of the parameters in linear time in the number of observations. This general class includes the OU process, as well as measurement error and missing data.
We exploit this algorithm to develop an efficient Bayesian inference framework that
relies on the use of an HMC sampler and allows for model selection
through marginal likelihood estimation.
We implement this framework in the \beast phylogenetic software package
\citep{Suchard2018}
for a sub-class of models, namely
the \OUa process with a diagonal selection strength matrix $\strengthm$,
that are of particular interest for the biological problems we study here.
%
%
In Section~\ref{sec:gradient}, we present the general likelihood and new gradient
computation algorithm
(with details in Appendices~\ref{app:post_pre_order} and~\ref{app:der}).
In Section~\ref{sec:statistical_inference}, we develop the Bayesian statistical
inference framework
(with details in Appendix~\ref{app:transformations}).
Finally, in Section~\ref{sec:bio}, we illustrate the method
on two recently published biological datasets as well as on simulations.

\section{Efficient Gradient Computations}\label{sec:gradient}

In this section, we show how the likelihood and its gradient with respect to
all the parameters in a general trait evolutionary model can be computed in linear time in the number of tips of a rooted phylogenetic tree.



\subsection{Statistical Model}\label{sec:gradient:model}
%
Conditioning on a tree $\tree$, we define the following general Gaussian model
of trait evolution:
\begin{mydef}[General Gaussian Model of Trait Evolution]
    \label{def:general_model}
    Let $\tree$ be a rooted 
    phylogenetic tree with $\ntaxa$ tips and
    $\nNodes$ 
    internal and external nodes.
    At each node $\iNode$, $1 \leq \iNode \leq \nNodes$, define a latent variable
    $\latentm{\iNode}$, and for each observation $\iObs$,
    $1 \leq \iObs \leq \nObsOne$, a measure $\datam{\iObs}$, both of dimension $\dimTrait$.
    The general Gaussian model of trait evolution on $\tree$ is then defined in a
    hierarchical way as follows:
    \begin{align}
            \latentm{\iRoot}
            & \sim \Normal{\rootMeanm}{\rootVariancem}
            &
            & \text{root;}
            \label{eq:general_model_root}
            \\
            \sachant{\latentm{\iNode}}{\latentm{\pa(\iNode)}}
            & \sim \Normal{
            \branchActualization{\iNode}
            \latentm{\pa(\iNode)}
            + \branchDisplacement{\iNode}
            }{
            \branchVariance{\iNode}
            }
            &
            & \text{propagation;}
            \label{eq:general_model_tree}
            \\
            \sachant{\datam{\iObs}}{\latentm{\pa(\iObs)}}
            & \sim \Normal{
                \latentm{\pa(\iObs)}
            }{
            \obsVariance{\iObs}
            }
            &
            & \text{observation;}
            \label{eq:general_model_obs}
    \end{align}
    where $\pa(\iNode)$ and $\pa(\iObs)$ denote, respectively,
    the unique parent node of node $\iNode$
    or latent tip associated with observation $\iObs$;
    %
    %
            $\rootMeanm$ and $\rootVariancem$ are the expectation and
            variance of the root variable $\latentm{\iRoot}$; 
            %
            for any node $\iNode$,
            $\branchActualization{\iNode}$, $\branchDisplacement{\iNode}$ and
            $\branchVariance{\iNode}$ are, respectively, the actualization,
            drift, and variance
            associated with the
            branch going from $\pa(\iNode)$ to $\iNode$; and
            for any observation $\iObs$, %
            $\obsVariance{\iObs}$ is the variance associated with it.
    We further denote by
    $\dataMatr = \transpose{(\datam{1}, \cdots, \datam{\nObsOne})}$
    and
    $\latentMatr = \transpose{(\latentm{1}, \cdots, \latentm{\nNodes})}$
    the matrices of observed and latent trait variables,
    and by
    $\completedMatr = \transpose{(\transpose{\dataMatr}, \transpose{\latentMatr})}$
    the \emph{complete} dataset.
    Furthermore, we assume that $\rootVariancem$,
    $\branchVariance{\iNode}$, and $\obsVariance{\iObs}$ are positive definite
    for any node $\iNode$ and observation $\iObs$.
\end{mydef}

This allows entertaining a highly generic framework that encompasses various evolutionary scenarios,
as underlined in the following three paragraphs.
%
\paragraph*{Stochastic Process Propagation}
    Equation~\eqref{eq:general_model_tree} describes the stochastic process
    that governs the evolution of the latent trait.
    It is similar to the \GLInv model described in \citet{Mitov2018PCM},
    or the generic formulation used in \citet{Bastide2017}.
    Both the \BMa and \OUa models
    can be cast into this framework, by setting:
    \begin{align}
        \branchActualization{\iNode}
        &= \identityMatrix{\dimTrait},
        &
        \branchDisplacement{\iNode}
        &= \branchLength_{\iNode}\driftm,
        &
        \branchVariance{\iNode}
        &= \branchLength_{\iNode} \variancem
        &
        \text{(\BMa)};
        \label{eq:general_to_BM}
        \\
        \branchActualization{\iNode}
        &= \actum{\branchLength_{\iNode}},
        &
        \branchDisplacement{\iNode}
        &= (\identityMatrix{\dimTrait} - \actum{\branchLength_{\iNode}})
        \optimm,
        &
        \branchVariance{\iNode}
        &= \stationaryVariancem
        - \actum{\branchLength_{\iNode}}\stationaryVariancem \actumt{\branchLength_{\iNode}}
        &
        \text{(\OUa)};
        \label{eq:general_to_OU}
    \end{align}
    where, as defined in the introduction, $\driftm$ and $\variancem$ are the
    constant drift and variance of a simple \BMa,
    and $\strengthm$, $\optimm$ and $\stationaryVariancem$ are the selection strength,
    optimal values and stationary variances of an \OUa.
    We refer to \citet{Mitov2018PCM} for more details concerning other models that
    can be described within this framework, including processes with shifts or jumps.

\paragraph*{Observation Model}
    Equation~\eqref{eq:general_model_obs} describes the observation model.
    %
    %
    The variance term $\obsVariance{\iObs}$ can have multiple significations, from
    a simple measurement error, to a \enquote{meta-analysis} effect
    (see \citealp{Hadfield2010} for a review on observation errors),
    or an \enquote{intra-specific} variance
    (see \eg \citealp{Goolsby2017}).
    In the simple case where the same observation error is assumed for all
    measures $\iObs$, $1 \leq \iObs \leq \nObsOne$,
    this term reduces to $\obsVariance{\iObs} = \obsVariance{}$.
    %
    %
    Note that, from a methodological point of view, Equation~\eqref{eq:general_model_obs}
    can be considered as a particular case of Equation~\eqref{eq:general_model_tree}. 
\paragraph*{Phylogenetic Factor and Repetitions}
We note that one may posit that $\latentm{\iObs}$ is of lower dimension
$\dimLatent < \dimTrait$ and link to $\datam{\iObs}$ through a linear combination
via a latent factor model \citep{Tolkoff2017}.
It would also be straightforward to include several measurements associated
with a single tip. 
For clarity, we omit these details in the main text, and refer to
Appendix~\ref{app:post_pre_order} for the derivations in this general framework.

\subsection{Likelihood Computations}\label{sec:gradient:likelihood}

The general model of Definition~\ref{def:general_model} is Gaussian, and it is
hence possible to write out the marginal distribution
$\cDensity{\dataMatr}{\allParams}$
of the measures $\dataMatr$ given the parameters $\thetav$ for some specific models
(see \eg \citealp{Clavel2015} for such formulations in the multivariate \OUa case).
However, this computation generally requires the inversion of a tree-induced
variance matrix, of dimension $\nObsOne \times \nObsOne$.
This is inefficient (worse than quadratic in $\nObsOne$, see \eg \citealp{Raz2003}),
and is ill-suited for handling large phylogenetic trees that now frequently confront practitioners (see \eg \citealp{Jetz2012,Blanquart2017}) or when the tree itself is random \citep{Pybus2012}.
To alleviate this issue, it is possible to write an efficient pruning-style algorithm
that is \emph{linear} in the number $\nObsOne$ of organisms.

\paragraph*{Pruning-Style Algorithm}
\citet{Felsenstein1973a} introduced the pruning algorithm into phylogenetics to compute the likelihood of a simple \BMa.
It draws from classical Gaussian conditional propagation ideas such
as the Kalman filter or other \enquote{forward-backward}
algorithms (see \eg \citealp{Rabiner1989} for a review).
Variants of this algorithm have been flowering in the literature,
sometimes under different names.
A non-exhaustive list of references for the \BMa case includes
\citet{Hadfield2010},
\citet{Pybus2012},
\citet{Fitzjohn2012} (Gaussian Elimination Method),
\citet{Freckleton2012},
\citet{Lartillot2014} (Phylogenetic Kalman Filter)
and
\citet{Cybis2015}.
%
\citet{Goolsby2016} and \citet{Hassler2019} have recently proposed adaptations
to handle missing data.
Finally, an extension of the algorithm to the general case as presented in
Definition~\ref{def:general_model}
(with missing data) was proposed by \citet{Bastide2017},
and also more recently by \citet{Mitov2018PCM}.


\paragraph*{Implementation in \beast}
All these methods
allow computing the likelihood in linear time in $\nObsOne$.
As is frequently the case for PCMs, many 
have been implemented in various independent software
packages, each likely with a specific use in mind
(with the notable exception of the
\citealp{Rstats}
package PCMbase,
see
\citealp{Mitov2018PCM}).
While this explains at least partly the numerous references to the algorithm, it limits the use of
several implementations beyond the specific models they consider.
%
Bayesian Evolutionary Analysis by Sampling Trees (\beast, \citealp{Suchard2018})
is a widely used, well established and versatile phylogenetic software package.
It encompasses a great variety of molecular sequence modeling tools, making it
possible to conduct a coherent joint inference of both the timed phylogenetic
tree and of the properties of the stochastic process,
without the need to resort to a two-step analysis as is
usually the case in previous methodologies (see
Section~\ref{sec:total_evidence} for more details).
%
%
We implemented the general algorithm presented in \citet{Bastide2017}
in this unified framework
(with improvements, see next paragraph),
allowing for its seamless integration
with the realistic analyses permitted by the software.

\paragraph*{Efficiency and Numerical Robustness}
When sampling the parameters in a wide region of the space, as
is typically done in a Bayesian analysis (see Section~\ref{sec:statistical_inference}),
numerical robustness is particularly important, as small divergences due to possibly
ill-conditioned matrices can accumulate over the tree traversal, and lead to diverging
results.
%
%
We tackled this issue using two independent developments.
First, we reduced the number of operations actually performed during the
traversal of the tree thanks to a careful analysis of the iteration steps,
making the algorithm both more efficient and more robust.
Second, we increased the numerical robustness by using a dedicated
linear algebra library
(the Efficient Java Matrix library, \citealp{EJML})
to conduct the computations.
Combined with the use of a Moore-Penrose pseudo inverse, this made our handling 
of the singular or near-singular matrices induced by the presence of missing data
(see \citealp{Bastide2017,Hassler2019})
more numerically stable.
%
%
These developments are presented in detail in Appendix~\ref{app:post_order}.

\subsection{Gradient Computation}\label{sec:gradient:gradient}
When performing statistical inference, either in a maximum likelihood or a
Bayesian framework, having access to the gradient of the likelihood at
relatively cheap computational cost facilitates faster and more accurate
algorithms (see Section~\ref{sec:statistical_inference}).
In this section, we present a novel algorithm to compute the gradient of the
likelihood with respect to any parameter in the general setting presented
in Definition~\ref{def:general_model}.
The algorithm relies on two main ingredients:
(1) as in \citet{Fisher2019}, we express the derivative of the likelihood as
the conditional expectation of a given function of the latent traits
$\latentm{\iNode}$ at the internal nodes $\iNode$, conditional on the observed
traits $\dataMatr$; and
(2) we use a pre-order algorithm inspired from the \enquote{downward}  phase in
\citet{Bastide2017} to compute this expectation in a linear time in
$\nObsOne$.

\paragraph*{Gradient as a Conditional Expectation}
We rely here on \emph{Fisher's identity} \citep{Cappe2005} that
links the gradient of the log-likelihood
$\log \cDensity{\dataMatr}{\allParams}$
of the observed variables to the conditional
expectation of the completed log-likelihood
$\log \cDensity{\latentMatr, \dataMatr}{\allParams}$:

\begin{proposition}[Fisher's Identity; \citealp{Cappe2005}]
    \label{prop:derivative_as_expectation}
    Under broad assumptions, that are verified for Gaussian densities, the following
    identity holds (Equation~10.12 in \citealp{Cappe2005}):
    \begin{equation}\label{eq:fisher_identity}
        \gradient{\allParams} \left[
            \log \cDensity{\dataMatr}{\allParams}
        \right]
        = \Espesq{
        \gradient{\allParams} \left[
            \log \cDensity{\completedMatr_s, \dataMatr}{\allParams}
        \right]
        }{
        \dataMatr
        } ,
    \end{equation}
    where $\completedMatr_{s}$ represents any subset taken from the complete data.
\end{proposition}
%
Applying this identity, we obtain the gradient of the likelihood with
respect to any parameter
\(
\allParams_{\iNode} =
\left(
    \branchActualization{\iNode},
    \branchDisplacement{\iNode},
    \branchVariance{\iNode}
\right)
\)
or
\(
\allParams_{\iObs} =
    \obsVariance{\iObs}
\)
of the model:

\begin{proposition}[Gradient with respect to Branch Parameters]
    \label{prop:derivative_generic}
    Under the general model of Definition~\ref{def:general_model},
    for any observation or node $\iAny$
    ($1 \leq \iAny \leq \nObsOne + \nNodes$),
    the following identity holds:
    \begin{multline}
        \label{eq:derivative_generic}
        \gradient{\allParams_{\iAny}} \left[
            \log \cDensity{\dataMatr}{\allParams_{\iAny}}
        \right]
        =
        \partialDer{\allParams_{\iAny}}{\nodePreMean{\iAny}}
        \cdot
        \nodePrePrecision{\iAny}
        (\fullConditionalMean{\iAny} - \nodePreMean{\iAny})
        \\
        +
        \partialDer{\allParams_{\iAny}}{\mtovh(\nodePrePrecision{\iAny})}
        \cdot
        \frac12 \mtovh
        \left(
            \nodePreVariance{\iAny}
            - (\fullConditionalMean{\iAny} - \nodePreMean{\iAny})
            \transpose{(\fullConditionalMean{\iAny} - \nodePreMean{\iAny})}
            - \fullConditionalVariance{\iAny}
        \right)
    \end{multline}
    with
    $\mtovh$ the symmetric vectorization operation \citep{Magnus1986}; and
    where
    $\nodePreMean{\iAny}$, $\fullConditionalMean{\iAny}$,
    $\nodePreVariance{\iAny}$ and $\fullConditionalVariance{\iAny}$
    are parameters, representing the expectations and
    variances of two Gaussian densities
    (see Equations~\ref{app:eq:pre_above_distribution} and~\ref{app:eq:pre_full_distribution}
    in Appendix~\ref{app:pre_order}),
    that can be computed in one pre-order traversal of the tree.
\end{proposition}

\begin{proof}[Proof of Proposition~\ref{prop:derivative_generic}]
    Let $\iAny$ be an observation or node index with associated trait variable
    $\completedm{\iAny}$ ($1 \leq \iAny \leq \nObsOne + \nNodes$).
    As in \citet{Fisher2019}, we decompose the observations $\dataMatr$ as
    $\dataMatr = (\dataBelowm{\iAny}, \dataAbovem{\iAny})$,
    where
    $\dataBelowm{\iAny}$ denotes the observations that are \enquote{below} node
    $\iAny$, \ie that have $\iAny$ as an ancestor, and
    $\dataAbovem{\iAny}$ denotes the observations that are \enquote{above} node
    $\iAny$, \ie that do not have $\iAny$ as an ancestor.
    %
    The tree conditional structure then induces the decomposition:
    \begin{equation*}
        \cDensity{\completedm{\iAny}, \dataBelowm{\iAny}, \dataAbovem{\iAny}}{\allParams_{\iAny}}
        =
        \cDensity{\dataBelowm{\iAny}}{\completedm{\iAny}}
        \cDensity{\completedm{\iAny}}{\dataAbovem{\iAny}, \allParams_{\iAny}}
        \density{\dataAbovem{\iAny}},
    \end{equation*}
    where only the middle term depends on parameters $\allParams_{\iAny}$ associated
    with the branch ending at node $\iAny$.
    Applying Proposition~\ref{prop:derivative_as_expectation}
    with $\completedMatr_{s} = \completedm{\iAny}$, 
    we get that:
    \begin{equation*}
        \gradient{\allParams_{\iAny}} \left[
            \log \cDensity{\dataMatr}{\allParams_{\iAny}}
        \right]
        =
        \Espesq{
            \gradient{\allParams_{\iAny}} \left[
                \log
                    \cDensity{\completedm{\iAny}}{\dataAbovem{\iAny}, \allParams_{\iAny}}
        \right]
        }{
        \dataMatr
        }.
    \end{equation*}
    Using the pre-order formulas presented in Appendix~\ref{app:pre:conditional_above}
    (Equation~\ref{app:eq:pre_above_distribution}), we can see that
    $\sachant{\completedm{\iAny}}{\dataAbovem{\iAny}}$
    is normally distributed, with expectation
    $\nodePreMean{\iAny}$
    and precision matrix
    $\nodePrePrecision{\iAny}$.
    Applying standard derivation formulas to a log Gaussian density
    (see \eg \citealp{Magnus1986}), we have:
    \begin{multline}
        \label{eq:derivative_partial}
        \gradient{\allParams_{\iAny}} \left[
            \log
            \cDensity{\completedm{\iAny}}{\dataAbovem{\iAny}, \allParams_{\iAny}}
        \right]
        =
        \partialDer{\allParams_{\iAny}}{\nodePreMean{\iAny}}
        \cdot
        \nodePrePrecision{\iAny}
        (\completedm{\iAny} - \nodePreMean{\iAny})
        \\
        +
        \partialDer{\allParams_{\iAny}}{\mtovh(\nodePrePrecision{\iAny})}
        \cdot
        \frac12 \mtovh
        \left(
            \nodePreVariance{\iAny}
            - (\completedm{\iAny} - \nodePreMean{\iAny})
            \transpose{(\completedm{\iAny} - \nodePreMean{\iAny})}
        \right),
    \end{multline}
    %
    %
    From Appendix~\ref{app:pre:conditional_full} (Equation~\ref{app:eq:pre_full_distribution}),
    we know that
    $\sachant{\completedm{\iAny}}{\dataMatr}$
    is normally distributed, with expectation
    $\fullConditionalMean{\iAny}$
    and variance
    $\fullConditionalVariance{\iAny}$.
    Equation~\eqref{eq:derivative_generic} is then obtained by 
    taking the conditional expectation of the above expression~\eqref{eq:derivative_partial}.
\end{proof}

\paragraph*{Chain Rule}
In Equation~\eqref{eq:derivative_generic}, we express the gradient
of the likelihood \wrt any branch parameter using only quantities that
can be computed in two traversals of the tree,
one post-order and one pre-order.
%
This provides the basis for an algorithm to compute the gradient of the likelihood
with respect to any parameter with a linear complexity in $\nObsOne$.
%
Indeed, one only needs to apply the derivation chain rule to:
(1) obtain the gradient of the branch parameters
$\nodePreMean{\iAny}$ and $\nodePrePrecision{\iAny}$ with respect to the
natural parameters of the process at hand;
and
(2) obtain the gradient of the likelihood with respect to parameters shared
between several branches.
We tackle task~(1) using the pre-order formulas in Appendix~\ref{app:der}.
Task~(2) is a straightforward application of the chain rule over all the
levels of the hierarchical model:
\begin{equation}
    \label{eq:chain_rule}
    \partialDer{\allParams}{\cDensity{\dataMatr}{\allParams, \tree}}
    =
    \sum_{\iAny = 1}^{\nObsOne + \nNodes}
    \partialDer{\allParams}{\allParams_{\iAny}}
    \gradient{\allParams_{\iAny}}\left[\cDensity{\dataMatr}{\allParams, \tree}\right].
\end{equation}

\paragraph*{Complexity}
Appendix~\ref{app:post_pre_order} implies that all the moments
$\nodePreMean{\iAny}$, $\fullConditionalMean{\iNode}$,
$\nodePreVariance{\iAny}$ and $\fullConditionalVariance{\iAny}$
($1\leq \iAny \leq \nObsOne + \nNodes$) appearing in Proposition~\ref{prop:derivative_generic}
can be computed in linear time in the number of observations.
Since formulas~\eqref{eq:derivative_generic} and~\eqref{eq:chain_rule} only involve
linear algebra operations in a space of the dimension of the parameters, the
total complexity remains linear in $\nObsOne$.
%
In addition, we note that the sum in Equation~\eqref{eq:chain_rule} does not need to follow
the tree order, as all the quantities are pre-computed, and hence can be parallelized
easily, reducing the actual computation time.



\section{Statistical Inference}\label{sec:statistical_inference}

In the previous section, we showed how both the likelihood and its gradient
with respect to all the parameters of the models
can be efficiently computed simultaneously.
These quantities are the cornerstone of many statistical analyses, and
allow for a wide range of analyses, from maximum likelihood to model selection
(for examples in trait evolution, see \eg \citealp{Clavel2018}).
Here, taking advantage of the comprehensive Bayesian inference framework made available
through the \beast phylogenetics software package, we propose a new Bayesian approach
that relies on the use of an efficient HMC sampler
to perform both posterior inference and marginal likelihood
estimation.

\subsection{Bayesian Phylogenetics and the Total Evidence Approach}\label{sec:total_evidence}
%
The likelihood and gradient algorithms presented below work conditionally on
a phylogenetic tree between sampled species being known without error.
However, the tree is generally a summary statistic resulting from a complex statistical analysis,
and is usually inferred from molecular sequences, which are the actual observed
data.
Many methods in the literature follow a two-step procedure and first infer
the tree from sequence data to then proceed with analyzing the continuous traits,
assuming that the phylogenetic tree is known and fixed
(see \eg \citealp{Harmon2019} for a recent review of such methods).
This approach suffers from two major drawbacks.
First, it ignores the uncertainty in the reconstruction of the tree, which,
given the difficulty of the task, can be substantial, and bias the subsequent
analyses
(see \citealp{Felsenstein2004} for a review, and
\citealp{Bastide2017PhD}, Section~5.1, for an example in a specific case).
Second, this approach does not allow for the complete use of the data available,
as it ignores continuous traits for the tree reconstruction.
Although, when present, sequence information tends to dominate over trait
information \citep{Baele2017DataIntegration}, it is not always available for
all sampled organisms. This is particularly true for ancient fossils, that might
bear some continuous trait data, but, because of the rapid degradation of
DNA molecules, can not be sequenced \citep{Leonardi2017}.
We refer to Section~\ref{sec:weasels} for an example of such a dataset, where
the continuous trait constitutes the only source of information available to reconstruct
the phylogenetic tree.

To overcome these limitations, we use a \emph{total evidence approach}
\citep{Ronquist2012}, that can analyze sequence and trait data jointly, using
all the information available in a Bayesian analysis.
Denote by $\seqData$ the sequence data (that might not be available for
all the sampled species), and by $\seqParams$ all the parameters associated with
the model of sequence evolution and the dating clock model,
(see \eg \citealp{Felsenstein2004} for a review of such models). 
The goal of Bayesian phylogenetics is then to learn about the posterior:
\(
    \cDensity{\allParams, \tree, \seqParams}{\dataMatr, \seqData}.
\)
One crucial assumption that we make is that, conditionally on the phylogenetic
tree $\tree$, the evolution of continuous traits and the sequences are
independent, such that:
%
%
\begin{equation}\label{eq:complete_posterior}
    \begin{aligned}
        \cDensity{\allParams, \tree, \seqParams}{\dataMatr, \seqData}
        & \propto
        \cDensity{\dataMatr, \seqData}{\allParams, \tree, \seqParams}
        \density{\allParams, \tree, \seqParams}
        \\
        & = 
        \cDensity{\dataMatr}{\allParams, \tree}
        \density{\allParams}
        \times
        \cDensity{\seqData}{\tree, \seqParams}
        \density{\tree, \seqParams}.
    \end{aligned}
\end{equation}
%
%
The term $ \cDensity{\seqData}{\tree, \seqParams} \density{\tree, \seqParams}$
has been the focus of an extensive literature,
and benefits from efficient methods readily available in \beast \citep{Suchard2018}.
Thanks to the tools presented in the previous section, we focus here on
$\cDensity{\dataMatr}{\allParams, \tree} \density{\allParams}$
that deals with the study of the distribution of continuous traits
among the population of species.

This conditional independence assumption, although limiting, is essential
from a computational point of view.
It has also proven to be useful and adequate to study a wide range of
biological questions, and it is widely spread in the field of PCMs,
applied to a fixed tree or in a total evidence approach
\citep[see \eg][for reviews]{Felsenstein2004,Harmon2019}.
As mentioned in the Introduction (see Section~\ref{sec:intro:scope}),
some attempts have been made to relax this assumption,
with fixed trees or discrete characters
(see \eg \citealt{Fitzjohn2012,Muller2017} and references therein).
However, the computational burden associated to these methods currently limits
their application to relatively small scale datasets.

\subsection{Hamiltonian Monte Carlo}\label{sec:hmc}

HMC is a powerful MCMC
sampling technique, that exploits the geometrical properties of the density to
be sampled through the use of Hamiltonian dynamics
\citep{Neal2012,Betancourt2017}.
It associates to a vector of parameters of interest $\allParams$,
viewed as the position of a particle in a $\dimParam$-dimensional space,
an auxiliary independent vector $\momentum$ of \enquote{momentum}, that
is typically chosen to be Gaussian:
$\momentum \sim \Normal{\vect{0}_{\dimParam}}{\identityMatrix{\dimParam}}$.
%
The log joint distribution of the parameter $(\allParams, \momentum)$
then represents the \enquote{total energy}
$\totalEnergy(\allParams, \momentum) = \potentialEnergy(\allParams) + \kineticEnergy(\momentum)$
of the particle, with
$\potentialEnergy(\allParams) = - \log \cDensity{\allParams}{\dataMatr, \tree}$
the \enquote{potential energy} set to be equal to the posterior density of interest,
and
$\kineticEnergy(\momentum) = \transpose{\momentum}\momentum / 2$
the \enquote{kinetic energy}.
The total energy is then invariant to the Hamiltonian dynamics:
\begin{equation*}
    \label{eq:hamiltonian_dynamic}
    \left\{
        \begin{aligned}
            \frac{\der \momentum}{\der t}
            &= - \gradient{\allParams} \potentialEnergy(\allParams)
            = \gradient{\allParams} \log \cDensity{\dataMatr}{\allParams, \tree}
            + \gradient{\allParams} \log \density{\allParams}
            \\
            \frac{\der \allParams}{\der t}
            &= + \gradient{\momentum} \kineticEnergy(\momentum)
            = \momentum.
        \end{aligned}
    \right.
\end{equation*}

The HMC sampling scheme exploits this property using proposals that
approximately follow these dynamics, as discretized by an appropriate numerical
scheme such as the leapfrog. 
This allows for a proposal that can have a small correlation with the current state,
while still having a high probability of acceptation \citep{Neal2012}.
Such an HMC sampler has already proven very successful in a phylogenetics
context \citep{Fisher2019,Ji2019}.
Our efficient and general algorithm for likelihood and gradient computation,
presented in Section~\ref{sec:gradient}, makes it now applicable to the wide variety
of models covered by Definition~\ref{def:general_model}.

\subsection{Confronting Constrained Natural Parameters}\label{sec:constrains}
Some of the parameters of the models, such as the variance of a \BMa or an \OUa,
live in constrained spaces with a non-trivial structure, that need to be
sampled adequately.
One standard way to deal with this structure \citep[see e.g.][]{Stan2017} is to map the
constrained natural parameters $\allParams$ to a vector of independent,
unconstrained parameters $\unconstrParams$ through a smooth transformation
$\transform$.
The density in the new, unconstrained space is then linked to the density in the
constrained space by a simple multiplication with the determinant of the Jacobian
matrix of the transformation \citep[see e.g.][]{LeGall2006},
such that:
%
\begin{equation*}
    \gradient{\unconstrParams}
    \log \density{\unconstrParams}
    =
    \partialDer{\unconstrParams}{\invTransform(\unconstrParams)}
    \gradient{\allParams} \log \density{\invTransform(\unconstrParams)}
    +
    \gradient{\unconstrParams} \log \Det{\partialDer{\unconstrParams}{\invTransform(\unconstrParams)}}.
\end{equation*}
This formula allows us to easily update the constrained parameters $\allParams$ from
movements in the unconstrained space of $\unconstrParams$.
We present the transformations and the associated priors used here
in detail in Appendix~\ref{app:transformations}.
In particular, we show that sampling the space of correlation matrices amounts to
sampling vectors in the half-euclidean sphere, which provides an original
and simple representation of the
classical LKJ transformation \citep{Lewandowski2009}.


\subsection{Model Selection and Marginal Likelihood Estimation}\label{sec:MLE}
We described above a general framework to efficiently infer the parameters of
a wide class of evolutionary models. 
When analyzing a dataset, a question that naturally arises is the choice
of the most suited model to interpret the evolutionary patterns 
in a specific problem.

\paragraph*{Bayesian Model Selection}
In a Bayesian setting, one natural way to compare a collection of models
$(\model_\iModel)_{1 \leq \iModel \leq \nModel}$ is to compute their
marginal likelihoods
$\cDensity{\dataMatr}{\model_\iModel}$
(see \eg \citealt{Oaks2019} for an introduction in a phylogenetic context).
The marginal likelihood, that integrates all the parameters against the prior,
takes the model complexity into account by design, \enquote{penalizing} complex
models, that otherwise mechanically have a higher likelihood.
Marginal likelihoods allow computing Bayes factors, which have a natural
comparison scale \citep{Jeffreys1935,Kass1995}.
However, because of the need to integrate over the potentially very large space of parameters,
this quantity is typically hard to compute, and  approximations
are required. 

\paragraph*{Generalized Stepping-Stone Sampling (GSS)}
In a phylogenetic context, 
the GSS approach
\citep{Fan2011}
has been successfully used to approximate
marginal likelihoods
\citep{Baele2016,Fourment2019}.
For a given model $\model$, it relies on the construction and sampling of a
path between the unnormalized posterior and a \enquote{working} prior distribution
$\workingPrior{\allParams}{\model}$:
\begin{equation}
    \label{eq:path_likelihood}
    \pathLikelihood{\pathParameter}{\allParams}
    =
    \left[
        \cDensity{\dataMatr}{\allParams, \tree, \model}
        \cDensity{\allParams}{\model}
    \right]^\pathParameter
    \left[
        \workingPrior{\allParams}{\model}
    \right]^{1 - \pathParameter}.
\end{equation}
When $\pathParameter = 1$, the path likelihood is proportional to the
classical posterior sampled in a standard MCMC analysis, while when
$\pathParameter = 0$, it reduces to the working prior
$\workingPrior{\allParams}{\model}$.
This working prior is chosen to match the empirical moments from a sample of the
posterior distribution, ensuring a less vague distribution that is closer to the posterior,
hence inducing a more accurate approximation for a reduced computational effort
than the standard stepping-stone sampling procedure \citep{Xie2011,Fan2011}.
As in \citet{Baele2016}, we adopt a kernel density estimator (KDE) for each
parameter, using a normal kernel, that is log-transformed for positive
parameters \citep[see e.g.][]{Jones2018}.

\paragraph*{Sampling the path with HMC}
The GSS estimation implies sampling from the path likelihood
$\pathLikelihood{\pathParameter}{\allParams}$
for a sequence of $\pathParameter$.
\citet{Xie2011} and \citet{Baele2016} show that choosing the path parameter as
evenly spaced quantiles of a Beta distribution with shape $0.3$ and
scale $1.0$, which allows for sampling more intensely regions where
$\pathParameter$ is small, and hence where the path likelihood is changing the most
rapidly, yields the best performance.
This sampling is usually done through a standard MCMC procedure.
Here, we use the efficient HMC approach presented in Section~\ref{sec:hmc},
which implies taking the gradient
of the log path likelihood~\eqref{eq:path_likelihood}:
\begin{equation*}
        \gradient{\allParams} \log \pathLikelihood{\pathParameter}{\allParams}
        =
        \pathParameter
        \gradient{\allParams}
        \log
        \left[
            \cDensity{\dataMatr}{\allParams, \tree, \model}
            \cDensity{\allParams}{\model}
        \right]
        +
        (1 - \pathParameter)
        \gradient{\allParams}
        \log \workingPrior{\allParams}{\model}.
\end{equation*}
%
%
This gradient involves a term proportional to the posterior that
we already dealt with in the HMC inference, and the working distribution,
that, as a product of independent KDE estimations, is straightforward to compute.
This makes it possible to use the efficient HMC sampling scheme in the
GSS marginal likelihood estimation framework already implemented and well
established in \beast \citep{Baele2016,Fourment2019}.

\section{Applications and Simulations}\label{sec:bio}

\subsection{Assumptions and Practical Implementation}
%
%
We showcase the usefulness of our inference framework using two recently published datasets,
one in ecology, and one in virology.
This led us to implement a subset from all the models made accessible by the method. 
Specifically, we limit the evolutionary model
to the \BMa and the \OUa with diagonal strength of selection $\strengthm$,
and 
a shared residual variance for all the measures,
possibly scaled by the tip heights
($\obsVariance{\iObs} = \obsVariance{}$
or
$\obsVariance{\iObs} = \nodeHeight_{\iObs}\obsVariance{}$
for any observation $\iObs$, $1 \leq \iObs \leq \nObsOne$).

\subsection{Phylogenetic Heritability}\label{sec:heritability}
The concept of phylogenetic heritability has been defined in the field of \PCMs to study
the relative importance of the evolution and observation models in the total
measured variance at the tip of the tree \citep{Lynch1991,Housworth2004}.
It is linked to the notion of phylogenetic signal \citep{Pagel1999}, and its
use has recently received considerable attention in studying infection traits in the field of virology
\citep{Alizon2010,Leventhal2016,Mitov2018}.
We introduce here a general definition of the phylogenetic heritability that
extends this notion to our general framework.
%
It relies on the expectation of the
population variances computed at the latent tip level
(for the
$\latentMatrTip = (\latentm{\iNode})_{1 \leq \iNode \leq \ntaxa}$),
and at the observation level
(for the
$\dataMatr = (\datam{\iObs})_{1 \leq \iObs \leq \nObsOne}$):
\begin{align}
    \popVar(\latentMatrTip)
    & =
    \Espe{
        \frac{1}{\ntaxa}
        \transpose{\left[\latentMatrTip - \expProcess[\latentMatrTip]\right]}
        \left[\latentMatrTip - \expProcess[\latentMatrTip]\right]
    }
    =
    \frac{1}{\ntaxa}
    \sum_{\iNode = 1}^{\ntaxa}
    \varProcess[\latentm{\iNode}],
    \label{eq:popVarLatent}
    \\
    \popVar(\dataMatr)
    & =
    \Espe{
        \frac{1}{\nObsOne}
        \left[\dataMatr - \expProcess[\dataMatr]\right]^T
        \left[\dataMatr - \expProcess[\dataMatr]\right]
    }
     =
    \frac{1}{\nObsOne}
    \sum_{\iObs = 1}^{\nObsOne}
    \varProcess[\datam{\iObs}],
    \label{eq:popVarObs}
\end{align}
where the expectation and variance are taken following the process
of evolution and observation defined in Definition~\ref{def:general_model} with parameters
$\allParams$.
These quantities have closed-form expressions for all the models considered here,
see \eg \citet{Clavel2015} for the general \OUa case.
The \enquote{heritability matrix} $\heritabilityMatrix$ can then be defined as:
\begin{equation}
    \label{eq:heritability}
    \heritability_{\iAny\iAnyBis} =
    \frac{
        \popVarOne_{\iAny\iAnyBis}(\latentMatrTip)
    }{
        \sqrt{
            \popVarOne_{\iAny\iAny}(\dataMatr)
            \popVarOne_{\iAnyBis\iAnyBis}(\dataMatr)
        }
    }.
\end{equation}
In the case of a standard univariate trait on an ultrametric tree with only one observation per tip,
this formula coincides with the classical definition found in
the literature \citep[see \eg][]{Mitov2018}.

\paragraph*{Population versus Empirical Variance}
    In equation~\eqref{eq:heritability}, we use the population variance,
    instead of the empirical one used for instance in
    \citet{Blanquart2017,Hassler2019}.
    We argue in Appendix~\ref{app:heritability} that, when the process
    is not a simple \BMa, the population variance is more appropriate,
    as the empirical variance might be impaired by confounding
    inter-group effects if the tips are expected to have different means
    under the trait evolution model, which is for instance the case for an \OUa
    model on a non-ultrametric tree.


\subsection{Morphological Evolution in the \weasels Superfamily}\label{sec:weasels}
We illustrate the total evidence approach
to study the evolution of some morphological features in the
\weasels superfamilly (including weasels and allies).



\subsubsection{Dataset and Analyses}\label{sec:weasels:dataset}

\begin{figure}[!ht]
\centering
\begin{knitrout}
\definecolor{shadecolor}{rgb}{0.969, 0.969, 0.969}
\input{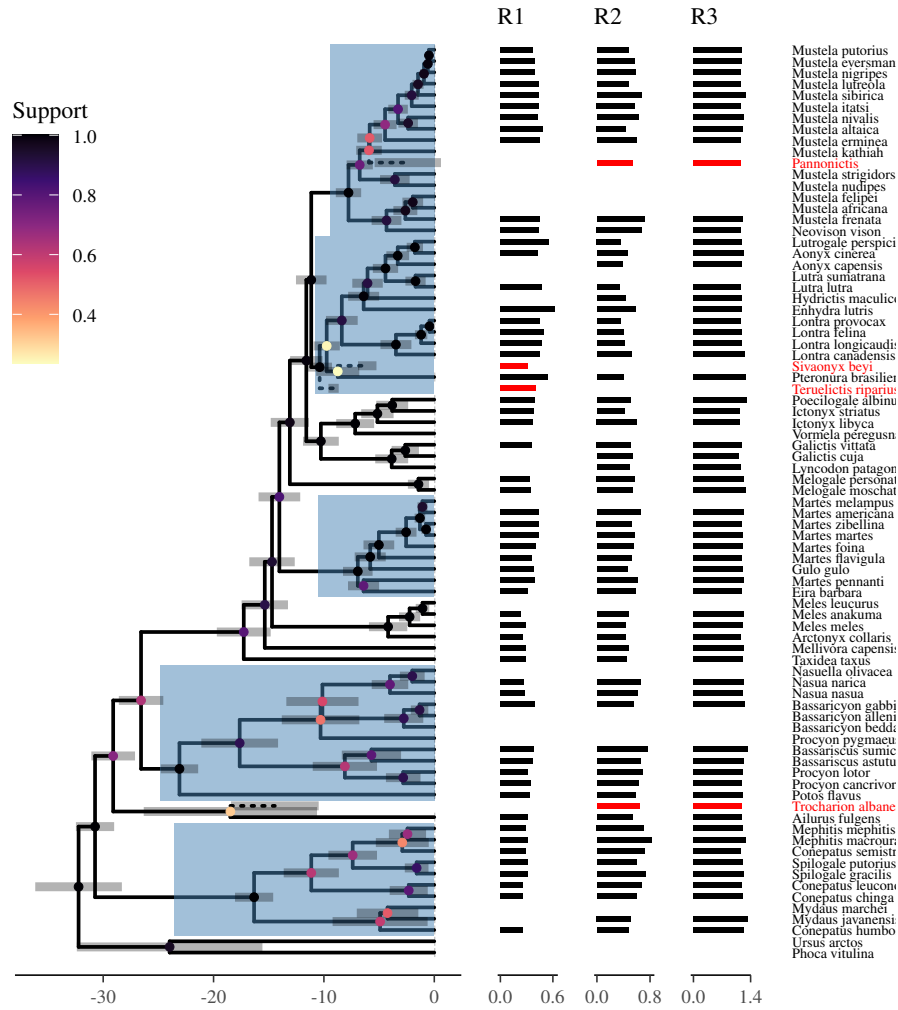}

\end{knitrout}
     \caption{
     Maximum clade credibility tree with continuous traits plotted at the tips.
     The tree is reconstructed from both the genetic data and the trait data,
     using an \OUa model with a diagonal selection strength.
     The time scale for the tree is in million of years, with $0$ indicating the present.
     Clades highlighted in blue are constrained to be monophyletic.
     Grey bars at nodes and fossils show the dating uncertainty.
     Node colors indicate the posterior support for each clade.
     Fossils are dotted, and highlighted in red.
     }
     \label{fig:weasels:tree_traits}
\end{figure}

\paragraph*{Dataset}
We reanalyze the dataset published by \citet{Schnitzler2017}, containing
$81$ taxa, including
$4$ fossils, and
$2$ outgroup species.
Aligned sequence data for all $77$ extant taxa are
available (containing 22 nuclear and 5 mitochondrial genes).
Three morphological traits are measured on
$65$ species, including fossils, with missing
data for some taxa (see Figure~\ref{fig:weasels:tree_traits}).
%
They are carnivorian ecometric traits, defined as meaningful ratios of
osteological measurements, and denoted by R1-3. 
%
Note that our Bayesian framework can readily handle this heterogeneous dataset,
and jointly analyse sequence and continuous traits with missing data on both.

\paragraph*{Questions}
We aim to address the following two questions.
First, does total evidence
(see Section~\ref{sec:total_evidence}) allow for better placement of the fossils?
Given that we can take trait data into account while performing
phylogenetic inference,
we might expect that this extra information leads to better
fossil placement estimates
compared with \citealt{Schnitzler2017}, who only considered sequence data and use monophyly constraints. 
%
Second, which model of trait evolution is most suited to explain the observed
trait distribution, and does this conclusion change when we include or set aside
fossil data, as suggested by \citet{Schnitzler2017}?

\paragraph*{Sequence Evolution and Dating}
We use the same sequence evolution model as \citet{Schnitzler2017} for all
27 partitions of the dataset,
with estimated base frequencies
and site rate heterogeneity modeled using a discretized gamma distribution with 6 rate categories;
an uncorrelated relaxed clock with an underlying gamma distribution;
and an exponential growth coalescent tree prior
\citep{Hasegawa1985,Tavare1986,Yang1994,Drummond2006}.
Following \citet{Schnitzler2017}, we constrain $5$ clades to be monophyletic,
and each fossil is \textit{a priori} assigned to one of these clades, except for
\taxon{Trocharion albanense}, which remains unconstrained
(see Figure~\ref{fig:weasels:tree_traits}).
We assume a normal prior on the time of the most recent common ancestor for each of those clades,
with means as in \citet{Schnitzler2017}, and standard deviation 1.
We assume a uniform prior on fossil dates, with maximum
ranges taken from \citet{Law2018} (Table S5),
except for \taxon{Teruelictis riparius}, for which dates were extracted
from the Paleobiology Database, relying on \citet{Salesa2013}.
Note that these assumptions differ slightly from \citet{Schnitzler2017},
 who provide insufficient information to reproduce their exact pipeline.

\paragraph*{Total Evidence Phylogenetic Inference}
%
%
We conduct phylogenetic inference using 3 different data integration scenarios:
no model of trait evolution (\ie the continuous traits are not used);
a \BMa model; and an \OUa with diagonal selection strength.  The latter two combine both sequence and trait evolution.
%
We run each analysis for 100 million iterations,
sample every 1000 steps,
and discard the first 10\% as burn-in.
%
We specified vague priors on the parameters of the continuous processes while
respecting biological constraints, as described in Appendix~\ref{app:transformations}.
%
The maximum clade credibility (MCC) tree is used to represent the evolutionary history.

\paragraph*{Fossil Placement Analysis}
We assess the uncertainty of fossil placement using a method introduced by
\citet{Klopfstein2019}.
Given a sample of trees from the posterior,
as the backbone tree is well resolved (see below),
the method amounts to computing, for each branch of the MCC tree,
the frequency a given fossil attaches to that branch.
We measure frequency vector concentration using
entropy; a fossil that is well resolved will be distributed over a small
number of branches with high frequency, and hence has a low entropy.

\paragraph*{Model Comparison}
As in \citet{Schnitzler2017}, we also conduct several
model comparisons, conditioning on a tree fixed to the MCC tree from one of the
previous analyses, with or without fossil species.
On each tree, we test several hypotheses about the nature of trait evolution,
by comparing (log) marginal likelihood estimates for the various models.
Models tested in this section are the BM and OU models,
but also the \enquote{trend} model, that is a BM with an added homogeneous
deterministic drift \citep{Hansen1996,Gill2016}.
See Supplementary Figure~\ref{app:fig:weasels:MLE} for the list of
all hypotheses tested.
%
We run each analysis for 100\,000 iterations,
sample every 10 steps,
and discard the first 10\% as burn-in.
50 steps of 1000 iterations each
are explored for the GSS estimation of the (log) marginal likelihood of each model.

\paragraph*{Analysis and Representation of the Results}
We use \beast, TreeAnnotator and Tracer to conduct the analyses
\citep{Suchard2018,Rambaut2018}.
Trees are imported into \code{R} and plotted using
\code{treeio} \citep{Wang2019},
\code{tidytree}
and \code{ggtree} \citep{Yu2017,Yu2018}.

\subsubsection{Results}\label{sec:weasels:results}
\paragraph*{Phylogenetic Inference}
Consistent with \citet{Schnitzler2017}, we estimate a well-resolved backbone tree,
with uncertainty mostly at the genus level, particularly in the
\taxon{Mephitidae} and \taxon{Procyonidae} families
(see Figure~\ref{fig:weasels:tree_traits}).
Including trait information does not dramatically change the inferred relationships
between extant species, confirming that molecular data
are generally more informative than trait data \citep{Baele2017DataIntegration}.

\paragraph*{Fossil Placement}
Taking into account trait information reduces entropy scores for each of
the fossils, with the \OUa model having the lowest entropy for $3/4$ of the  fossils (see Table~\ref{table:weasel:entropy} and
Supplementary Figure~\ref{app:fig:applis:weasels:fossils}).
Continuous trait measurements bear different amounts of information for each
fossil, which leads to different entropy score behavior.
%
The fossil \taxon{Pannonictis} is evenly distributed over all the branches of the
clade (\taxon{Mustelinae}) where it is assigned.
It has a high entropy that does not decrease much when traits are taken into account.
This result is not surprising given that the traits vary little among all the species
of this clade (see Figure~\ref{fig:weasels:tree_traits}),
and hence yield little information with respect to fossil placement.
Fossils \taxon{Sivaonyx beyi} and \taxon{Teruelictis riparius} are both assigned
to the same clade (\taxon{Lutrinae}).
Figure~\ref{fig:weasels:tree_traits} illustrates that their R1 trait is relatively lower
compared to other members of the clade.
Taking this trait into account is thus informative, and entropy decreases,
with the fossil estimated to lie at the root of the tree
(see Supplementary Figure~\ref{app:fig:applis:weasels:fossils}).
Finally, the species \taxon{Trocharion albanense} is not assigned to any clade.
Taking  traits into account concentrates this fossil as a sister lineage either to
\taxon{Ailurus fulgens}, or to the whole \taxon{Mephitidae} clade, which,
in the assumed time range, have the most similar traits.

\begin{table}[!ht]
\begin{center}

\begin{tabular}{lrrrr}
\toprule
\em{ } & \em{Pannonictis} & \em{Sivaonyx beyi} & \em{Teruelictis riparius} & \em{Trocharion albanense}\\
\midrule
No Traits & 2.84 & 1.80 & 1.24 & 2.28\\
BM & \textbf{2.81} & 1.22 & 1.22 & 2.01\\
OU & 2.92 & \textbf{1.20} & \textbf{1.10} & \textbf{1.86}\\
\bottomrule
\end{tabular}

\caption{Entropy of the fossil position for each fossil and inference method
(over 1000 trees from the posterior).
Entropy should decrease if the fossils are better resolved.
Maximal entropy (no information) is $5.02$.
OU models appear to reduce entropy for three fossils out of four.}
\label{table:weasel:entropy}
\end{center}
\end{table}

\paragraph*{Model Comparisons}
We find that the favored model, for all tested trees, with or without fossils,
is a simple \BMa for the first trait (R1), and an \OUa with diagonal
selection strength but full correlation for the two other traits (R2 and R3),
with R1 evolving independently from R2 and R3
(see Supplementary Figure~\ref{app:fig:weasels:MLE}).
The parameter estimates are consistent with those from \citet{Schnitzler2017}
(see Supplementary Figure~\ref{app:fig:applis:weasels:BM_OU_OU_estimates}).
\citet{Schnitzler2017} fitted the three traits independently and
used a simple penalized likelihood approach
(using the Akaike Information Criterion, \citealt{Akaike1974}) to demonstrate that a
\enquote{trend} model is favored to the simple \BMa model for R1 when
fossils are included.
In contrast, our method is robust to the addition of these fossils that, given
the missing data, amounts to the addition of two data points in the analysis
(see Figure~\ref{fig:weasels:tree_traits}).
The selected model has a log Bayes factor
of at least $1.5$ compared to the second
best fitting model in all the scenarios, providing \enquote{substantial evidence} \citep{Kass1995} against the simple \BMa model.



\subsection{Virulence Heritability in Human Immunodeficiency Viruses (HIV)}\label{sec:hiv}
New challenges for \PCMs have recently emerged in infectious disease research,
more specifically on the extent to which virulence is a heritable trait in
HIV.
Here, we employ our new modeling framework to perform
a fine-grained analysis to gain insight into this problem.


\subsubsection{Dataset and Analyses}\label{sec:hiv:dataset}

\begin{figure}[!ht]
\centering
\begin{knitrout}
\definecolor{shadecolor}{rgb}{0.969, 0.969, 0.969}
\input{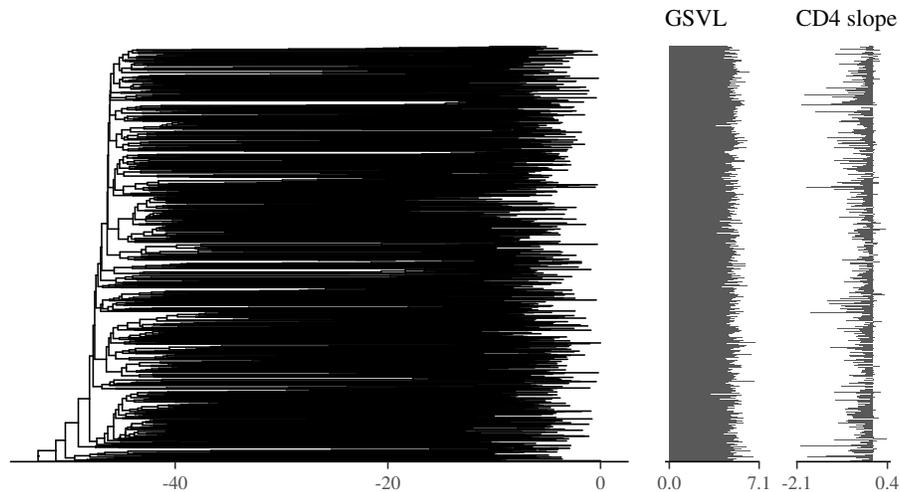}

\end{knitrout}
\caption{
HIV-1 dataset from \citet{Blanquart2017} and \citet{Hassler2019}.
A maximum-likelihood tree estimate depicts the phylogenetic relationships for all HIV-1 subtype B viruses from MSM patients.
The time scale is in calendar years (0 being the present, putting the root at around 1960).
The two traits are the GSVL and the CD4 count slope decrease.
}
\label{fig:hiv:tree_traits}
\end{figure}

\paragraph*{Dataset}
We revisit the most comprehensive dataset on HIV-1 heritability published in
\citet{Blanquart2017} and further analysed in \citet{Hassler2019}.
We focus on subtype B and the measurements available for male subjects who have
sex with men (MSM), which comprises a dataset of $1171$ viral samples.
Two traits associated with HIV virulence \citep{Alizon2010, Blanquart2017}
are measured for each sample:
(i) the \enquote{gold standard viral load} (GSVL) that is a standardized measure of
the viral load, taken on a single sample between 6 and 24 months after infection
and before initiation of antiretroviral therapy;
and (ii) the CD4 cell count slope decline
(see Figure~\ref{fig:hiv:tree_traits}).
%
A dated maximum likelihood phylogeny for this dataset has recently been
presented by \citet{Hassler2019}.
Following a similar methodology, we use it as a fixed tree in our analyses,
which focus on continuous trait model selection and heritability
estimations.

\paragraph*{Questions}
Disease progression varies greatly among patients.
Similar to other rapidly evolving human pathogens, it is
challenging to determine to what extent this variance is due to the host or virulence of the
viral genotype.
The pioneering application of \PCMs by \citet{Alizon2010} to estimate the
heritability of HIV virulence using set-point viral load (spVL), 
has stimulated the generation of comprehensive
data sets \citep{Blanquart2017}, but also led to a discussion concerning the
underlying models \citep{Mitov2018,Bertels2018}.
Depending on the method and datasets used, the heritability of the spVL has
been quite controversial \citep{Leventhal2016}, with estimates ranging from
about $50\%$ \citep{Alizon2010}, to around $30\%$ \citep{Vrancken2015} and to
as low as about $6\%$ \citep{Hodcroft2014}.
We explore here the fit of several models of trait evolution
and individual variation (Equations~\ref{eq:general_model_tree} and~\ref{eq:general_model_obs})
to study their impact on heritability estimation and other parameters of interest.
As the virus-host interactions are a major source of trait variation, we expect
the individual variation layer to be particularly important in these models.

\paragraph*{Trait Evolution Models}
We use three different evolution models for the two traits on the tree:
a multivariate \BMa, a multivariate \OUa (with diagonal selection strength),
and a mixed multivariate \enquote{\OUBMa} model, that has an \OUa model on the
GSVL, and a \BMa model on the CD4 slope, the two still being correlated.
%
This last model illustrates the flexibility of our framework in model specification.
It is motivated by the results presented in \citet{Blanquart2017}, and by the
data distribution (see Figure~\ref{fig:hiv:tree_traits}), with the CD4 slope
being much more spread out than the GSVL
(with respective quartile coefficients of dispersion of
$-0.56$ and
$0.09$).

\paragraph*{Individual Variation Models}
One major driver of diversity for the two traits is the interaction of the
virus with its host, that is independent from the viral phylogeny.
This individual variation is captured through our observation model
layer. 
We take this variation to be either identically distributed, or scaled by the
tip heights, with or without trait correlation.
Scaling the independent noise by the tip heights is an empirical model inspired
from Pagel's $\lambda$ model \citep{Pagel1999} and is well suited for an
environmental contribution that increases linearly with sampling time \citep{Leventhal2016}.

\paragraph*{Model Comparison and Model Fit}
%
As in the previous example using a fixed tree,
and using the same computational tools,
priors (see Appendix~\ref{app:transformations}),
and chain settings,
we estimate (log) marginal likelihoods for all the models under study.
%
%

\subsubsection{Results}\label{sec:hiv:results}

\begin{figure}[!ht]
\centering
\begin{knitrout}
\definecolor{shadecolor}{rgb}{0.969, 0.969, 0.969}
\input{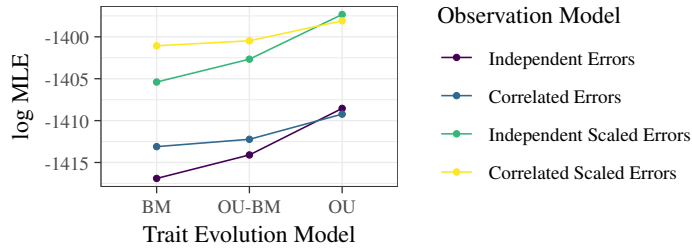}

\end{knitrout}
\caption{Log Marginal Likelihood Estimation (log MLE)
for three different trait evolution models
(BM on both traits, OU on GSVL and BM on CD4, and OU on both traits)
and four observation models, either independent or correlated,
and scaled by tip heights or not. 
The best fitting model is the OU with independent
scaled errors, with a log Bayes factor of
$0.77$
compared to the second best model, the more complex
OU with correlated scaled errors.
%
}
\label{fig:hiv:mle}
\end{figure}

%

\paragraph*{Model Selection}
The model favored according to the log marginal likelihood estimation is the OU model
on both traits, with an independent, scaled observation matrix
(see Figure~\ref{fig:hiv:mle}).
Compared to the OU-BM model with the same error structure, it has a log Bayes factor
support of
$3.14$,
indicating a strong support for the more complex OU model \citep{Kass1995}.
In general, the scaled error models appear to be much better supported than the
non-scaled ones.

\paragraph*{Heritability}
%
Under the best fitted model, the heritability is estimated to be, respectively,
$0.3$ ($95\%$ highest posterior density interval: $[0.16, 0.45]$)
for the GSVL, and
$0.36$ ($95\%$ HPDI $[0.15, 0.65]$)
for the CD4 slope.
This is in line with the selection strength estimates,
with
a phylogenetic half-life (in percentage of the tree height) of
$0.27$ ($95\%$ HPDI $[0.1, 0.61]$)
for the GSVL, and
$0.12$ ($95\%$ HPDI $[0.05, 0.26]$)
for the CD4 slope.
The CD4 slope has a higher selection strength, so that the phylogenetic model
allows for more individual variation (see \eg \citealp{Bastide2017}), and hence
the heritability, which is the relative importance of this phylogenetic model
in the total variation, is mechanically higher.
%
%
See Appendix~\ref{app:appli:sec:hiv} for a complete presentation of the results.


\paragraph*{Discussion}
The results presented above are surprising on two accounts.
First, the heritability results under the best supported OU model are different from
the ones reported in \citet{Blanquart2017}, who find a heritability that is
larger for the GSVL than for the CD4, with estimates of
$0.31$ ($95\%$ HPDI $[0.15, 0.43]$)
for the GSVL, and
$0.1$ ($95\%$ HPDI $[0.01, 0.27]$)
for the CD4 slope.
We note however that the selection strength parameter in \citet{Blanquart2017} is poorly estimated,
with a wide confidence interval that abuts against the lower and upper
limits that were arbitrarily imposed: the half-life is estimated to
$0.08$ ($95\%$ HPDI $[0.07, 0.29]$)
for the GSVL, and
$7.3$ ($95\%$ HPDI $[0.0693, \ensuremath{6.3\times 10^{5}}]$)
for the CD4 slope.
In contrast, the priors we set,
that reflect biologically reasonable
assumptions, might help us regularize the estimation of this notoriously hard to infer
parameter (see \eg \citealp{Uyeda2014,Bastide2016}). 

Second, the log MLE favors a model with an OU on the CD4 slope, while
\citet{Blanquart2017} favor a BM on this trait, which seems more reasonable from
a biological point of view.
It is interesting to note that when we use
the less supported OU-BM model,
we find estimates of the heritability to be more in line with the literature,
while still on the upper range:
$0.35$ ($95\%$ HPDI $[0.22, 0.51]$)
for the GSVL, and
$0.21$ ($95\%$ HPDI $[0.12, 0.32]$)
for the CD4 slope. 
This might raise some concerns on the ability of the MLE model selection procedure
to select for the best suited model in this setting. 
%
We address both concerns with a simulation study in the next section.



\subsection{Exploration of Model Selection in a Heritability Estimation Context}\label{sec:simus}
In the previous section, we estimated the heritability to be lower than $50\%$.
This means that the phylogenetic model \eqref{eq:general_model_tree} accounts for at most
half of the total variation observed in the dataset, while individual variation
\eqref{eq:general_model_obs}
takes up the remaining part.
In the absence of strong phylogenetic signal in the trait, it might be challenging to
uncover the true underlying trait evolution model.
Using a simulation scheme that is inspired by the empirical dataset, we explore the
limits of the MLE model selection procedure in this setting.


\subsubsection{Setting}\label{sec:simus:setting}

\paragraph*{Base Evolutionary Scenario}
We used the same fixed
HIV tree as in the previous section,
normalized so that it had a maximum root to tip height of one. 
We then simulated a bivariate trait according to a multivariate correlated \OUBMa model.
By analogy, the two traits are named GSVL and CD4, and the parameters of the
process were taken to be similar to the ones inferred in the previous section.
For the \enquote{GSVL} trait, we took
a half-life of $0.3\%$ of the tree height,
a stationary variance $\variance/(2\strength)$ of $0.1$,
and an optimal value (equal to the root conditional value) of $1$.  
For the \enquote{CD4} trait, we took
a selection strength of $0$ (\BMa),
a variance of $0.01$,
and a root conditional value of $-0.1$.
The correlation between the two traits was set to $-0.9$. 

\paragraph*{Independent Variations}
On top of this evolutionary model, we added independent individual variations
at each tip.
Under the base scenario, the variance of this extra noise was taken to be equal
to the variance of the process
($0.1$ on the GSVL and $0.01$ on the CD4).
This reflects a heritability of about $50\%$
(computed to be of
$49.4\%$ for the GSVL and
$45.9\%$ for the CD4).
%
We then multiplied this noise variance by a factor $\factorNoise$ varying
between $0.1$ and $3$,
leading to a maximal heritability of
$0.91$ and
$0.89$;
and a minimal heritability of
$0.25$ and
$0.22$
for both traits, respectively.
The estimates for the empirical data set imply a
scenario where this factor $\factorNoise$ is high.

\paragraph*{Simulation and Inference}
We simulated the \OUBMa process using the \code{R} package \code{PhylogeneticEM}
\citep{Bastide2016,Bastide2017}.
Each scenario was repeated $50$ times.
On each of these datasets, we performed an analysis similar to the previous section,
fitting three evolution models (the \BMa, the true \OUBMa and the \OUa),
with independent identically distributed individual variations at each tip.

\paragraph*{Questions}
We analysed the results to address the two following questions:
(i) does the MLE model selection procedure recover the true generative model?
and (ii) to what extent is heritability correctly estimated?

\subsubsection{Results}\label{sec:simus:results}

\paragraph*{Model Selection}
The proportion of each model being selected over the $50$
repetitions is presented in Table~\ref{table:simus:model_selection}.
When the individual variation variance is low or equal to the evolutionary
variance, the correct \OUBMa model is selected in all or most of the cases.
When this noise increases however, the proportion drops considerably, with the
correct model being selected less than $75\%$ of the cases when $\factorNoise = 3$.

\begin{table}[!ht]
\begin{center}

\begin{tabular}{rrrrrrrr}
\toprule
\em{Noise Level} & \em{0.1} & \em{0.25} & \em{0.5} & \em{0.75} & \em{1} & \em{2} & \em{3}\\
\midrule
BM & 0.00 & 0.00 & 0.00 & 0.02 & 0.02 & 0.00 & 0.04\\
\textbf{OU-BM} & \textbf{0.96} & \textbf{0.98} & \textbf{1.00} & \textbf{0.88} & \textbf{0.94} & \textbf{0.78} & \textbf{0.68}\\
OU & 0.04 & 0.02 & 0.00 & 0.10 & 0.04 & 0.22 & 0.28\\
\bottomrule
\end{tabular}

     \caption{Proportion of times each model is selected over the
     $50$ replicates.
     When the independent noise is small compared to the phylogenetic signal,
     the right model (OU-BM, bold) is almost always selected.
     When the independent noise become overwhelming, the phylogenetic signal
     is lost, and the model selection is less efficient.
     The HIV example explored in Section~\ref{sec:hiv} falls under the latter category,
     with a high noise (or low heritability), which might explain the somewhat
     unexpected results provided by the model selection in the HIV example.
     }
     \label{table:simus:model_selection}
\end{center}
\end{table}

\begin{figure}[!ht]
\centering
\begin{knitrout}
\definecolor{shadecolor}{rgb}{0.969, 0.969, 0.969}
\input{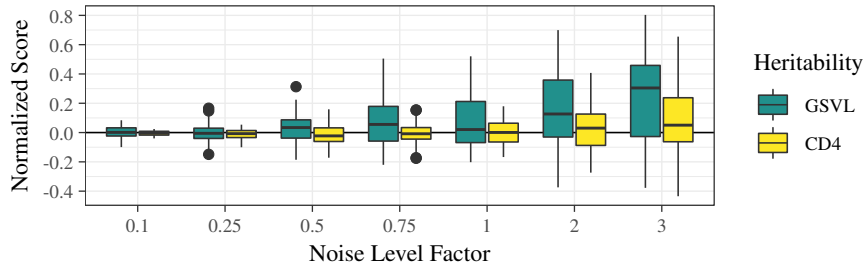}

\end{knitrout}
     \caption{Estimation of the heritability when inferred with the
              correct (OU-BM) model for various levels of noise.
              The values are normalized by the true value used in the simulation,
              so they should converge to 0 (horizontal line).
              The distribution of the normalized individual estimates
              over the $50$ repetitions
              are summarized by a box-plot within a violin plot.
              When the noise level increases, the estimates are more variable,
              and the heritability tends to be over-estimated.}
     \label{fig:applis:simus:OUBM_heritability_plot}
\end{figure}

\paragraph*{Estimation of the Heritability}
In Figure~\ref{fig:applis:simus:OUBM_heritability_plot}, we show the normalized
estimated heritability when we use the correct \OUBMa model.
When the noise level factor $\factorNoise$ increases, the variance of the estimates
over the $50$ repetitions increases substantially,
with higher levels of noise leading to an over-estimation of the heritability for
both traits.
Note that the estimate is more variable for the trait under the \OUa model (GSVL).
This is consistent with the fact that it relies on the estimate of the
strength of selection, which is difficult to infer
(see Supplementary Figure~\ref{app:fig:applis:simus:all_estimates_plot}).
The empirical coverage level of the $95\%$ HPD interval remains however
relatively high, never dropping below $80\%$
(see Supplementary Figure~\ref{app:fig:applis:simus:all_coverage_plot}).

\paragraph*{Selection Strength}
When the \OUa model is incorrectly selected over the \OUBMa (which can happen almost
one fourth of the time under high levels of noise), then the selection strength
on the CD4 trait, that is simulated without selection, is estimated to be
generally higher than the selection strength of the GSVL
(see Supplementary Figure~\ref{app:fig:applis:simus:all_estimates_plot}).

\paragraph*{Discussion and Caveat}
When the noise level is high, we observe three main artifacts in the estimation:
(i) the \OUa model is often wrongly selected over the \OUBMa;
(ii) the heritability tends to be over-estimated;
(iii) when the \OUa model is selected, the selection strength for the selection-free
trait is estimated to be larger than for the trait under selection.
These three artifacts are precisely the points that raised questions from a
biological point of view in the previous section.
This simulation study therefore illustrates that complex models of trait evolution
such as the \OUa, that have recently been advocated in the context of heritability studies
\citep{Mitov2018,Bertels2018},
should be treated with caution when applied to a dataset that is burdened with high levels of noise.
%
In our case, we simulated the data according to the exact same model used for the
statistical inference (an \OUBMa with residual variance).
The performance of the model could be further affected by other sources of variation in a real-world example that are
not accounted for in our framework.
Examples of such possible mechanisms are discussed below
(see Section~\ref{sec:discussion:model}).


\subsection{Computational Efficiency of the HMC sampler}\label{sec:timing}
In all of the applications above, we used the standard HMC sampler described
in Section~\ref{sec:hmc}, that relies on the efficient gradient computation
algorithm of Section~\ref{sec:gradient:gradient}.
In the case of a simple BM, \citet{Hassler2019} developed a
Gibbs sampler on the variance parameter.
This approach uses a simplified version of the efficient likelihood computation
algorithm described in Section~\ref{sec:gradient:likelihood} to analytically
integrate missing values.
It was shown to be much more efficient than a previous approach
based on a numerical integration of these missing values
\citep{Cybis2015}, with a minimum 25-fold speed-up in all of the
configuration tested \citep{Hassler2019}.

When applying an OU process, such a Gibbs sampler cannot be used anymore,
as the joint distribution of the trait values at the tips of the tree
cannot be expressed as a simple Kronecker product
\citep[see \eg][for analytical expressions]{Clavel2015}.
Using the transformations described in Appendix~\ref{app:transformations},
it is however straightforward to derive an MCMC sampler based on a classical
random walk (RW) Metropolis-Hasting algorithm on the space of constrained parameters.
Such a sampler still relies on the efficient likelihood computation
algorithm of Section~\ref{sec:gradient:likelihood}.
Contrary to the Gibbs sampler, that can only be used with a set of
restricted priors, it has the same flexibility as the HMC sampler and can hence 
be targeted toward the exact same posterior distribution,
allowing for fair comparisons with the HMC.

Given the literature \citep{Neal2012,Ji2019,Fisher2019}, we expect
HMC to be more efficient than a simple RW, with better scalability
in the number of possibly correlated parameters.
To explore computational gains in this context,
we reran analyses on the two datasets explored in the previous
section.
We selected the models with, respectively, the least number of parameters
(the BM with no residual error) and the highest number of parameters
(the diagonal OU with correlated residual errors).
We ran each analysis 10 times and compared the average effective sample
size (ESS) per minute for each parameter.
We found that HMC delivered appreciable speed-up, with a median
$4.37$-fold increase in ESS per minute over all the parameters and
configuration tested.
The speed-up was particularly relevant for the most complex
model on the large HIV tree, with an approximate 5 to 10-fold increase
in ESS per minute for the variance and selection strength parameters,
that are known to be correlated and hence particularly difficult to
estimate with a classical RW sampler.
We refer to Appendix~\ref{app:timing} for the detailed set up and results
of this analysis.

We note that, as both the HMC and RW samplers use the same likelihood computation
algorithm (although the RW does not make use of the gradient), the speed-ups
reported here are mostly due to the sampling technique, and not the
algorithms described in this work.
In particular, we here used a classical HMC, with a fixed number of steps and
step sizes in the numerical approximation scheme.
Further speed-ups may be obtained by using more refined versions of the HMC sampler,
for instance using the No-U-Turn Sampler \citep{Hoffman2014} for optimal
exploration of the space,
or through the preconditioning of the posterior using an adequate mass matrix
\citep{Girolami2011,Neal2012,Ji2019}.
These improvements could be the focus of future work.

\section{Discussion}\label{sec:discussion} 

\subsection{Fossil Placement Using Continuous Traits}\label{sec:discussion:fossils}
%
The subject of combining morphological traits with molecular data, which are
typically not available for fossils, has received considerable attention over the last
few years.
Several studies focusing on discrete morphological characters showed that
combining both sources of information could improve fossil placement
\citep{Wiens2009,Wiens2010},
either using a fixed tree through maximum likelihood \citep{Berger2010},
or with a total evidence approach \citep{Ronquist2012,Gavryushkina2017}.
A recent trend appears to favor continuous morphological characters
over discretized ones, but these may be more challenging to use in an
inference framework \citep{Parins-Fukuchi2018a}.
This motivated the development of several methods for
fossil placement \citep{Revell2015,Parins-Fukuchi2018}
or divergence time estimation \citep{Alvarez-Carretero2019}
on known phylogenetic trees, using quantitative traits.
Given the potentially highly informative value of continuous morphological characters,
some attempts have also been made to infer phylogenies without any molecular data,
although with limited success \citep{VaronGonzalez2020}.
This is consistent with previous findings that sequence information generally tends
to dominate over trait information in a joint inference framework \citep{Baele2017DataIntegration}.

%
In Section~\ref{sec:weasels}, we showed how the total evidence framework could be used to inform fossil
placement.
Overall, we observed limited accuracy gains 
when using continuous trait information (see Section~\ref{sec:weasels:results}).
However, in this example, only three continuous characters were available,
offering limited information.
It would be interesting to test this framework on more extensive morphometric
datasets that include several dozens of traits \citep{Alvarez-Carretero2019},
and for which the associated signal might be stronger.
Combined with phylogenetic factor analysis \citep{Tolkoff2017},
it might provide a good alternative to fixed tree approaches previously
mentioned.

\subsection{Heritability of Virulence Estimation}\label{sec:discussion:hiv}
In line with previous work \citep{Leventhal2016,Mitov2018}, we
found in Sections~\ref{sec:hiv} and~\ref{sec:simus}
that the value and quality of the
heritability estimate was strongly dependent on our ability to,
first, select for the right model of evolution and,
second, to infer the parameters of this model with sufficient precision.
As shown in our simulation study, both of these tasks are however
challenging when the level of individual variation is high, \ie the
level of heritability is low.
This result may not be surprising from a statistical point of view.
Indeed, when the level of individual variation is high, then the evolutionary
model on the phylogeny only explains a small proportion of the observed tip
variation, and it hence becomes difficult to discriminate the signature
of one particular model over another in the observed data.

These results call for extreme caution when interpreting not only the heritability
estimates obtained in this \PCMa framework, but also all the other parameters
inferred from the model, such as the correlation between the traits
(see Appendix~\ref{app:appli:sec:hiv}).
The quality of the estimates, in particular for the
\OUa model, can depend on many factors, including
the shape and size of the phylogenetic tree \citep{Cooper2016}.
In line with previous recommendations
\citep[see \eg][]{Pennell2015,Cooper2016},
we found here that it could be useful
to complement the empirical analysis of a dataset with tailored
simulation studies,
designed to explore the potential limits or blind spots of a
given configuration.

In the HIV example presented here, the main conclusion that emerges,
irrespective of the exact estimate values and favored models,
is that the heritability of these virulence traits appears to be limited.
%
%
Overall, this indicates that host factors constitute
an important contribution to the virulence of HIV infections, and that mechanisms underlying complex
host-pathogen interactions remain to be explored in more details
\citep{Bartha2017}.

\subsection{Comparison with Maximum Likelihood Approaches}\label{sec:discussion:ml}
Numerous maximum likelihood tools exist to fit and compare
complex models of evolution on a fixed tree.
Such tools are widely used, usually fast, and have proven useful in many situations.
For instance, both case studies cited in this article
\citep{Schnitzler2017,Blanquart2017} used a maximum likelihood framework for their
analyses.
%
%
In addition, the \code{R} \citep{Rstats} \PCMs ecosystem is well developed, so that there exist
many specialized packages that cover a wide range of models with
various sets of assumptions (see Section~\ref{sec:intro:scope}).

Although usually more computationally demanding, Bayesian methods complement
maximum likelihood approaches in a number of ways. 
%
In particular, the priors' regularizing effect ensures that the parameters estimates remain biologically
reasonable (see Appendix~\ref{app:transformations}).
This feature proved useful in both our applications, where, in contrast, maximum likelihood
methods had difficulties in estimating some of the parameters, 
with estimates lying on arbitrarily fixed upper or lower bounds, yielding poor
biological interpretation.

The Bayesian framework also allows for the computation of the marginal likelihood
of a model,
that provides a theoretically consistent way to perform model selection.
Some penalised likelihood methods, such as the BIC criterion, can be seen as
approximations of this gold standard, although rather coarse \citep{Lebarbier2006}.
The GSS estimation used here, although computationally intensive, has been shown on the other
hand to be one of the most precise methods in a phylogenetic context \citep{Fourment2019}.

Finally, one further strength of our framework is that it is integrated, and
allows for the use of all of the extensive \beast modeling features.
Many \code{R} packages, on the other hand, although very useful in some situations,
have often been developed independently, and their strengths cannot be
combined into one global analysis.
This is particularly true when one wishes to use a total evidence approach.
Most maximum likelihood methods assume a fixed tree,
while our Bayesian framework can combine state-of-the-art phylogenetic reconstruction
methods with the complex and realistic continuous trait models described here.

\subsection{Modeling Assumptions}\label{sec:discussion:model}
%
As detailed in the introduction (see Section~\ref{sec:intro:scope}),
the framework described here relies on several simplifying assumptions, that
are common to most standard PCMs.
Relaxing those assumptions usually comes at the cost of a substantial increase
in computational complexity, and requires the development of specific algorithms,
that are not covered here.

Several simulation studies have been designed to challenge those standard
assumptions.
They usually rely on a complex trait simulator, that is tailored to a given
biological system, and aimed at producing realistic datasets.
For instance, in a macro-evolutionary context,
\citet{Duchen2020asymmetrical} study the impact of asymmetrical inheritance
on the ability of classical models to describe the produced patterns of
trait distribution among species, and show that they can be flawed in some
cases.
Similarly, in the field of virology, several epidemic models have already been
proposed to test the accuracy of virulence heritability estimations
\citep{Leventhal2016,Mitov2018}.
Exploring the behavior of our framework when applied to such realistic
simulated datasets could be the focus of future work.


\subsection{Concluding Remarks}
Motivated by the need to accommodate \OUa processes,
we have presented an efficient inference procedure for a
broad class of trait evolution models in a Bayesian inference framework.
Using two empirical examples and a
simulation study, we have demonstrated its applicability
to answer a large spectrum of biological questions, in fields ranging from
paleontology to virology.

At the core of the inference procedure, the likelihood and now gradient computation
algorithms are linear in the number of observations, making them efficient
on large trees, and applicable to a broad class of Gaussian evolutionary
processes, including but not limited to the popular \OUa model.
This algorithm however has worse than quadratic complexity in the number of latent
traits propagated on the tree. On the other hand, using techniques such as
phylogenetic factor analysis \citep{Tolkoff2017}, this latent dimension can
be reduced to a manageable size.
%
In this work, we made use of efficient gradient computation algorithm in a
Bayesian context. Note that this gradient could be more broadly exploited
in other settings, such as in a maximum likelihood inference.

All the formulas are written here for the general model of
Definition~\ref{def:general_model}.
Specific formulas are however only implemented for a sub-set of
the possible models, namely \OUa models with diagonal selection strength,
with a constant or time-scaled noise, and with only one observation for each
tip.
Extensions to more general models will be required in order to study specific datasets and
answer relevant biological questions. 
Although the main core mechanism remains unchanged, some derivations may
still be needed to propagate the gradient in such complex models.
For instance, dealing with the \OUa with a general selection strength
 implies taking the derivative of a matrix exponential with respect to
a matrix, which is a notoriously difficult problem, and may require
some approximations \citep{Al-Mohy2010}.
However, in the quest for ever more complex models, particular care should be
taken concerning practical and theoretical identifiability issues, as illustrated by
our simulation study.

\section{Data and Scripts}\label{sec:data_scripts}
All the scripts and data used in this manuscript are publicly available
as a GitHub repository: \url{https://github.com/pbastide/HMC_OU}.

\section{Acknowledgments}
PB conducted this research as a postdoctoral fellow funded by the
Fonds Wetenschappelijk Onderzoek (FWO, Belgium).
The research leading to these results has received funding from the European Research Council
under the European Union's Horizon 2020 research and innovation programme (grant agreement no. 725422-ReservoirDOCS).
The Artic Network receives funding from the Wellcome Trust through project 206298/Z/17/Z.
PL acknowledges support by the Research Foundation -- Flanders (`Fonds voor Wetenschappelijk Onderzoek -- Vlaanderen', G066215N, G0D5117N and G0B9317N).
GB acknowledges support from the Interne Fondsen KU Leuven / Internal Funds KU Leuven under grant agreement C14/18/094, and the Research Foundation -- Flanders (`Fonds voor Wetenschappelijk Onderzoek -- Vlaanderen', G0E1420N).
LSTH was supported by startup funds from Dalhousie University, the Canada Research Chairs program, the
NSERC Discovery Grant RGPIN-2018-05447, and the NSERC Discovery Launch Supplement DGECR-2018-00181.
MAS acknowledges support from National Institutes of Health grant U19 AI135995 and U01 AI151812.
We are grateful to the INRAE MIGALE bioinformatics facility (MIGALE, INRAE,
2020. Migale bioinformatics Facility, doi: 10.15454/1.5572390655343293E12) for
providing computing resources.
PB thanks Pierre Gloaguen for an enlightening discussion about Fisher's identity.
The authors thank Jan Schnitzler for sharing the alignment data to reproduce the
\weasels analyses,
as well as Jeffrey S Morris and two anonymous reviewers for their useful comments
that helped improve this manuscript.

\begin{supplement}
    \textbf{\nameref{app:post_pre_order}}.
    Formal and detailed description of the post and pre-order algorithms used to
    compute the likelihood and its gradient.
\end{supplement}
\begin{supplement}
    \textbf{\nameref{app:der}}.
    Formal derivation of the gradient formulas \wrt natural parameters.
\end{supplement}
\begin{supplement}
    \textbf{\nameref{app:transformations}}.
    Description of the smooth transformations used to map the constrained parameters
    to an unconstrained space.
\end{supplement}
\begin{supplement}
    \textbf{\nameref{app:heritability}}.
    Case study of the new population variance phylogenetic heritability on a toy
    example.
\end{supplement}
\begin{supplement}
    \textbf{\nameref{app:applications}}.
    Supplementary figures for biological applications and simulations.
\end{supplement}

\appendix

\renewcommand{\thefigure}{S\arabic{figure}}
\setcounter{figure}{0}

\section{Post and Pre-Order Algorithms}\label{app:post_pre_order}
In this appendix, we show how all the quantities used in the main text can
be computed in only two traversals of the tree:
one post-order (from the tips to the root) for likelihood computations,
and one pre-order (from the root to the tips) for gradient computations.

\subsection{Post-Order Algorithm For Likelihood Computation}\label{app:post_order}
The post-order propagation formulas with missing data have been presented in
\citet{Bastide2017} and \citet{Mitov2018PCM}.
We start by writing a slightly modified version of the formulas found in the
first reference, before showing how they can be made more efficient and robust
by reducing the number of operations per propagation iteration.

\subsubsection{Gaussian Propagation Formulas}\label{app:post:propagation}
Here, we re-write the propagation formulas found in \citet{Bastide2017}
(Appendix~2.2), using the notation conventions found in \citet{Hassler2019},
which provide a similar framework, but limited to Brownian diffusions.

\paragraph*{Model}
Using the notation of Definition~\ref{def:general_model}, we re-write
the model as a propagation on a
directed acyclic graph (DAG),
that mirrors the tree structure with an additional
layer of external nodes that represent observations
linked to tips of the tree, as shown Figure~\ref{fig:general_model}.
%
Equations~\eqref{eq:general_model_tree} and~\eqref{eq:general_model_obs}
can then be merged into a unique generic propagation step:
\begin{equation}\label{app:eq:general_model}
    \sachant{\completedm{\iAny}}{\completedm{\pa(\iAny)}}
    \sim \Normal{
        \genBranchActualization{\iAny}
        \completedm{\pa(\iAny)}
        + \genBranchDisplacement{\iAny}
        }{
        \genBranchVariance{\iAny}
    },
\end{equation}
where $\pa(\iAny)$ denotes the unique parent of node $\iAny$
in the underlying DAG of Figure~\ref{fig:general_model},
\ie the parent of node $\iAny$ in the phylogenetic tree,
or the latent tip associated with an observation.
This generic step encompasses both Equations, by taking:
\begin{equation}\label{app:eq:cor_main_app}
    \begin{aligned}
        \completedm{\iAny}
        & = \datam{\iAny},
        &
        \genBranchActualization{\iAny}
        &= \identityMatrix{\dimTrait}
        &
        \genBranchDisplacement{\iAny}
        & = \vect{0}_{\dimTrait}
        &
        \genBranchVariance{\iAny}
        & = \obsVariance{\iAny},
        &
        1 \leq \iAny \leq \nObs;
        \\
        \completedm{\iAny}
        & = \latentm{\iAnyPrime},
        &
        \genBranchActualization{\iAny}
        &= \branchActualization{\iAnyPrime},
        &
        \genBranchDisplacement{\iAny}
        & = \branchDisplacement{\iAnyPrime},
        &
        \genBranchVariance{\iAny}
        & = \branchVariance{\iAnyPrime},
        &
        \nObs + 1 \leq \iAny \leq \nObs + \nNodes,
    \end{aligned}
\end{equation}
with $\iAnyPrime = \iAny - \nObs$.
Note that, using the definition of $\pa$ as the parent of a
node on the underlying DAG,
if a node $\iAny$ is observed, \ie $1 \leq \iAny \leq \nObs$,
then its parent is such that
$\nObs + 1 \leq \pa(\iAny) \leq \nObs + \nNodes$,
and applying the second equation of~\eqref{app:eq:cor_main_app},
$\completedm{\pa(\iAny)}  = \latentm{\pa(\iAny) - \nObs}$.

\begin{remark}
    Several aspects of this definition are worth noting:
    \begin{itemize}
        \item Here, there can be several measurements
            $\datam{\iObs}$ associated to the same tip $\latentm{\pa(\iObs)}$,
            so that we have $\nObs \geq \ntaxa$ observations.
        \item In all the derivations below, we do not assume that
            the actualization matrices $\genBranchActualization{k}$ are invertible.
            This allows to encompass the phylogenetic factor model in this
            framework (with non-square actualization matrices that represent
            loading matrices).
        \item%
            All the theoretical developments presented in the main text and in this
            appendix stand for a general, possibly non-binary tree
            (a node can have more than two children).
            However, the \beast phylogenetic software \citep{Suchard2018},
            is restricted to binary trees.
            The practical implementation of this algorithm is hence limited to
            binary trees for the moment.
        \item%
            Similarly, this framework in theory allows for observations to be
            attached to internal nodes of the tree, and not just tips as in
            Figure~\ref{fig:general_model}. 
            In practice however, such a measurement requires a specific data
            structure (that, for instance, guaranties that the internal node always 
            exists, even when the tree is integrated out), that is currently not
            implemented in the released version of \beast.
    \end{itemize}
\end{remark}

\begin{figure}[!ht]
    \begin{center}
\begin{tikzpicture}[x=1pt,y=1pt]
\definecolor{fillColor}{RGB}{255,255,255}
\path[use as bounding box,fill=fillColor,fill opacity=0.00] (0,0) rectangle (325.21,143.09);
\begin{scope}
\path[clip] (  1.32,  1.32) rectangle (323.89,141.77);
\definecolor{drawColor}{RGB}{0,0,0}

\path[draw=drawColor,line width= 1.2pt,line join=round,line cap=round] ( 46.45, 52.04) -- ( 97.51, 52.04);

\path[draw=drawColor,line width= 1.2pt,line join=round,line cap=round] ( 97.51, 32.53) -- (123.04, 32.53);

\path[draw=drawColor,line width= 1.2pt,line join=round,line cap=round] (123.04, 19.53) -- (148.57, 19.53);

\path[draw=drawColor,line width= 1.2pt,line join=round,line cap=round] (123.04, 45.54) -- (199.62, 45.54);

\path[draw=drawColor,line width= 1.2pt,line join=round,line cap=round] ( 97.51, 71.55) -- (225.15, 71.55);

\path[draw=drawColor,line width= 1.2pt,line join=round,line cap=round] ( 46.45,110.56) -- (123.04,110.56);

\path[draw=drawColor,line width= 1.2pt,line join=round,line cap=round] (123.04, 97.56) -- (174.10, 97.56);

\path[draw=drawColor,line width= 1.2pt,line join=round,line cap=round] (123.04,123.57) -- (250.68,123.57);

\path[draw=drawColor,line width= 1.2pt,line join=round,line cap=round] ( 46.45, 52.04) -- ( 46.45,110.56);

\path[draw=drawColor,line width= 1.2pt,line join=round,line cap=round] ( 97.51, 32.53) -- ( 97.51, 71.55);

\path[draw=drawColor,line width= 1.2pt,line join=round,line cap=round] (123.04, 19.53) -- (123.04, 45.54);

\path[draw=drawColor,line width= 1.2pt,line join=round,line cap=round] (123.04, 97.56) -- (123.04,123.57);

\node[text=drawColor,anchor=base west,inner sep=0pt, outer sep=0pt, scale=  1.00] at (151.12, 15.72) {\bfseries $\latentm{5}$};

\node[text=drawColor,anchor=base west,inner sep=0pt, outer sep=0pt, scale=  1.00] at (202.18, 41.74) {\bfseries $\latentm{4}$};

\node[text=drawColor,anchor=base west,inner sep=0pt, outer sep=0pt, scale=  1.00] at (227.70, 67.75) {\bfseries $\latentm{3}$};

\node[text=drawColor,anchor=base west,inner sep=0pt, outer sep=0pt, scale=  1.00] at (176.65, 93.76) {\bfseries $\latentm{2}$};

\node[text=drawColor,anchor=base west,inner sep=0pt, outer sep=0pt, scale=  1.00] at (253.23,119.77) {\bfseries $\latentm{1}$};

\node[text=drawColor,anchor=base east,inner sep=0pt, outer sep=0pt, scale=  1.00] at ( 91.29, 40.77) {\bfseries $\latentm{7}$};

\node[text=drawColor,anchor=base east,inner sep=0pt, outer sep=0pt, scale=  1.00] at (116.82, 21.26) {\bfseries $\latentm{8}$};

\node[text=drawColor,anchor=base east,inner sep=0pt, outer sep=0pt, scale=  1.00] at (116.82, 99.29) {\bfseries $\latentm{9}$};

\node[text=drawColor,anchor=base east,inner sep=0pt, outer sep=0pt, scale=  1.00] at ( 39.85, 78.80) {\bfseries $\completedm{14} = \latentm{6}$};

\node[text=drawColor,anchor=base west,inner sep=0pt, outer sep=0pt, scale=  1.00] at (282.81,127.58) {$\datam{1} = \completedm{1}$};

\node[text=drawColor,anchor=base west,inner sep=0pt, outer sep=0pt, scale=  1.00] at (180.70, 10.53) {$\datam{8}$};

\node[text=drawColor,anchor=base west,inner sep=0pt, outer sep=0pt, scale=  1.00] at (180.70, 23.54) {$\datam{7}$};

\node[text=drawColor,anchor=base west,inner sep=0pt, outer sep=0pt, scale=  1.00] at (231.75, 43.04) {$\datam{6}$};

\node[text=drawColor,anchor=base west,inner sep=0pt, outer sep=0pt, scale=  1.00] at (257.28, 62.55) {$\datam{5}$};

\node[text=drawColor,anchor=base west,inner sep=0pt, outer sep=0pt, scale=  1.00] at (257.28, 75.56) {$\datam{4}$};

\node[text=drawColor,anchor=base west,inner sep=0pt, outer sep=0pt, scale=  1.00] at (206.22, 95.06) {$\datam{3}$};

\node[text=drawColor,anchor=base west,inner sep=0pt, outer sep=0pt, scale=  1.00] at (282.81,114.57) {$\datam{2}$};

\path[draw=drawColor,line width= 1.2pt,line join=round,line cap=round] (163.88, 19.53) -- (174.10, 13.02);

\path[draw=drawColor,line width= 1.2pt,line join=round,line cap=round] (163.88, 19.53) -- (174.10, 26.03);

\path[draw=drawColor,line width= 1.2pt,line join=round,line cap=round] (214.94, 45.54) -- (225.15, 45.54);

\path[draw=drawColor,line width= 1.2pt,line join=round,line cap=round] (240.47, 71.55) -- (250.68, 65.04);

\path[draw=drawColor,line width= 1.2pt,line join=round,line cap=round] (240.47, 71.55) -- (250.68, 78.05);

\path[draw=drawColor,line width= 1.2pt,line join=round,line cap=round] (189.41, 97.56) -- (199.62, 97.56);

\path[draw=drawColor,line width= 1.2pt,line join=round,line cap=round] (266.00,123.57) -- (276.21,117.06);

\path[draw=drawColor,line width= 1.2pt,line join=round,line cap=round] (266.00,123.57) -- (276.21,130.07);
\end{scope}
\end{tikzpicture}
        \caption{
            General model of trait evolution on a tree $\tree$
            with $\ntaxa = 5$ tips and $\nObs = 8$ observations.
            A latent trait $\latentm{\iNode}$ of dimension $\dimLatent$ is
            associated with each node $\iNode$
            ($1 \leq \iNode \leq \nNodes = 9$, internal and external) of the tree.
            One or several observations $\datam{\iObs}$ ($1 \leq \iObs \leq \nObsOne$)
            of dimension $\dimTrait$ are associated to each of the tips.
            $\completedMatr = (\dataMatr,\latentMatr)$ represents the completed dataset with both
            observed and latent traits.
        \label{fig:general_model}}
    \end{center}
\end{figure}

\paragraph*{Pseudo-Gaussian Propagation}
As in \citet{Hassler2019}, we define the \enquote{pseudo-Gaussian} density
of mean $\muv$ and precision $\matr{P}$ of dimension $\dimTrait$ as the function:
\begin{equation*}
    \log\altPhi{\bigcdot}{\muv}{\matr{P}}:
    \left\{
        \begin{aligned}
            \R^{\dimTrait} &\mapsto \R_+\\
            \vect{x}
            & \to
            - \frac12 \rank(\matr{P}) \log(2\pi)
            + \frac12 \log\altDet{\matr{P}}
            - \frac12 \transpose{(\vect{x} - \muv)}\matr{P}(\vect{x} - \muv),
        \end{aligned}
    \right.
\end{equation*}
where $\altDet{\matr{P}}$ is the product of all non-zero singular values of $\matr{P}$.
Note that when $\matr{P}$ is positive definite, then this coincides with the standard
Gaussian distribution.

The algorithm presented in \citet{Bastide2017} shows that for any index $\iAny$, the density
of $\dataBelowm{\iAny}$ (\ie all the measurements that have $\iAny$ as an
ancestor) conditionally on $\completedm{\iAny}$ is proportional to a pseudo-Gaussian:
\begin{equation}
    \log \cDensity{\dataBelowm{\iAny}}{\completedm{\iAny},\allParams}
    =
    \log \nodeRemainder{\iAny}
    +
    \log\altPhi{\completedm{\iAny}}{\nodeMean{\iAny}}{\nodePrecision{\iAny}},
\end{equation}
with the following propagation formulas:
\begin{align}
    \nodePrecision{\iAny}
    &=
    \sumChildren{\iAnyBis}{\iAny}
    \transpose{\genBranchActualization{\iAnyBis}}
    \deflatedNodePrecision{\iAnyBis}
    \genBranchActualization{\iAnyBis}
    \label{app:eq:post_precision}
    \\
    \nodeMean{\iAny}
    &=
    \specialInverse{\nodePrecision{\iAny}}
    \sumChildren{\iAnyBis}{\iAny}
    \transpose{\genBranchActualization{\iAnyBis}}
    \deflatedNodePrecision{\iAnyBis}
    (\nodeMean{\iAnyBis} - \genBranchDisplacement{\iAnyBis})
    \label{app:eq:post_mean}
\end{align}
\begin{subequations}
    \label{app:eq:post_remainder}
    \begin{align}
        \log \nodeRemainder{\iAny}
        &=
        \sumChildren{\iAnyBis}{\iAny}
        \log \nodeRemainder{\iAnyBis}
        \\
        & \quad
        + \frac12
        \rank(\nodePrecision{\iAny})\log(2\pi)
        - \frac12 \sumChildren{\iAnyBis}{\iAny}
        \rank(\deflatedNodePrecision{\iAnyBis})\log(2\pi)
        \label{app:subeq:post_remainder_rank}
        \\
        & \quad
        - \frac12 \log\altDet{\nodePrecision{\iAny}}
        + \frac12 \sumChildren{\iAnyBis}{\iAny}
        \log\altDet{\deflatedNodePrecision{\iAnyBis}}
        \label{app:subeq:post_remainder_det}
        \\
        & \quad
        + \frac12
        \transpose{\nodeMean{\iAny}}
        \nodePrecision{\iAny}
        \nodeMean{\iAny}
        - \frac12 \sumChildren{\iAnyBis}{\iAny}
        \transpose{(\nodeMean{\iAnyBis} - \genBranchDisplacement{\iAnyBis})}
        \deflatedNodePrecision{\iAnyBis}
        (\nodeMean{\iAnyBis} - \genBranchDisplacement{\iAnyBis}),
        \label{app:subeq:post_remainder_ss}
    \end{align}
\end{subequations}
where, given a node $\iAny$, $\children(\iAny)$ is the set of all the
direct children of $\iAny$,
and $\deflatedNodePrecision{\iAny}$ is defined in the next paragraph.


\paragraph*{Computation of $\deflatedNodePrecision{\iAny}$}
To compute $\deflatedNodePrecision{\iAny}$, we distinguish between observations
and tree nodes:
\begin{align}
    \deflatedNodePrecision{\iAny}
    &=
    \specialInverse{(\indMissing{\iAny}\obsVariance{\iAny}\indMissing{\iAny})}
    & \text{for an observation } 1 \leq \iAny \leq \nObs
    \label{app:eq:deflated_precision_obs}
    \\
    \deflatedNodePrecision{\iAny}
    &=
    \nodePrecision{\iAny}
    - \nodePrecision{\iAny}
    \inverse{(\nodePrecision{\iAny} + \inverse{\branchVariance{\iAny}})}
    \nodePrecision{\iAny}
    & \text{for a node } \nObs + 1 \leq \iAny \leq \nObs + \nNodes.
    \label{app:eq:deflated_precision_nodes}
\end{align}
where $\indMissing{\iAny}$ is a diagonal matrix, with a $0$ if the
trait is missing, and a $1$ otherwise.
Note that we use the Moore-Penrose pseudo inverse $\specialInverse{(\bigcdot)}$
in Equations~\eqref{app:eq:post_mean} and~\eqref{app:eq:deflated_precision_obs},
which can be computed using a singular value decomposition of the matrix.
In Equation~\eqref{app:eq:deflated_precision_nodes}, the regular inverse
can be used, as $\branchVariance{\iAny}$ is assumed to be positive definite
for any node $\iAny$.


\paragraph*{Initialization}
Given Equation~\eqref{app:eq:deflated_precision_obs}, we only need to
initialize $\nodeMean{\iObs}$ and $\nodeRemainder{\iObs}$ for all the observations
$\iObs$.
As in \citet{Hassler2019}, we take:
$\nodeMean{\iObs} = \indMissing{\iObs}\datam{\iObs}$ and
$\nodeRemainder{\iObs} = 1$,
so that $\nodeMean{\iObs}$ is equal to the observation when present, and $0$
otherwise.

\paragraph*{Root and Likelihood}
Once at the root $\iRoot$, we get the likelihood of the observed data given
the root trait:
\(
\log \cDensity{\dataBelowm{\iRoot}}{\completedm{\iRoot},\allParams}
=
\log \cDensity{\dataMatr}{\completedm{\iRoot},\allParams}.
\)
An extra integration on the root trait, using Equation~\eqref{eq:general_model_root},
gives the likelihood as
\(
    \log \cDensity{\dataMatr}{\allParams}
    =
    \log \nodeRemainder{\iTot},
\)
with:
\begin{align}
    \nodePrecision{\iTot}
    &=
    \nodePrecision{\iRoot}
    - \nodePrecision{\iRoot}
    \inverse{(\nodePrecision{\iRoot} + \inverse{\rootVariancem})}
    \nodePrecision{\iRoot}
    \\
    \nodeMean{\iTot}
    &=
    \specialInverse{\nodePrecision{\iTot}}
    \left(
        \nodePrecision{\iRoot}
        \nodeMean{\iRoot}
        +
        \inverse{\rootVariancem}
        \rootMeanm
    \right)
    \\
    \log \nodeRemainder{\iTot}
    &=
    \log \nodeRemainder{\iRoot}
    - \frac12 \rank(\nodePrecision{\iRoot})\log(2\pi)
    \\
    & \quad
    - \frac12 \log\altDet{\nodePrecision{\iTot}}
    + \frac12 \log\altDet{\nodePrecision{\iRoot}}
    - \frac12 \log\altDet{\rootVariancem}
    \nonumber\\
    & \quad
    + \frac12
    \transpose{\nodeMean{\iTot}}
    \nodePrecision{\iTot}
    \nodeMean{\iTot}
    - \frac12
    \transpose{\nodeMean{\iRoot}}
    \nodePrecision{\iRoot}
    \nodeMean{\iRoot}
    - \frac12
    \transpose{\rootMeanm}
    \inverse{\rootVariancem}
    \rootMeanm
    \nonumber
\end{align}
Note that, if the root is fixed ($\rootVariancem = \matr{0}_{\dimTrait\dimTrait}$),
then this expression simplifies to:
\begin{equation}
    \begin{aligned}
        \log \cDensity{\dataMatr}{\allParams}
        & =
        \log \nodeRemainder{\iRoot}
        - \frac12 \rank(\nodePrecision{\iRoot})\log(2\pi)
        \\
        & \quad
        + \frac12 \log\altDet{\nodePrecision{\iRoot}}
        - \frac12
        \transpose{(\rootMeanm - \nodeMean{\iRoot})}
        \nodePrecision{\iRoot}
        (\rootMeanm - \nodeMean{\iRoot})
\end{aligned}
\end{equation}

\paragraph*{Algorithmic Complexity}
In the generic case, the complexity of this algorithm is
$\Or(\nObs \dimTrait^3 + \nNodes \dimTrait^3)$.
However, the computations on the observations (Equation~\ref{app:eq:deflated_precision_obs})
could easily be parallelized, reducing the computational overhead.
In addition, if the observation model is parametrized in term of the
precision matrix $\obsPrecision{\iObs}$ (as is standard in Bayesian analyses),
then Equation~\eqref{app:eq:deflated_precision_obs} can be replaced with:
\(
\deflatedNodePrecision{\iObs}
=
\indMissing{\iObs}\obsPrecision{\iObs}\indMissing{\iObs},
\)
reducing the complexity to
$\Or(\nObs \dimTrait^2 + \nNodes \dimTrait^3)$.
Finally, if we assume 
that the observation variance is
diagonal, then the complexity reduces to
$\Or(\nObs \dimTrait + \nNodes \dimTrait^3)$.
%

\begin{remark}
A necessary condition for guaranteeing the correctness of this algorithm is that
$
\nodePrecision{\iAny} \nodeMean{\iAny} =
    \sumChildren{\iAnyBis}{\iAny}
    \transpose{\genBranchActualization{\iAnyBis}}
    \deflatedNodePrecision{\iAnyBis}
    (\nodeMean{\iAnyBis} - \genBranchDisplacement{\iAnyBis})
$
(see Equations~\ref{app:eq:post_precision} and~\ref{app:eq:post_mean}),
which is proven by the Lemma below.
\end{remark}

\begin{lemma}
\[
\nodePrecision{\iAny} \specialInverse{\nodePrecision{\iAny}}
\left [\sumChildren{\iAnyBis}{\iAny}
    \transpose{\genBranchActualization{\iAnyBis}}
    \deflatedNodePrecision{\iAnyBis}
    (\nodeMean{\iAnyBis} - \genBranchDisplacement{\iAnyBis}) \right ]=
    \sumChildren{\iAnyBis}{\iAny}
    \transpose{\genBranchActualization{\iAnyBis}}
    \deflatedNodePrecision{\iAnyBis}
    (\nodeMean{\iAnyBis} - \genBranchDisplacement{\iAnyBis}).
\]
\label{lem:guarantee}
\end{lemma}

\newcommand{\unknown}{\mathbf{x}}
\newcommand{\avec}{\mathbf{v}}
\newcommand{\amat}{\mathbf{U}}
\newcommand{\zero}{\mathbf{0}}

\begin{proof}
It is sufficient to prove that the equation $\nodePrecision{\iAny} \unknown =  \sumChildren{\iAnyBis}{\iAny}
    \transpose{\genBranchActualization{\iAnyBis}}
    \deflatedNodePrecision{\iAnyBis}
    (\nodeMean{\iAnyBis} - \genBranchDisplacement{\iAnyBis})$ has a solution.
By Theorem 1 in \citet{ben1969linear}, we only need to prove that $\transpose{\nodePrecision{\iAny}} \avec = \zero$ implies $\transpose{\avec} \left [\sumChildren{\iAnyBis}{\iAny}
    \transpose{\genBranchActualization{\iAnyBis}}
    \deflatedNodePrecision{\iAnyBis}
    (\nodeMean{\iAnyBis} - \genBranchDisplacement{\iAnyBis}) \right ] = 0$
    for any real vector $\avec$ of appropriate dimension.

Indeed, since $\transpose{\nodePrecision{\iAny}} \avec = \zero$, we have
\[
0 = \transpose{\avec} \nodePrecision{\iAny} \avec = \transpose{\avec} \left ( \sumChildren{\iAnyBis}{\iAny} \transpose{\genBranchActualization{\iAnyBis}} \deflatedNodePrecision{\iAnyBis} \genBranchActualization{\iAnyBis} \right ) \avec.
\]
This implies $\transpose{\avec} \transpose{\genBranchActualization{\iAnyBis}} \deflatedNodePrecision{\iAnyBis} \genBranchActualization{\iAnyBis} \avec = 0$ for all $\iAnyBis \in \children(\iAny)$.

For each child $\iAnyBis$ of $\iAny$, $\iAnyBis \in \children(\iAny)$, there exists an orthogonal matrix $\amat_\iAnyBis$ such that $\deflatedNodePrecision{\iAnyBis} = \transpose{\amat_\iAnyBis} \diag(\lambda^{(\iAnyBis)}_1, \ldots, \lambda^{(\iAnyBis)}_k, 0, \ldots, 0) \amat_\iAnyBis$ where $\lambda^{(\iAnyBis)}_1, \ldots, \lambda^{(\iAnyBis)}_k$ are positive eigenvalues of $\deflatedNodePrecision{\iAnyBis}$.
Since $\transpose{\avec} \transpose{\genBranchActualization{\iAnyBis}} \deflatedNodePrecision{\iAnyBis} \genBranchActualization{\iAnyBis} \avec = 0$, we derive that the first $k$ coordinates of $\amat_\iAnyBis \genBranchActualization{\iAnyBis} \avec$ are $0$.
Hence, $\transpose{\avec} \transpose{\genBranchActualization{\iAnyBis}} \deflatedNodePrecision{\iAnyBis} = 0$.
Therefore, $\transpose{\avec} \transpose{\genBranchActualization{\iAnyBis}} \deflatedNodePrecision{\iAnyBis} (\nodeMean{\iAnyBis} - \genBranchDisplacement{\iAnyBis}) = 0$ for every $\iAnyBis \in \children(\iAny)$, which completes the proof.
\end{proof}

\begin{remark}
    In \citet{Bastide2017} and \citet{Hassler2019},
    the authors use a \enquote{low-dimensional} generalized inverse to deal with missing
    values.
    The generalized inverse $\lowDimInverse{\anyMatrix}$ of a
    matrix $\anyMatrix$ is defined as follow:
    (1) find the indices $i \in \setIndices$
    such that $\anyMatrix_{ii}$ is not infinite nor zero;
    (2) invert the sub-matrix $\subMatrix{\anyMatrix}{\setIndices}$ with only rows and
    columns of $\anyMatrix$ that are in $\setIndices$;
    (3) for any $i$ in $\setIndices$ set
    $\lowDimInverse{\anyMatrix}_{ii} = (\inverse{\subMatrix{\anyMatrix}{\setIndices}})_{ii}$;
    (4) set all other coefficients of $\lowDimInverse{\anyMatrix}$ to zero or infinite
    (with non-standard rule $1/\infty = 0$, and $1 / 0 = \infty$).
    Given the special structure of the matrix to invert in Equation~\eqref{app:eq:deflated_precision_obs},
    the \enquote{infinite variance} values amount to marking the missing values, and the
    low-dimensional invert has the same result as the Moore-Penrose invert.
    However, it is easy to see that
    $\anyMatrix\lowDimInverse{\anyMatrix}\anyMatrix \neq \anyMatrix$
    so that this low-dimensional inverse is not a pseudo-inverse.
    Moreover, it is unclear that Lemma \ref{lem:guarantee} holds for this inverse.
    For consistency and numerical robustness (see below), we use here the more
    standard Moore-Penrose pseudo-inverse.
\end{remark}

\subsubsection{A More Robust Propagation}\label{app:post:robust}
In this section, we make the post-order traversal more efficient and numerically
robust by reducing the number of operations needed at each step and using more
numerically stable matrix algebra tools.

\paragraph*{Reducing computations}
\newcommand{\noNode}[1]{\mathcal{F}_{-#1}}
Here, we show that we can reduce the number of operations at each step of the
propagation, by eliminating matching terms over successive steps.
Let $\iNode$ be any node that is not the root, with unique parent $\pa(\iNode)$.
Then, in the computation of $\log \nodeRemainder{\pa(\iNode)}$
(Equation~\ref{app:eq:post_remainder}), keeping only the
terms that depend on $\iNode$ or $\pa(\iNode)$, we get:
\begin{align*}
    \log \nodeRemainder{\pa(\iNode)}
    &=
    \noNode{\iNode} +
    \log \nodeRemainder{\iNode}
    \\
    & \quad
    + \frac12
    \rank(\nodePrecision{\pa(\iNode)})\log(2\pi)
    - \frac12
    \rank(\deflatedNodePrecision{\iNode})\log(2\pi)
    \\
    & \quad
    - \frac12 \log\altDet{\nodePrecision{\pa(\iNode)}}
    + \frac12 \log\altDet{\deflatedNodePrecision{\iNode}}
    \\
    & \quad
    + \frac12
    \transpose{\nodeMean{\pa(\iNode)}}
    \nodePrecision{\pa(\iNode)}
    \nodeMean{\pa(\iNode)}
    - \frac12
    \transpose{(\nodeMean{\iNode} - \genBranchDisplacement{\iNode})}
    \deflatedNodePrecision{\iNode}
    (\nodeMean{\iNode} - \genBranchDisplacement{\iNode})
    \\
    &=
    \noNode{\iNode}' +
    \\
    & \quad
    + \frac12 \rank(\nodePrecision{\iNode})\log(2\pi)
    - \frac12
    \rank(\deflatedNodePrecision{\iNode})\log(2\pi)
    \\
    & \quad
    - \frac12 \log\altDet{\nodePrecision{\iNode}}
    + \frac12 \log\altDet{\deflatedNodePrecision{\iNode}}
    \\
    & \quad
    + \frac12
    \transpose{\nodeMean{\iNode}}
    \nodePrecision{\iNode}
    \nodeMean{\iNode}
    - \frac12
    \sss{
        \nodeMean{\iNode} - \genBranchDisplacement{\iNode}
        }{
        \deflatedNodePrecision{\iNode}
    },
\end{align*}
where $\noNode{\iNode}$ and $\noNode{\iNode}'$ are quantities that do not depend
directly on $\iNode$.
From Equation~\eqref{app:eq:deflated_precision_nodes}, we get:
\begin{equation*}
    \log\altDet{\deflatedNodePrecision{\iNode}}
    =
    \log\altDet{
        \nodePrecision{\iNode}
    }
    - \log\altDet{
            \identityMatrix{\dimTrait}
            - \inverse{(\nodePrecision{\iNode} + \inverse{\branchVariance{\iNode}})}
            \nodePrecision{\iNode}
    },
\end{equation*}
so that, using Lemma~\ref{app:lemma:rank} 
(see below),
this expression then simplifies to:
\begin{align*}
    \log \nodeRemainder{\pa(\iNode)}
    &=
    \noNode{\iNode}' +
    \\
    & \quad
    - \frac12 \log\altDet{
        \identityMatrix{\dimTrait} -
        \inverse{(\nodePrecision{\iNode} + \inverse{\branchVariance{\iNode}})}
        \nodePrecision{\iNode}
    }
    \\
    & \quad
    + \frac12
    \transpose{\nodeMean{\iNode}}
    \nodePrecision{\iNode}
    \nodeMean{\iNode}
    - \frac12
    \sss{
        \nodeMean{\iNode} - \genBranchDisplacement{\iNode}
        }{
        \deflatedNodePrecision{\iNode}
    }.
\end{align*}
Changing the initialization for observations to
\begin{equation}
    \nodeRemainder{\iObs}
    =
    - \frac12 \rank(\deflatedNodePrecision{\iObs})\log(2\pi)
    + \frac12 \log\altDet{\deflatedNodePrecision{\iObs}}
    \quad\text{for } 1 \leq \iObs \leq \nObs,
\end{equation}
we can then replace the propagation Equation~\eqref{app:eq:post_remainder} by the
following:
\begin{equation}
    \begin{aligned}
        \log \nodeRemainder{\iAny}
        &=
        \sumChildren{\iAnyBis}{\iAny}
        \log \nodeRemainder{\iAnyBis}
        - \frac12 \log\altDet{
            \identityMatrix{\dimTrait} -
            \inverse{(\nodePrecision{\iAny} + \inverse{\branchVariance{\iAny}})}
            \nodePrecision{\iAny}
        }
        \\
        & \quad
        + \frac12
        \transpose{\nodeMean{\iAny}}
        \nodePrecision{\iAny}
        \nodeMean{\iAny}
        - \frac12 \sumChildren{\iAnyBis}{\iAny}
        \transpose{(\nodeMean{\iAnyBis} - \genBranchDisplacement{\iAnyBis})}
        \deflatedNodePrecision{\iAnyBis}
        (\nodeMean{\iAnyBis} - \genBranchDisplacement{\iAnyBis}).
    \end{aligned}
\end{equation}
Note that this propagation formula is more efficient on several accounts.
First, it implies less computations, as it requires only one determinant
computation, instead of the number of children of $\iAny$ plus one.
Second, by analytically reducing the matching quantities between iterations
(rank and determinant),
we prevent numerical errors from propagating up the tree, and hence obtain a more
numerically robust algorithm.

\begin{lemma}
    \label{app:lemma:rank}
    For any node $\iNode$, we have:
    $\rank(\deflatedNodePrecision{\iNode}) = \rank(\nodePrecision{\iNode})$.
\end{lemma}

\begin{proof}
    From Equation~\eqref{app:eq:deflated_precision_nodes}, we get:
    \begin{equation*}
        \deflatedNodePrecision{\iNode}
        =
        \nodePrecision{\iNode}
            - \nodePrecision{\iNode} \inverse{(\nodePrecision{\iNode} + \inverse{\branchVariance{\iNode}})}
            \nodePrecision{\iNode}
         = \inverse{\branchVariance{\iNode}} \inverse{(\nodePrecision{\iNode} + \inverse{\branchVariance{\iNode}})}
            \nodePrecision{\iNode}.
    \end{equation*}
    Since $\inverse{\branchVariance{\iNode}}$ and $\inverse{(\nodePrecision{\iNode} + \inverse{\branchVariance{\iNode}})}$ are full rank, $\rank(\deflatedNodePrecision{\iNode}) = \rank(\nodePrecision{\iNode})$.

\end{proof}

\paragraph*{Robust Linear Algebra}
We mentioned above that we replaced the \enquote{low-dimensional} inverse used in
previous studies by the standard Moore-Penrose inverse.
In addition to simplifying the formula, the use of this standard pseudo-inverse allows
us to use efficient and well established tools to perform the computations
\citep{EJML}.
It also protects us from ill-conditioned variance matrices $\branchVariance{\iNode}$
which, depending on the values taken by the branch lengths $\branchLength_{\iNode}$,
and, in the case of an \OUa process, on the values of
the selection strength $\strengthm$, can become close to
singular matrices (see Equations~\ref{eq:general_to_BM} and~\ref{eq:general_to_OU}).

\subsection{Pre-Order Algorithm For Gradient Computation}\label{app:pre_order}
In this section, we show how to compute the moments appearing in
Proposition~\ref{prop:derivative_generic} in one extra pre-order traversal of the tree.
These formulas extend the developments found in \citet{Fisher2019} from the
simple BM case to the general case of Definition~\ref{def:general_model}.

\subsubsection{
    Conditional Moments \texorpdfstring{$\nodePreMean{\iAny}$}{nk}
    and \texorpdfstring{$\nodePrePrecision{\iAny}$}{Qk}
}
\label{app:pre:conditional_above}
Let $\iAny$ be any index. We write the distribution of
$\sachant{\completedm{\iAny}}{\dataAbovem{\iAny}}$ as a pseudo-Gaussian with mean
$\nodePreMean{\iAny}$ and precision $\nodePrePrecision{\iAny}$:
\begin{equation}
    \label{app:eq:pre_above_distribution}
    \cDensity{\completedm{\iAny}}{\dataAbovem{\iAny}, \allParams}
    =
    \altPhi{\completedm{\iAny}}{\nodePreMean{\iAny}}{\nodePrePrecision{\iAny}}.
\end{equation}
If $\iAny = \iRoot$ is the root of the tree, then as, by definition, all the
observations have the root as an ancestor,
$
\cDensity{\completedm{\iRoot}}{\dataAbovem{\iRoot}, \allParams}
=
\cDensity{\completedm{\iRoot}}{\allParams}
$
and is given by Equation~\eqref{eq:general_model_root} from
Definition~\ref{def:general_model}.
Otherwise, denote by $\iPar = \pa(\iAny)$ the unique parent of $\iAny$, and by
$\siblings(\iAny) = \children(\pa(\iAny)) \setminus \set{\iAny}$
all the siblings of $\iAny$
(\ie the direct children of $\pa(\iAny)$ different from $\iAny$).
Using the graphical independence structure, we get:
\begin{equation}
    \label{app:eq:int_pre_order}
    \cDensity{\completedm{\iAny}}{\dataAbovem{\iAny}, \allParams}
    = \int_{\R^{\dimTrait}}
    \cDensity{\completedm{\iAny}}{\completedm{\iPar}, \allParams}
    \times
    \cDensity{\completedm{\iPar}}{\dataAbovem{\iAny}, \allParams}
    \der \completedm{\iPar}.
\end{equation}
The first term in the integral is a Gaussian density, that we know from the
propagation Equation~\eqref{app:eq:general_model} along the tree.
To obtain the second term, we note that the observations not descending from node
$\iAny$ are exactly the observations not descending from parent node
$\iPar$, plus the observations that do descend from sibling nodes
$\siblings(\iAny)$:
$\dataAbovem{\iAny} =
(\dataAbovem{\iPar}, [\dataBelowm{\iAnyBis}]_{\iAnyBis \in \siblings(\iAny)})$.
Hence, using the graphical independence structure:
\begin{equation}
    \cDensity{\completedm{\iPar}}{\dataAbovem{\iAny}, \allParams} \propto
    \cDensity{\completedm{\iPar}}{\dataAbovem{\iPar}, \allParams}
    \times
    \prod_{\iAnyBis \in \siblings(\iAny)}
    \cDensity{\dataBelowm{\iAnyBis}}{\completedm{\iPar}, \allParams}.
\end{equation}
We know the first term of the product from the recursion.
The second term is similar to the quantities we have to deal with in the post-order
traversal. We get that it is proportional to a Gaussian density, with precision
$\deflatedNodePrecisionSibling{\iAny}$ and mean
$\deflatedNodeMeanSibling{\iAny}$ that are such that:
\begin{equation}
    \left\{
        \begin{aligned}
            \deflatedNodePrecisionSibling{\iAny}
            & =
            \sum_{\iAnyBis \in \siblings(\iAny)}
            \transpose{\genBranchActualization{\iAnyBis}}
            \deflatedNodePrecision{\iAnyBis}
            \genBranchActualization{\iAnyBis}
            \\
            \deflatedNodePrecisionSibling{\iAny}\deflatedNodeMeanSibling{\iAny}
            & =
            \sum_{\iAnyBis \in \siblings(\iAny)}
            \transpose{\genBranchActualization{\iAnyBis}}
            \deflatedNodePrecision{\iAnyBis}
            (\nodeMean{\iAnyBis} - \genBranchDisplacement{\iAnyBis}),
        \end{aligned}
    \right.
\end{equation}
where $\deflatedNodePrecision{\iAny}$ is computed during the post-order, and defined in
Equations~\eqref{app:eq:deflated_precision_obs} and~\eqref{app:eq:deflated_precision_nodes}.
Applying standard Gaussian combination rules, we hence get:
\begin{equation}
    \cDensity{\completedm{\iPar}}{\dataAbovem{\iAny}, \allParams} \propto
    \altPhi{\completedm{\iPar}}
    {\deflatedNodePreMean{\iAny}}{\deflatedNodePrePrecision{\iAny}} ,
\end{equation}
with:
\begin{equation}
    \left\{
        \begin{aligned}
            \deflatedNodePrePrecision{\iAny}
            & = \deflatedNodePrecisionSibling{\iAny} + \nodePrePrecision{\iPar}
            \\
            \deflatedNodePreMean{\iAny}
            & =
            \deflatedNodePreVariance{\iAny}
            (\deflatedNodePrecisionSibling{\iAny}\deflatedNodeMeanSibling{\iAny} +
            \nodePrePrecision{\iPar}\nodePreMean{\iPar}).
        \end{aligned}
    \right.
\end{equation}
Finally, from the integral of Equation~\eqref{app:eq:int_pre_order}, we get:
\begin{equation}
    \label{app:eq:pre_moments}
    \left\{
        \begin{aligned}
            \nodePrePrecision{\iAny}
            & = \inverse{(\genBranchActualization{\iAny}
            \deflatedNodePreVariance{\iAny}
            \transpose{\genBranchActualization{\iAny}}
            + \genBranchVariance{\iAny})}
            \\
            \nodePreMean{\iAny}
            & = \genBranchActualization{\iAny} \deflatedNodePreMean{\iAny}
            + \genBranchDisplacement{\iAny}
        \end{aligned}
    \right.
\end{equation}
Note, as a sanity check, that we indeed recover the formulas of \citet{Fisher2019} in the
case of a BM with no drift.
Note also that, as $\genBranchVariance{\iAny}$ is assumed to be positive
definite for all node $\iAny$, $\nodePrePrecision{\iAny}$ is also positive definite,
and the regular inverse can be used in the formula above.

\begin{remark}
    \label{app:remark:pre_moments_fixed_root}
    In the special case where the root is fixed
    ($\rootVariancem = \matr{0}_{\dimTrait\dimTrait}$),
    for any children node of the root $\iAny$ (such that $\pa(\iAny) = \iRoot$),
    the propagation formulas~\eqref{app:eq:pre_moments} simplify to:
    \begin{equation}
        \label{app:eq:pre_moments_fixed_root}
        \left\{
            \begin{aligned}
                \nodePrePrecision{\iAny}
                & = \inverse{\genBranchVariance{\iAny}}
            \\
            \nodePreMean{\iAny}
            & = \genBranchActualization{\iAny} \rootMeanm
            + \genBranchDisplacement{\iAny}
        \end{aligned}
    \right.
\end{equation}
\end{remark}

\subsubsection{
    Conditional Moments \texorpdfstring{$\fullConditionalMean{\iAny}$}{Mk}
    and \texorpdfstring{$\fullConditionalVariance{\iAny}$}{Vk}
}
\label{app:pre:conditional_full}
As in \citet{Fisher2019}, we compute the full conditional moments of
\begin{equation}
    \label{app:eq:pre_full_distribution}
    \cDensity{\completedm{\iAny}}{\dataMatr, \allParams}
    =
    \altPhi{\completedm{\iAny}}{\fullConditionalMean{\iAny}}{\specialInverse{\fullConditionalVariance{\iAny}}}
\end{equation}
as a combination of the moments of
$\cDensity{\dataBelowm{\iAny}}{\completedm{\iAny}, \allParams}$
and
$\cDensity{\completedm{\iAny}}{\dataAbovem{\iAny}, \allParams}$
computed in the previous two sections in the post and pre-order traversals of the tree
(Equations~\ref{app:eq:post_precision} -- \ref{app:eq:post_remainder}
and~\ref{app:eq:pre_moments}).
We distinguish between three cases, depending on the possible missing values.

\paragraph*{Latent or Unobserved trait}
We assume here that $\completedm{\iAny}$ is either a latent trait
(\ie a trait for a tree node, $\nObs \leq \iAny \leq \nObs + \nNodes$)
or a measurement that is completely missing.
Using Bayes rule, we write:
\begin{equation}
    \cDensity{\completedm{\iAny}}{\dataMatr, \allParams}
    \propto
    \cDensity{\dataBelowm{\iAny}}{\completedm{\iAny}, \allParams}
    \cDensity{\completedm{\iAny}}{\dataAbovem{\iAny}, \allParams},
\end{equation}
so that:
\begin{equation}
    \left\{
        \begin{aligned}
            \fullConditionalVariance{\iAny}
            & =
            \specialInverse{\left[
                    \nodePrecision{\iAny} + \nodePrePrecision{\iAny}
            \right]}
            \\
            \fullConditionalMean{\iAny}
            & =
            \fullConditionalVariance{\iAny}
            \left(
                \nodePrecision{\iAny} \nodeMean{\iAny}
                + \nodePrePrecision{\iAny} \nodePreMean{\iAny}
            \right).
            \\
        \end{aligned}
    \right.
\end{equation}

\paragraph*{Completely Observed Trait}
We assume here that $\completedm{\iAny} = \datam{\iAny}$ is a completely observed
measurement ($1 \leq \iAny \leq \nObs$).
Then
$
\cDensity{\completedm{\iAny}}{\dataMatr, \allParams}
=
\cDensity{\datam{\iAny}}{\datam{\iAny}, \allParams}
$,
so that:
\begin{equation}
    \left\{
        \begin{aligned}
            \fullConditionalVariance{\iAny}
            & =
            \matr{0}_{\dimTrait,\dimTrait}
            \\
            \fullConditionalMean{\iAny}
            & =
            \nodeMean{\iAny}
            =
            \datam{\iAny}.
            \\
        \end{aligned}
    \right.
\end{equation}

\paragraph*{Partially Observed Trait}
We assume here that $\completedm{\iAny} = \datam{\iAny}$ is a partially observed measurement,
such that
$\datam{\iAny} = \permObs{\iAny} \dataObsm{\iAny} + \permMiss{\iAny} \dataMissm{\iAny}$,
where $\dataObsm{\iAny}$ and $\dataMissm{\iAny}$ are the vectors of observed and missing
data at measurement $\iAny$, with dimension $\dimTraitObs{\iAny}$ and $\dimTraitMiss{\iAny}$
($\dimTraitObs{\iAny} + \dimTraitMiss{\iAny} = \dimTrait$);
and $\permObs{\iAny}$ and $\permMiss{\iAny}$ are permutations of dimensions
$\dimTrait \times \dimTraitObs{\iAny}$ and
$\dimTrait \times \dimTraitMiss{\iAny}$ that trace the observed and missing
indices to their right places.
Using Gaussian conditioning, we get:
\begin{equation}
    \left\{
        \begin{aligned}
            \fullConditionalVariance{\iAny}
            & =
            \permMiss{\iAny}
            \fullConditionalVarianceMiss{\iAny}
            \transpose{\permMiss{\iAny}}
            \\
            \fullConditionalMean{\iAny}
            & =
            \permObs{\iAny} \dataObsm{\iAny}
            +
            \permMiss{\iAny}
            \fullConditionalMeanMiss{\iAny}
        \end{aligned}
    \right.
\end{equation}
with:
\begin{equation*}
    \left\{
        \begin{aligned}
            \fullConditionalVarianceMiss{\iAny}
            & =
            \specialInverse{
                \left(
                    \transpose{\permMiss{\iAny}}
                    \nodePrePrecision{\iAny}
                    \permMiss{\iAny}
                \right)
            }
            \\
            \fullConditionalMeanMiss{\iAny}
            & =
                \transpose{\permMiss{\iAny}} \nodePreMean{\iAny}
                -
                \fullConditionalVarianceMiss{\iAny}
                \transpose{\permMiss{\iAny}}
                \nodePrePrecision{\iAny}
                \permObs{\iAny}
                \left(
                    \dataObsm{\iAny}
                    -
                    \transpose{\permObs{\iAny}}
                    \nodePreMean{\iAny}
                \right).
        \end{aligned}
    \right.
\end{equation*}

\section{Gradients and Chain Rules Formulas}\label{app:der}
In this appendix, we show how derivation chain rules can be used
in combination with Proposition~\ref{prop:derivative_generic} to compute the
gradient of the likelihood \wrt any of the natural parameters of a \BMa or an
\OUa, as defined in Section~\ref{sec:intro:model}.

\subsection{Gradients with respect to Generic Model Parameters}\label{app:der:generic}
In this section, we use the general model of Definition~\ref{def:general_model},
with generic propagation Equation~\eqref{app:eq:general_model},
and exploit the pre-order formulas~\eqref{app:eq:pre_moments} to get the derivative of
the pre-order moments $\nodePreVariance{\iAny}$ and $\nodePreMean{\iAny}$
that appear in Equation~\eqref{eq:derivative_generic} \wrt the generic
propagation parameters
$\genBranchActualization{\iAny}$,
$\genBranchDisplacement{\iAny}$ and
$\genBranchVariance{\iAny}$, for any index $\iAny$, $1 \leq \iAny \leq \nObs + \nNodes$.
In the rest of this appendix, $\anyMatrix$, $\anySymMatrix$ and $\anyVector$ are,
respectively,
an arbitrary test matrix, symmetric test matrix and test vector of adequate dimensions,
on which the derivatives are applied.
%
%
\subsubsection{Generic Formulas}\label{app:der:generic:formulas}
From formulas~\eqref{app:eq:pre_moments}, and using standard symmetric matrix
derivative formulas, we get:
\begin{align}
    &
    \left\{
        \begin{aligned}
            \partialDer{\mtov(\genBranchActualization{\iAny})}{\mtovh(\nodePreVariance{\iAny})}
            \mtovh(\anySymMatrix)
            & =
            2 \mtov(\genBranchActualization{\iAny} \deflatedNodePreVariance{\iAny} \anySymMatrix),
            \\
            \partialDer{\mtov(\genBranchActualization{\iAny})}{\nodePreMean{\iAny}}
            \anyVector
            & =
            \mtov(\anyVector \transpose{[\deflatedNodePreMean{\iAny}]}),
            \label{app:eq:der_actu}
        \end{aligned}
    \right.
    \\
    &
    \left\{
        \begin{aligned}
            \partialDer{\genBranchDisplacement{\iAny}}{\mtovh(\nodePreVariance{\iAny})}
            \mtovh(\anySymMatrix)
            & =
            \vect{0},
            \\
            \partialDer{\genBranchDisplacement{\iAny}}{\nodePreMean{\iAny}}
            \anyVector
            & =
            \anyVector,
            \label{app:eq:der_displacement}
        \end{aligned}
    \right.
    \\
    &
    \left\{
        \begin{aligned}
            \partialDer{\mtovh(\genBranchVariance{\iAny})}{\mtovh(\nodePreVariance{\iAny})}
            \mtovh(\anySymMatrix)
            & =
            \mtovh(\anySymMatrix),
            \\
            \partialDer{\mtovh(\genBranchVariance{\iAny})}{\nodePreMean{\iAny}}
            \anyVector
            & =
            \vect{0}.
            \label{app:eq:der_variance}
        \end{aligned}
    \right.
\end{align}
where $\mtov$ and $\mtovh$ are, respectively, the standard and symmetric
vectorization operators \citep{Magnus1986}.

\subsubsection{Missing Data}\label{app:der:generic:missing_data}
In Equation~\eqref{app:eq:der_variance}, we take the derivative \wrt the
symmetrically vectorized version of the variance matrix $\genBranchVariance{\iAny}$
in order to account for only actual and unique parameters.
However, when there is missing data, the variance terms associated to the
missing dimensions are not relevant anymore, and must be excluded from the
parameters.
To account for this, we introduce the \enquote{symmetric with missing values}
vectorization operator $\mtovhmn{\iAny}$ that, for any node $\iAny$, maps the matrix
$\genBranchVariance{\iAny}$ to a vector with only observed dimensions:
\begin{equation}
    \mtovh(\genBranchVariance{\iAny})
    =
    \duplicationMatrixMissing{\iAny}\mtovhmn{\iAny}(\genBranchVariance{\iAny})
\end{equation}
where, if $\dimTraitObs{\iAny}$ is the number of dimensions that are
observed in at least one of the descendants of $\iAny$ (\ie the dimensions $d$ such that
$[\nodePrecision{\iAny}]_{dd} \neq 0$), $\duplicationMatrixMissing{\iAny}$ is the
$
[\dimTraitObs{\iAny} (\dimTraitObs{\iAny} + 1) / 2]
\times
[\dimTrait (\dimTrait+ 1) / 2]
$
matrix that maps the symmetric matrix to its components matching with the observed
dimension.
Using this notation, Equation~\eqref{app:eq:der_variance} becomes:
\begin{align}
    \partialDer{\mtovhmn{\iAny}(\genBranchVariance{\iAny})}{\mtovh(\nodePreVariance{\iAny})}
    \mtovh(\anySymMatrix)
    & =
    \mtovhmn{\iAny}(\anySymMatrix),
    &
    \partialDer{\mtovhmn{\iAny}(\genBranchVariance{\iAny})}{\nodePreMean{\iAny}}
    \anyVector
    & =
    \vect{0},
    \label{app:eq:der_variance_missing}
\end{align}
where the derivative is taken according to the actual parameters only.

\subsubsection{Derivative \wrt the Inverse}\label{app:der:generic:inverse}
Note that, for convenience, we show the formulas for the derivative of
$\nodePreVariance{\iAny}$, but going back to $\nodePrePrecision{\iAny}$ is
straightforward, using the chain rule, and the following derivation formula
\citep{Magnus1986}:
\begin{equation}
    \label{app:eq:der_inverse}
    \partialDer{\mtovh(\nodePreVariance{\iAny})}{\mtovh(\nodePrePrecision{\iAny})}
    \mtovh(\anySymMatrix)
    =
    - \mtovh(\nodePrePrecision{\iAny} \anySymMatrix \nodePrePrecision{\iAny})
\end{equation}


\subsection{Gradients \wrt Natural Parameters}\label{app:der:natural}
The last step in the chained derivative is to link the generic parameters
$\genBranchActualization{\iAny}$,
$\genBranchDisplacement{\iAny}$ and
$\genBranchVariance{\iAny}$
to the natural parameters of the process in use.
Note that, until this step, the equations as well as the implementation are very
general, and valid for any model that can be cast in the framework of
Definition~\ref{def:general_model}.
In this section, we show how to link these generic formulas to actual models,
that can be used in a phylogenetic analysis.
One of the strengths of this framework is however to be quite easily extendable:
for any new model of interest, one only has to express the generic parameters
$\genBranchActualization{\iAny}$,
$\genBranchDisplacement{\iAny}$ and
$\genBranchVariance{\iAny}$
and their derivatives in terms of the natural parameters of the model in
order to use the general machinery described here.

\subsubsection{Brownian Motion with Drift}\label{app:der:BM}
Using Equations~\eqref{eq:general_to_BM}, for any node $\iNode$ that is not
an observation nor the root, we get the derivative of
$\branchActualization{\iNode}$,
$\branchDisplacement{\iNode}$ and
$\branchVariance{\iNode}$
with respect to the variance and drift parameters $\variancem$ and $\driftm$
of the \BMa:
\begin{align}
    &
    \left\{
        \begin{aligned}
            \partialDer{\driftm}{\mtov(\branchActualization{\iNode})}
            \mtov(\anyMatrix)
            &=
            \vect{0},
            \\
            \partialDer{\driftm}{\branchDisplacement{\iNode}}
            \anyVector
            &=
            \branchLength_{\iNode} \anyVector,
            \\
            \partialDer{\driftm}{\mtovh(\branchVariance{\iNode})}
            \mtovh(\anySymMatrix)
            &=
            \vect{0},
        \end{aligned}
    \right.
    \label{app:eq:der_wrt_drift_BM}
    \\
    &
    \left\{
        \begin{aligned}
            \partialDer{\mtovh(\variancem)}{\mtov(\branchActualization{\iNode})}
            \mtov(\anyMatrix)
            &=
            \vect{0},
            \\
            \partialDer{\mtovh(\variancem)}{\branchDisplacement{\iNode}}
            \anyVector
            &=
            \vect{0},
            \\
            \partialDer{\mtovh(\variancem)}{\mtovh(\branchVariance{\iNode})}
            \mtovh(\anySymMatrix)
            &=
            \branchLength_{\iNode}\mtovh(\anySymMatrix).
        \end{aligned}
    \right.
    \label{app:eq:der_wrt_variance_BM}
\end{align}


\subsubsection{\OU}\label{app:der:OU}
Using Equations~\eqref{eq:general_to_OU}, for any node $\iNode$ that is not
an observation nor the root, we get the derivative of
$\branchActualization{\iNode}$,
$\branchDisplacement{\iNode}$ and
$\branchVariance{\iNode}$
with respect to the optimal value and variance parameters
$\optimm$ and $\variancem$
of the \OUa:
\begin{align}
    &
    \left\{
        \begin{aligned}
            \partialDer{\optimm}{\mtov(\branchActualization{\iNode})}
            \anyVector
            &=
            \vect{0},
            \\
            \partialDer{\optimm}{\branchDisplacement{\iNode}}
            \anyVector
            &=
            (\identityMatrix{\dimLatent} - \actum{\branchLength_{\iNode}})
            \anyVector,
            \\
            \partialDer{\optimm}{\mtovh(\branchVariance{\iNode})}
            \mtovh(\anySymMatrix)
            &=
            \vect{0},
        \end{aligned}
    \right.
    \label{app:eq:der_wrt_optim_OU}
    \\
    &
    \left\{
        \begin{aligned}
            \partialDer{\mtovh(\variancem)}{\mtov(\branchActualization{\iNode})}
            \anyVector
            &=
            \vect{0},
            \\
            \partialDer{\mtovh(\variancem)}{\branchDisplacement{\iNode}}
            \anyVector
            &=
            \vect{0},
            \\
            \partialDer{\mtovh(\variancem)}{\mtovh(\branchVariance{\iNode})}
            \mtovh(\anySymMatrix)
            &=
            -
            \mtovh\left\{
                \passage \left[
                    \derivativeVarianceVariance(\strengthEigen)
                    \hadamard
                    \passage^{-1}
                    \anySymMatrix
                    \passage^{-T}
                \right] \passage^T
            \right\},
        \end{aligned}
    \right.
    \label{app:eq:der_wrt_variance_OU}
\end{align}
where we use the eigen-decomposition of the attenuation matrix
$\strengthm = \passage \strengthEigen \passage^{-1}$, and:
\begin{equation}
    \derivativeVarianceVariance_{\iNode}(\strengthEigen)
    =
    \left(
        \frac{
            1 -
            e^{-(\strengthEigenValue_l + \strengthEigenValue_r)
            \branchLength_{\iNode}}
            }{
            \strengthEigenValue_l + \strengthEigenValue_r
        }
    \right)_{1 \leq l, r \leq \dimLatent}.
\end{equation}
Note that, when any $\strengthEigenValue_l$ goes to zero,
$\derivativeVarianceVariance_{\iNode}(\strengthEigen)_{ll}$ goes to
$\branchLength_{\iNode}$,
and the derivative in Formulas~\eqref{app:eq:der_wrt_variance_OU} on this
dimension converges to BM Formulas~\eqref{app:eq:der_wrt_variance_BM}.

To get the derivative \wrt the selection strength $\strengthm$, we make the
extra assumption that it is diagonal, so that the actualization terms
$\branchActualization{\iNode}$ are also diagonal, and we get:
\begin{equation}
    \left\{
        \begin{aligned}
            \partialDer{\diag(\strengthm)}{\diag(\branchActualization{\iNode})}
            \anyVector
            &=
            -\branchLength_{\iNode}\branchActualization{\iNode} \anyVector
            \\
            \partialDer{\diag(\strengthm)}{\branchDisplacement{\iNode}}
            \anyVector
            &=
            -\branchLength_{\iNode}\diag(\branchActualization{\iNode}\anyVector\transpose{\optimm}),
            \\
            \partialDer{\diag(\strengthm)}{\mtovh(\branchVariance{\iNode})}
            \mtovh(\anySymMatrix)
            &=
            \left(
                \sum_{l = 1}^\dimLatent
                \derivativeVarianceAttenuation_{\iNode}(\strengthEigen)_{lr}
                \variancem_{lr}
                \anySymMatrix_{lr}
            \right)_{1 \leq r \leq \dimLatent},
        \end{aligned}
    \right.
\end{equation}
with:
\begin{equation}
    \derivativeVarianceAttenuation_{\iNode}(\strengthEigen)_{lr}
    =
    - 2
    \frac{
        1 -
        [1 + (\strengthEigenValue_l + \strengthEigenValue_r) \branchLength_{\iNode}]
        e^{-(\strengthEigenValue_l + \strengthEigenValue_r) \branchLength_{\iNode}}
    }
    {
        (\strengthEigenValue_l + \strengthEigenValue_r)^2
    }.
\end{equation}
Note that, when any $\strengthEigenValue_l$ goes to zero,
$\derivativeVarianceAttenuation_{\iNode}(\strengthEigen)_{ll}$ goes to
$\branchLength^2_{\iNode}$.


\subsubsection{Simple Error Model}\label{app:der:error}
For a simple error model, with no dimension jump and the same error variance
matrix for all the observations, we get:
\begin{equation}
    \obsVariance{\iObs}
    = \obsVariance{},
\end{equation}
and the derivatives are straightforward to obtain.

\section{Constrained Parameters}\label{app:transformations}

\subsection{The LKJ Transformation by Sampling Spheres}\label{app:trans:LKJ}
The LKJ \citep{Lewandowski2009} transformation is a popular tool to handle 
variance matrices in a Bayesian analysis. 
It relies on the decomposition of the variance matrix 
$\variancem = \diagVariance \correlation \diagVariance$
as the product of the diagonal matrix of standard deviations
$\diagVariance$ and the correlation matrix $\correlation$.
This allows for sampling both parameters independently, avoiding scaling effects
that can occur when sampling the variance matrix directly, using for instance
a Wishart distribution \citep{Barnard2000}.
The LKJ transformation and distribution act on the constrained space of 
correlation matrices $\corSpace$, defined as the space of squared matrices
of dimension $\dimMat$, that are symmetric positive definite with diagonal
values equal to one.

\subsubsection{The LKJ Transformation}\label{app:trans:LKJ:trans}
We show here that sampling in the space of correlation amounts to sampling
vectors in the half euclidean positive sphere.

\paragraph*{Cholesky Representation}
As in \citet{Stan2017}, we use the Cholesky decomposition of the correlation matrix 
$\correlation = \transpose{\cholCorrelation} \cholCorrelation$, with 
$\cholCorrelation$ upper triangular.
To ensure identifiability, we further assume that all the diagonal coefficients of
$\cholCorrelation$ are positive.
Instead of sampling the space $\corSpace$ of correlation matrices, we subsequently
sample the space $\cholCorSpace$ of Cholesky matrices, defined as the space of 
real upper triangular matrices with positive diagonal values such that 
$\transpose{\cholCorrelation} \cholCorrelation \in \corSpace$.
This transformation has two advantages.
First, as we will see below, this space has actually a relatively simple
structure.
Second, having the Cholesky decomposition of the correlation matrix is useful for 
all subsequent operations involving this matrix, such as taking the inverse.
In practice, in the implementation we will never compute the actual matrix
$\correlation$, inducing better numerical performances.

\paragraph*{Structure of the Cholesky Space}
The following proposition shows that sampling the Cholesky space $\cholCorSpace$
amounts to sampling $\dimMat$ vectors of dimensions $1$ to $\dimMat$ in the 
half-euclidean sphere.
\begin{proposition}
    \label{app:prop:chol_to_sphere}
    Let $\cholCorSpace$ be the space of Cholesky matrices of correlation matrices, and,
    for any $\dimAny$, $1 \leq \dimAny \leq \dimMat$,
    $\halfSphereAny$ the half Euclidean sphere defined by
    \(
    \halfSphereAny 
    = 
    \set{
        \vect{x} \in \R^{\dimAny}
        ~|~
        x_{\dimAny} > 0 
        \text{ and } 
        \sum_{\iOne = 1}^{\dimAny} x_{\iOne}^2 = 1 
    }.
    \)
    Then $\cholCorSpace$ is diffeomorphic to the Cartesian product of these half spheres:
    \begin{equation}
        \cholCorSpace 
        \diffeo 
        \bigtimes_{\dimAny = 1}^{\dimMat} \halfSphereAny.
    \end{equation}
\end{proposition}
\begin{proof}
    Let $\cholCorrelation \in \cholCorSpace$, and $1 \leq \dimAny \leq \dimMat$ a dimension.
    As $\cholCorrelation$ is upper-triangular, only the first $\dimAny$ coefficients
    of its $\dimAny^{th}$ column vector are non zero.
    Denote by $\cholCorrelation_\dimAny \in \R^{\dimAny}$ the vector with these non-zero
    coefficients. 
    Then, from the definition of $\cholCorSpace$, we get that 
    $\cholCorrelationCoef_{\dimAny\dimAny} > 0$, and
    $\transpose{\cholCorrelation_{\dimAny}}\cholCorrelation_{\dimAny} = 1$,
    \ie that
    $\cholCorrelation_{\dimAny}$ is in the half sphere $\halfSphereAny$.
    The function that to $\cholCorrelation$ associates the vectors
    $(\cholCorrelation_1, \dotsc, \cholCorrelation_{\dimMat})$ is then a 
    diffeomorphism between
    $\cholCorSpace$ and 
    $\bigtimes_{\dimAny = 1}^{\dimMat} \halfSphereAny$.
\end{proof}

\paragraph*{Sampling the Half Euclidean Sphere}
Sampling the half Euclidean sphere $\halfSphereAny$ is then relatively standard.
We start by mapping it to the underlying Euclidean ball $\ballAny$,
then go to the infinite norm ball $\ballInfAny$,
before reaching the unconstrained space $\R^{\dimAny - 1}$. 
As, for any $\vect{x} \in \halfSphereAny$, 
$x_{\dimAny} = \sqrt{1 - \sum_{\iOne = 1}^{\dimAny} x_{\iOne}^2}$,
the first step of going from 
$\halfSphereAny$
to
$\ballAny$ is straightforward.
The last step of mapping 
$\ballInfAny$
to
$\R^{\dimAny - 1}$
is also standard, using a \enquote{Fisher Z} transformation, or area hyperbolic
tangent.
To cover the missing step, we use the following transformation
$\ballInfToBallLKJ : \ballInfAny \to \ballAny$, adapted from
\citet{Lewandowski2009}, and defined for any $\vect{x} \in \ballInfAny$
and $1 \leq \iOne \leq \dimAny$ by:
\begin{equation}
    \ballInfToBallLKJ_\iOne(\vect{x}) =
    x_{\iOne}\prod_{\iTwo = 1}^{\iOne-1}\sqrt{1 - x_{\iTwo}^2}, 
\end{equation}
with the convention that the product over the empty set is equal to one.
Note that we use these three transformations to be consistent with the 
classical LKJ transformation (see below), but one could craft other ways
to map the half sphere to the unconstrained space. For instance, the 
standard transformation that to a vector $\vect{x} \in \ballAny$ associates
the vector $\vect{y} = (1 - \norm{\vect{x}})^{-1/2} \cdot \vect{x} $ in $\R^{\dimAny - 1}$
could also be used instead.

\paragraph*{The LKJ Transformation}
The LKJ transformation, as defined in \citet{Lewandowski2009} and detailed
\eg in \citet{Stan2017}, is then just equivalent to the joint transformation
of all the column-vectors $\cholCorrelation_{\dimAny}$ of the Cholesky transformation
$\cholCorrelation$ ($1 \leq \dimAny \leq \dimMat$) from the Euclidean half sphere
$\halfSphereAny$ to the unconstrained space $\R^{\dimAny - 1}$.

\subsubsection{The LKJ Distribution}\label{app:trans:LKJ:distr}
The LKJ distribution with parameter $\paramLKJ$ is defined in
\citet{Lewandowski2009} as a distribution over the space of correlation matrices
$\corSpace$, with density proportional to their determinant:
\begin{equation}
    \label{app:eq:LKJ_pdf}
    \LKJ(\sachant{\correlation}{\paramLKJ}) 
    =
    \constantLKJ \Det{\correlation}^{\paramLKJ - 1},
\end{equation}
with $\constantLKJ$ defined below in Equation~\eqref{app:eq:constant_LKJ}.
Note that when $\paramLKJ = 1$, it represents the uniform distribution over
correlation matrices. Without any expert information, that is the default
uninformative prior we use in our analyses.
It can be rewritten as an equivalent distribution over the Cholesky space
$\cholCorSpace$ \citep{Stan2017}:
\begin{equation}\label{eq:LKJChol_distribution}
    \LKJChol(\sachant{\cholCorrelation}{\eta})
    = \constantLKJ \prod_{\iOne = 1}^\dimMat \left(
        1 - \sum_{\iTwo = 1}^{\iOne-1} \cholCorrelationCoef_{\iTwo\iOne}^2
    \right)^{(\dimMat - \iOne + 2\paramLKJ - 2) / 2}.
\end{equation}

\paragraph*{The Spherical Beta Distribution}
\citet{Lewandowski2009} define the following elliptically contoured distribution
over the Euclidean ball (see Lemma 7 in the aforementioned paper):
\begin{mydef}[Spherical Beta Density]
    For any positive integer $\dimAny$ and positive real $\paramSB > 0$,
    the spherical beta distribution is defined by the following density,
    for any $\vect{x} \in \halfSphereAny$:
    \begin{equation}
        \sphericalBeta{\dimAny}{\paramSB}{x} 
        =
        \constantSB{\dimAny}{\paramSB} (1 - \sum_{\iNode = 1}^{\dimAny - 1}x_{\iOne}^2)^{\paramSB-1}
    \end{equation}
    with:
    \begin{equation}
        \constantSB{\dimAny}{\paramSB} = \Gamma(\paramSB + \dimAny/2) \pi^{-\dimAny/2} \inverse{\Gamma(\paramSB)}.
    \end{equation}
\end{mydef}
As in the previous section and using Proposition~\ref{app:prop:chol_to_sphere},
the LKJ distribution can be recovered by jointly applying this spherical beta
distribution to all the column vectors of the Cholesky matrix with adequate parameters:
\begin{equation}
    \LKJChol(\sachant{\cholCorrelation}{\paramLKJ}) 
    = \prod_{\dimAny = 2}^{\dimMat} 
    \sphericalBeta{\dimAny}{\paramLKJ + \frac{\dimMat-\dimAny}{2}}{\cholCorrelation_\dimAny}.
\end{equation}
It is easy to check that the induced constants are indeed the same:
\begin{equation}
    \label{app:eq:constant_LKJ}
    \begin{aligned}
        \prod_{\dimAny = 2}^{\dimMat} 
        \constantSB{\dimAny}{\eta + \frac{\dimMat-\dimAny}{2}}
        &= \prod_{\dimAny = 2}^{\dimMat} \frac{
        \Gamma\left(\eta + (\dimMat-\dimAny)/2 + (\dimAny-1)/2\right) \pi^{-(\dimAny-1)/2}}{
        \Gamma\left(\eta + (\dimMat-\dimAny)/2\right)
        }\\
        &= \Gamma\left(\eta + (\dimMat-1)/2\right)^{\dimMat-1}
        \left[\prod_{\dimAny = 1}^{\dimMat-1}
            \pi^{\dimAny/2}\Gamma\left(\eta + \frac{\dimMat-\dimAny-1}{2}\right)
        \right]^{-1},
    \end{aligned}
\end{equation}
where the last expression matches with the definition of the LKJ distribution
\citep[see][Section~3.3]{Lewandowski2009}.

\subsection{Other Parameters of the Models}\label{app:trans:other}
We detail below the transformations and priors used on the other parameters
of the model.

\subsubsection{Variance Parameters}\label{app:trans:var}
In addition to the correlation matrix, from the decomposition above we also
need to sample the diagonal variance term.
These are just constrained to be positive, so we use a standard $\log$
transformation on them,
and a vague half-Student prior 
(with default degree of freedom $1$ and scale $2.5$, for a normalized tree).

\subsubsection{Mean and Optimal Values Vectors}\label{app:trans:beta}
The mean and optimal values vectors are unconstrained in the general case, so 
no transformation is needed, and we use a default normal prior on them
(with default expectation $0$ and standard deviation $5$).

\subsubsection{Diagonal Selection Strength}\label{app:trans:strength}
When the selection strength $\strengthm$ is diagonal, all its diagonal terms
$(\strength_1, \dotsc, \strength_\dimTrait)$
are constrained to be positive. 
We hence use a standard log transformation on them.
For the prior, we use a vague half-normal, with standard deviation set so that, 
under the prior distribution on a normalized tree of unit height,
the phylogenetic half-life 
$\halfLife = \log(2) / \strength_\dimAny$
\citep{Hansen1997}
is larger than $5\%$ of the tree height $95\%$ of the time
(i.e. 
$\text{sd} = \log(2) / (5 / 100) /  q^{\text{half-normal}}_{95} \approx 7.07$).


\section{The Heritability Statistics}\label{app:heritability}
In Section~\ref{sec:heritability}, we introduced a new heritability statistic
based on the population variance, instead of the empirical variance as 
done in previously published studies \citep{Hassler2019}.
We argue here that, when the evolutionary process is not a simple
\BMa process (as the one they consider in \citealt{Hassler2019}),
using the population variance is more appropriate, and can avoid some
bias in the analysis.

\begin{figure}[!ht]
    \begin{minipage}{.49\textwidth}
        \begin{center}
\begin{tikzpicture}[x=1pt,y=1pt]
\definecolor{fillColor}{RGB}{255,255,255}
\path[use as bounding box,fill=fillColor,fill opacity=0.00] (0,0) rectangle (289.08,216.81);
\begin{scope}
\path[clip] ( 54.12, 67.32) rectangle (261.36,162.69);
\definecolor{drawColor}{gray}{0.50}

\path[draw=drawColor,line width= 1.2pt,line join=round,line cap=round] ( 61.80, 90.48) -- ( 79.56, 90.48);

\path[draw=drawColor,line width= 1.2pt,line join=round,line cap=round] ( 79.56, 80.66) -- ( 97.33, 80.66);

\path[draw=drawColor,line width= 1.2pt,line join=round,line cap=round] ( 79.56,100.29) -- ( 97.33,100.29);
\definecolor{drawColor}{RGB}{0,0,0}

\path[draw=drawColor,line width= 1.2pt,line join=round,line cap=round] ( 61.80,129.72) -- ( 79.56,129.72);

\path[draw=drawColor,line width= 1.2pt,line join=round,line cap=round] ( 79.56,119.91) -- ( 97.33,119.91);

\path[draw=drawColor,line width= 1.2pt,line join=round,line cap=round] ( 79.56,139.53) -- ( 97.33,139.53);

\path[draw=drawColor,line width= 1.2pt,line join=round,line cap=round] ( 61.80,110.10) -- ( 61.80,129.72);
\definecolor{drawColor}{gray}{0.50}

\path[draw=drawColor,line width= 1.2pt,line join=round,line cap=round] ( 79.56, 80.66) -- ( 79.56,100.29);
\definecolor{drawColor}{RGB}{0,0,0}

\path[draw=drawColor,line width= 1.2pt,line join=round,line cap=round] ( 79.56,119.91) -- ( 79.56,139.53);
\definecolor{drawColor}{gray}{0.50}

\path[draw=drawColor,line width= 1.2pt,line join=round,line cap=round] ( 61.80, 90.48) -- ( 61.80,110.10);

\node[text=drawColor,anchor=base west,inner sep=0pt, outer sep=0pt, scale=  1.00] at (100.88, 76.86) {\bfseries $\latentm{4}$};

\node[text=drawColor,anchor=base west,inner sep=0pt, outer sep=0pt, scale=  1.00] at (100.88, 96.49) {\bfseries $\latentm{3}$};
\definecolor{drawColor}{RGB}{0,0,0}

\node[text=drawColor,anchor=base west,inner sep=0pt, outer sep=0pt, scale=  1.00] at (100.88,116.11) {\bfseries $\latentm{2}$};

\node[text=drawColor,anchor=base west,inner sep=0pt, outer sep=0pt, scale=  1.00] at (100.88,135.73) {\bfseries $\latentm{1}$};
\definecolor{drawColor}{gray}{0.50}

\node[text=drawColor,anchor=base west,inner sep=0pt, outer sep=0pt, scale=  1.00] at (139.47, 78.17) {$\datam{4}$};

\node[text=drawColor,anchor=base west,inner sep=0pt, outer sep=0pt, scale=  1.00] at (139.47, 97.79) {$\datam{3}$};
\definecolor{drawColor}{RGB}{0,0,0}

\node[text=drawColor,anchor=base west,inner sep=0pt, outer sep=0pt, scale=  1.00] at (139.47,117.42) {$\datam{2}$};

\node[text=drawColor,anchor=base west,inner sep=0pt, outer sep=0pt, scale=  1.00] at (139.47,137.04) {$\datam{1}$};
\definecolor{drawColor}{gray}{0.50}

\path[draw=drawColor,line width= 1.2pt,line join=round,line cap=round] (118.65, 80.66) -- (132.87, 80.66);

\path[draw=drawColor,line width= 1.2pt,line join=round,line cap=round] (118.65,100.29) -- (132.87,100.29);
\definecolor{drawColor}{RGB}{0,0,0}

\path[draw=drawColor,line width= 1.2pt,line join=round,line cap=round] (118.65,119.91) -- (132.87,119.91);

\path[draw=drawColor,line width= 1.2pt,line join=round,line cap=round] (118.65,139.53) -- (132.87,139.53);

\node[text=drawColor,anchor=base,inner sep=0pt, outer sep=0pt, scale=  1.00] at ( 79.56,155.95) {$\variance{} = 0.01$};

\node[text=drawColor,anchor=base,inner sep=0pt, outer sep=0pt, scale=  1.00] at (125.76,155.95) {$\obsVarianceUni{} = 1.0$};
\definecolor{drawColor}{gray}{0.50}

\path[draw=drawColor,line width= 1.2pt,line join=round,line cap=round] ( 63.57, 86.55) -- ( 63.57, 94.40);

\node[text=drawColor,anchor=base,inner sep=0pt, outer sep=0pt, scale=  1.00] at ( 63.57, 73.72) {$\shift$};
\end{scope}
\end{tikzpicture} 
        \end{center}
    \end{minipage}
    \hfill
    \begin{minipage}{.49\textwidth}
        \begin{center}
\begin{tikzpicture}[x=1pt,y=1pt]
\definecolor{fillColor}{RGB}{255,255,255}
\path[use as bounding box,fill=fillColor,fill opacity=0.00] (0,0) rectangle (144.54,144.54);
\begin{scope}
\path[clip] (  0.00,  0.00) rectangle (144.54,144.54);
\definecolor{drawColor}{RGB}{255,255,255}
\definecolor{fillColor}{RGB}{255,255,255}

\path[draw=drawColor,line width= 0.5pt,line join=round,line cap=round,fill=fillColor] (  0.00,  0.00) rectangle (144.54,144.54);
\end{scope}
\begin{scope}
\path[clip] ( 34.90, 27.73) rectangle (139.54,139.54);
\definecolor{fillColor}{RGB}{255,255,255}

\path[fill=fillColor] ( 34.90, 27.73) rectangle (139.54,139.54);
\definecolor{drawColor}{gray}{0.92}

\path[draw=drawColor,line width= 0.3pt,line join=round] ( 34.90, 44.78) --
	(139.54, 44.78);

\path[draw=drawColor,line width= 0.3pt,line join=round] ( 34.90, 70.40) --
	(139.54, 70.40);

\path[draw=drawColor,line width= 0.3pt,line join=round] ( 34.90, 96.02) --
	(139.54, 96.02);

\path[draw=drawColor,line width= 0.3pt,line join=round] ( 34.90,121.65) --
	(139.54,121.65);

\path[draw=drawColor,line width= 0.3pt,line join=round] ( 55.51, 27.73) --
	( 55.51,139.54);

\path[draw=drawColor,line width= 0.3pt,line join=round] ( 87.22, 27.73) --
	( 87.22,139.54);

\path[draw=drawColor,line width= 0.3pt,line join=round] (118.93, 27.73) --
	(118.93,139.54);

\path[draw=drawColor,line width= 0.5pt,line join=round] ( 34.90, 31.96) --
	(139.54, 31.96);

\path[draw=drawColor,line width= 0.5pt,line join=round] ( 34.90, 57.59) --
	(139.54, 57.59);

\path[draw=drawColor,line width= 0.5pt,line join=round] ( 34.90, 83.21) --
	(139.54, 83.21);

\path[draw=drawColor,line width= 0.5pt,line join=round] ( 34.90,108.83) --
	(139.54,108.83);

\path[draw=drawColor,line width= 0.5pt,line join=round] ( 34.90,134.46) --
	(139.54,134.46);

\path[draw=drawColor,line width= 0.5pt,line join=round] ( 39.65, 27.73) --
	( 39.65,139.54);

\path[draw=drawColor,line width= 0.5pt,line join=round] ( 71.36, 27.73) --
	( 71.36,139.54);

\path[draw=drawColor,line width= 0.5pt,line join=round] (103.07, 27.73) --
	(103.07,139.54);

\path[draw=drawColor,line width= 0.5pt,line join=round] (134.78, 27.73) --
	(134.78,139.54);
\definecolor{drawColor}{RGB}{248,118,109}

\path[draw=drawColor,line width= 0.6pt,line join=round] ( 39.65, 32.81) --
	( 45.99, 58.06) --
	( 52.34, 90.69) --
	( 58.68,108.89) --
	( 65.02,118.30) --
	( 71.36,123.49) --
	( 77.70,126.58) --
	( 84.05,128.55) --
	( 90.39,129.87) --
	( 96.73,130.80) --
	(103.07,131.47) --
	(109.42,131.98) --
	(115.76,132.37) --
	(122.10,132.67) --
	(128.44,132.91) --
	(134.78,133.11);
\definecolor{drawColor}{RGB}{0,191,196}

\path[draw=drawColor,line width= 0.6pt,line join=round] ( 39.65, 32.98) --
	( 45.99, 32.98) --
	( 52.34, 32.98) --
	( 58.68, 32.98) --
	( 65.02, 32.98) --
	( 71.36, 32.98) --
	( 77.70, 32.98) --
	( 84.05, 32.98) --
	( 90.39, 32.98) --
	( 96.73, 32.98) --
	(103.07, 32.98) --
	(109.42, 32.98) --
	(115.76, 32.98) --
	(122.10, 32.98) --
	(128.44, 32.98) --
	(134.78, 32.98);
\definecolor{drawColor}{RGB}{0,0,0}

\path[draw=drawColor,line width= 0.6pt,line join=round] ( 34.90,134.46) -- (139.54,134.46);
\definecolor{drawColor}{gray}{0.20}

\path[draw=drawColor,line width= 0.5pt,line join=round,line cap=round] ( 34.90, 27.73) rectangle (139.54,139.54);
\end{scope}
\begin{scope}
\path[clip] (  0.00,  0.00) rectangle (144.54,144.54);
\definecolor{drawColor}{gray}{0.30}

\node[text=drawColor,anchor=base east,inner sep=0pt, outer sep=0pt, scale=  0.73] at ( 30.40, 29.24) {0.00};

\node[text=drawColor,anchor=base east,inner sep=0pt, outer sep=0pt, scale=  0.73] at ( 30.40, 54.87) {0.25};

\node[text=drawColor,anchor=base east,inner sep=0pt, outer sep=0pt, scale=  0.73] at ( 30.40, 80.49) {0.50};

\node[text=drawColor,anchor=base east,inner sep=0pt, outer sep=0pt, scale=  0.73] at ( 30.40,106.11) {0.75};

\node[text=drawColor,anchor=base east,inner sep=0pt, outer sep=0pt, scale=  0.73] at ( 30.40,131.74) {1.00};
\end{scope}
\begin{scope}
\path[clip] (  0.00,  0.00) rectangle (144.54,144.54);
\definecolor{drawColor}{gray}{0.20}

\path[draw=drawColor,line width= 0.5pt,line join=round] ( 32.40, 31.96) --
	( 34.90, 31.96);

\path[draw=drawColor,line width= 0.5pt,line join=round] ( 32.40, 57.59) --
	( 34.90, 57.59);

\path[draw=drawColor,line width= 0.5pt,line join=round] ( 32.40, 83.21) --
	( 34.90, 83.21);

\path[draw=drawColor,line width= 0.5pt,line join=round] ( 32.40,108.83) --
	( 34.90,108.83);

\path[draw=drawColor,line width= 0.5pt,line join=round] ( 32.40,134.46) --
	( 34.90,134.46);
\end{scope}
\begin{scope}
\path[clip] (  0.00,  0.00) rectangle (144.54,144.54);
\definecolor{drawColor}{gray}{0.20}

\path[draw=drawColor,line width= 0.5pt,line join=round] ( 39.65, 25.23) --
	( 39.65, 27.73);

\path[draw=drawColor,line width= 0.5pt,line join=round] ( 71.36, 25.23) --
	( 71.36, 27.73);

\path[draw=drawColor,line width= 0.5pt,line join=round] (103.07, 25.23) --
	(103.07, 27.73);

\path[draw=drawColor,line width= 0.5pt,line join=round] (134.78, 25.23) --
	(134.78, 27.73);
\end{scope}
\begin{scope}
\path[clip] (  0.00,  0.00) rectangle (144.54,144.54);
\definecolor{drawColor}{gray}{0.30}

\node[text=drawColor,anchor=base,inner sep=0pt, outer sep=0pt, scale=  0.73] at ( 39.65, 17.79) {0};

\node[text=drawColor,anchor=base,inner sep=0pt, outer sep=0pt, scale=  0.73] at ( 71.36, 17.79) {5};

\node[text=drawColor,anchor=base,inner sep=0pt, outer sep=0pt, scale=  0.73] at (103.07, 17.79) {10};

\node[text=drawColor,anchor=base,inner sep=0pt, outer sep=0pt, scale=  0.73] at (134.78, 17.79) {15};
\end{scope}
\begin{scope}
\path[clip] (  0.00,  0.00) rectangle (144.54,144.54);
\definecolor{drawColor}{RGB}{0,0,0}

\node[text=drawColor,anchor=base,inner sep=0pt, outer sep=0pt, scale=  0.91] at ( 87.22,  6.94) {$\shift$};
\end{scope}
\begin{scope}
\path[clip] (  0.00,  0.00) rectangle (144.54,144.54);
\definecolor{drawColor}{RGB}{0,0,0}

\node[text=drawColor,rotate= 90.00,anchor=base,inner sep=0pt, outer sep=0pt, scale=  0.91] at ( 11.80, 83.63) {Heritability};
\end{scope}
\begin{scope}
\path[clip] (  0.00,  0.00) rectangle (144.54,144.54);
\definecolor{fillColor}{RGB}{255,255,255}

\path[fill=fillColor] ( 63.26, 55.87) rectangle (132.10,111.40);
\end{scope}
\begin{scope}
\path[clip] (  0.00,  0.00) rectangle (144.54,144.54);
\definecolor{drawColor}{RGB}{0,0,0}

\node[text=drawColor,anchor=base west,inner sep=0pt, outer sep=0pt, scale=  0.91] at ( 68.26, 98.63) {Method};
\end{scope}
\begin{scope}
\path[clip] (  0.00,  0.00) rectangle (144.54,144.54);
\definecolor{fillColor}{RGB}{255,255,255}

\path[fill=fillColor] ( 68.26, 76.77) rectangle ( 84.16, 92.66);
\end{scope}
\begin{scope}
\path[clip] (  0.00,  0.00) rectangle (144.54,144.54);
\definecolor{drawColor}{RGB}{248,118,109}

\path[draw=drawColor,line width= 0.6pt,line join=round] ( 69.85, 84.71) -- ( 82.57, 84.71);
\end{scope}
\begin{scope}
\path[clip] (  0.00,  0.00) rectangle (144.54,144.54);
\definecolor{fillColor}{RGB}{255,255,255}

\path[fill=fillColor] ( 68.26, 60.87) rectangle ( 84.16, 76.77);
\end{scope}
\begin{scope}
\path[clip] (  0.00,  0.00) rectangle (144.54,144.54);
\definecolor{drawColor}{RGB}{0,191,196}

\path[draw=drawColor,line width= 0.6pt,line join=round] ( 69.85, 68.82) -- ( 82.57, 68.82);
\end{scope}
\begin{scope}
\path[clip] (  0.00,  0.00) rectangle (144.54,144.54);
\definecolor{drawColor}{RGB}{0,0,0}

\node[text=drawColor,anchor=base west,inner sep=0pt, outer sep=0pt, scale=  0.73] at ( 89.16, 81.99) {Empirical};
\end{scope}
\begin{scope}
\path[clip] (  0.00,  0.00) rectangle (144.54,144.54);
\definecolor{drawColor}{RGB}{0,0,0}

\node[text=drawColor,anchor=base west,inner sep=0pt, outer sep=0pt, scale=  0.73] at ( 89.16, 66.09) {Population};
\end{scope}
\end{tikzpicture} 
        \end{center}
    \end{minipage}
    \caption{
        Comparison of the heritability statistics obtained with the 
        empirical \citep{Hassler2019} and population (this paper)
        approach (right) computed on for a \BMa model on a simple tree,
        with a \BMa variance of $\variance = 0.01$, an observation
        variance of $\obsVarianceUni{} = 1.0$, and a shift $\shift$ in the process
        value affecting the two lower tips (left), so that the expectation of the 
        tips in black is $0$, and the expectation of the tips in grey is $\shift$.
        As the observation error is much higher than the process variance, 
        the heritability is low, as found by both method when $\shift = 0$.
        When $\shift$ increases, the \enquote{inter-group} variance between
        grey and black tips blurs the empirical variance, leading to 
        an inflated empirical heritability, that converges to $1$.
        The population heritability is robust to this model change,
        as it uses the process parameters directly.
    \label{app:fig:heritability_comparison}}
\end{figure}
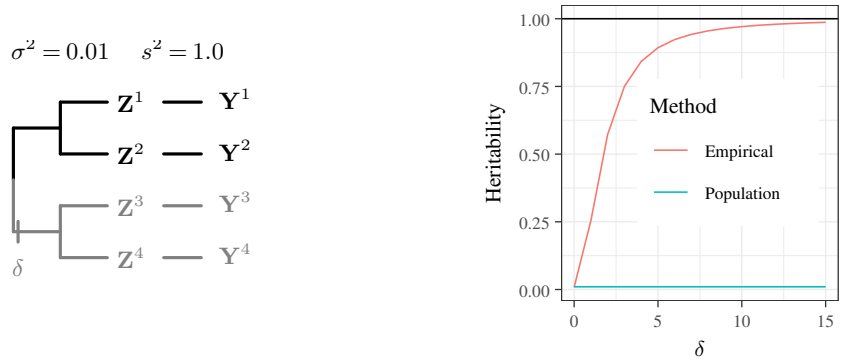

The main point of the argument is that, under any process that is not
a simple \BMa (\eg a \BMa with shifts or drift, or an \OUa), all the tips
do not have the same expected trait values, and hence the empirical mean 
is not a good estimate of the population mean.

To see this, we study the simple example of a \BMa on a four taxon tree with
one shift $\shift$ affecting half of the species
(see Figure~\ref{app:fig:heritability_comparison}, left).
The tree is taken to be ultrametric, with total height $\treeHeight = 1.0$,
and all branches of length $0.5$.
To simplify the analysis, we consider the case of a univariate process,
with variance $\variance = 0.01$, ancestral mean $\rootMean = 0$,
and a uniform observation process with variance $\obsVarianceUni{} = 1.0$.
Note that, as the observation variance is much larger than the process variance,
we expect the heritability to be low.
In that case, using the definition and notations found in \citet{Hassler2019},
the \enquote{empirical} heritability is given by:
\begin{equation}
    \heritability_{\text{emp}}
    =
    \frac{
        c_{\sigma} \variance + \shift^2/4
    }
    {
        c_{\sigma} \variance + c_{\gamma} \obsVarianceUni{} + \shift^2/4
    },
\end{equation}
where the extra term $\shift^2/4$ comes from the difference of expectations
at the tips 
(see \citealp{Hassler2019}, Formula~(5) in Supplementary Section~2).
From this formula, we can see that the empirical heritability will
converge to $1.0$ when $\shift$ becomes big enough, whatever the values of 
$\variance$ and $\obsVarianceUni{}$ 
(see Figure~\ref{app:fig:heritability_comparison}, right).

In contrast, the \enquote{population} heritability defined here
(see Equation~\ref{eq:heritability}) reduces, in this simple
case, to:
\begin{equation}
    \heritability_{\text{pop}}
    =
    \frac{
        \variance
    }
    {
        \variance + \obsVarianceUni{}
    },
\end{equation}
which does not depend on the value of the shift $\shift$.

Beyond this toy example, where the effects of the 
inter-group and intra-group variances is clearly marked, 
we would like to point out that the 
empirical variance is going to be biased as soon as
all the tips are not in the same \enquote{group}, \ie 
as soon as they do not have the same expected values under the
model. This happens in most of the models, such as the 
\BMa with drift \citep{Gill2016}, or the \OUa 
on a non-ultrametric tree \citep{Clavel2015}.
Using our population heritability might be a first step towards 
addressing this issue.

\newpage
\section{Supplementary Figures}\label{app:applications}


\subsection{Morphological Evolution in the \weasels Superfamily}\label{app:appli:sec:weasels}
\FloatBarrier



\begin{figure}
\centering
\begin{knitrout}
\definecolor{shadecolor}{rgb}{0.969, 0.969, 0.969}
\input{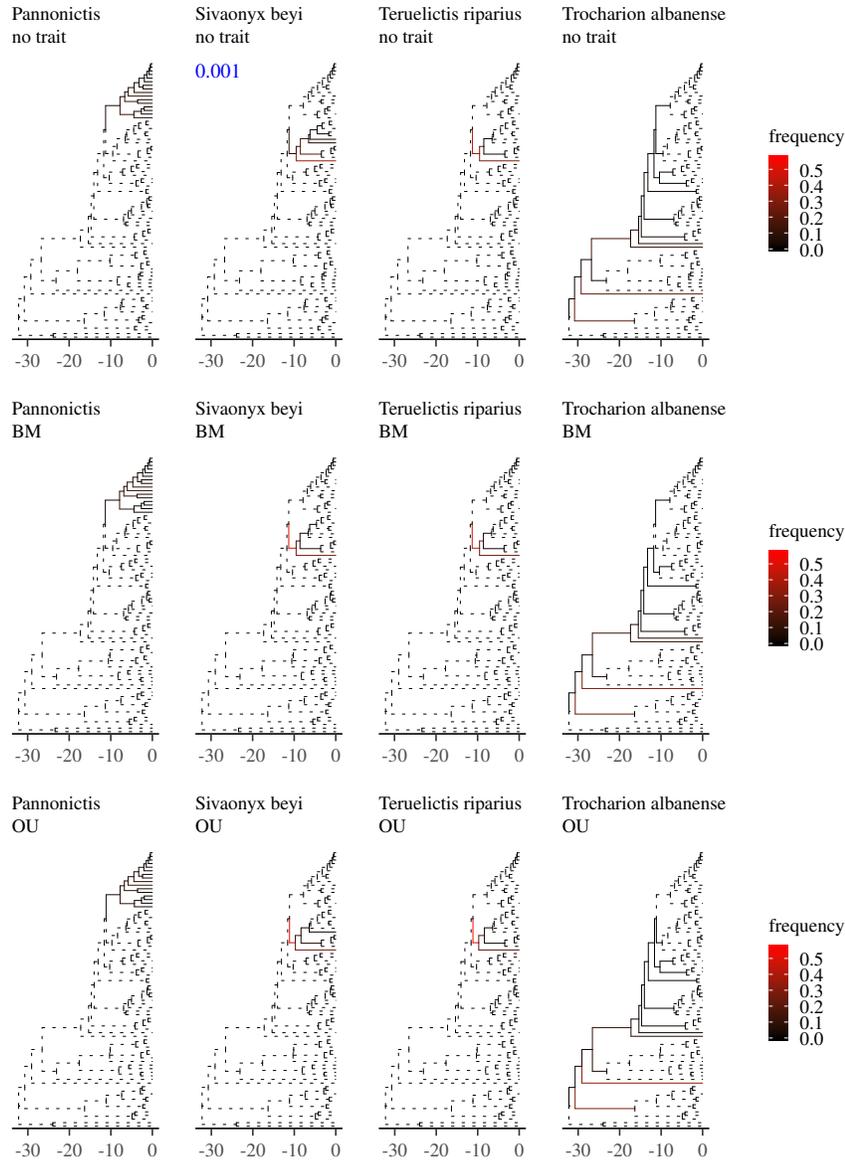}

\end{knitrout}
  \caption{
    Uncertainty in the placement of the fossils.
    Dotted branches are never associated with any fossil.
    Frequency over 1000 randomly sampled trees from the posterior.
    Trees are sufficiently similar that all sister clades to the fossil in backbone MCC tree almost always exist in all the trees
    (except in one case, when only 1 of the 1000 trees have a sister clade to the fossil not present in the MCC tree).
    Computations of frequencies done with function \code{get.rogue.placement} from \citet{Klopfstein2019}.
  }
\label{app:fig:applis:weasels:fossils}
\end{figure}

\begin{figure}
\centering
\begin{knitrout}
\definecolor{shadecolor}{rgb}{0.969, 0.969, 0.969}
\input{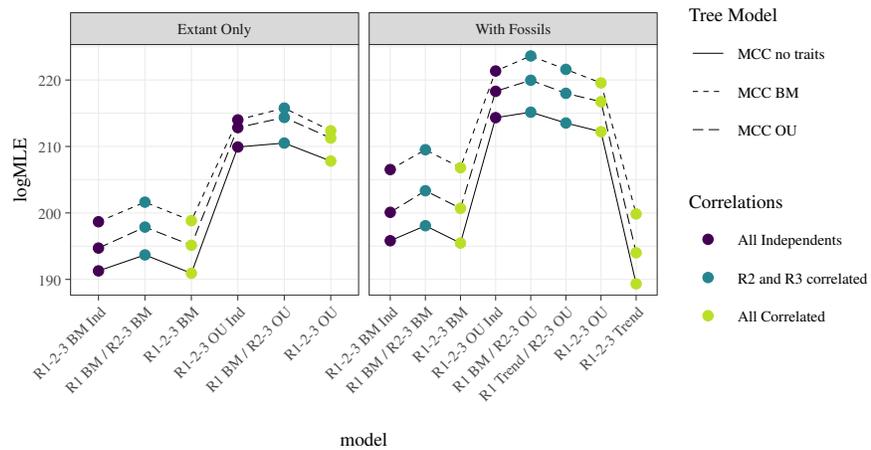}

\end{knitrout}
     \caption{GSS log Marginal Likelihood Estimation (logMLE) for various models (x axis),
     on MCC trees obtained with the models used during the complete evidence approach (line type).
     Fitted models are either the BM, the \enquote{trend} model (i.e. a BM with drift)
     or the OU.
     The three traits are either all independent, all correlated, or R2 and R3 correlated,
     but R1 independent (colors of the points).
     The model where R1 follows a BM and is independent from R2 and R3 following an OU,
     with no error, seems to be favored in every setting.
     Including measurement error does not seem to improve the fit in this setting (data not shown).
     The \enquote{trend} model, that is favored for R1 when fossils are included
     in \citet{Schnitzler2017} comes second in this analysis, with a log marginal
     likelihood difference of $1.97$,
     \ie a log Bayes factor of $1.97$,
     which, according to the guidelines found in \citet{Kass1995},
     can be considered as \enquote{substantial evidence} against the trend model.}
     \label{app:fig:weasels:MLE}
\end{figure}

\begin{figure}
\centering
\begin{knitrout}
\definecolor{shadecolor}{rgb}{0.969, 0.969, 0.969}
\input{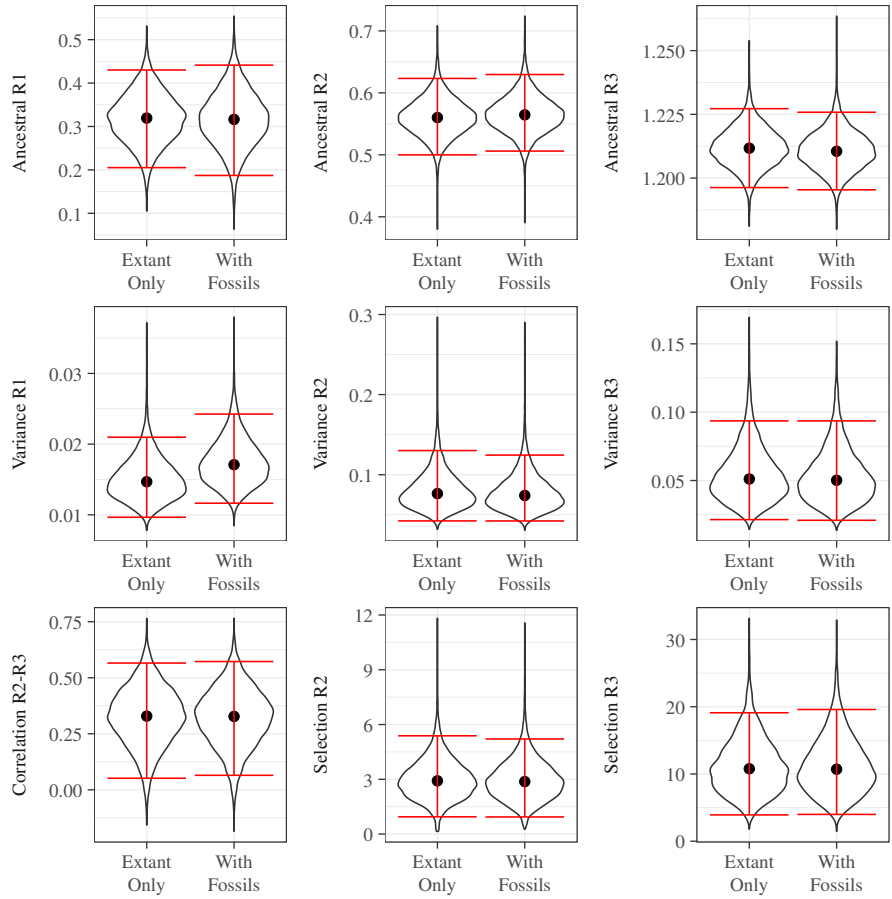}

\end{knitrout}
     \caption{Comparison between the best model fit on a tree with or without fossils.
     The model is a \BMa on R1, and an \OUa on R2 and R3, with correlations
     only between R2 and R3.
     Black points are the median, and red bars the 95\% HPDI.
     Including the fossils does not seem to have a dramatic effect on the estimates.}
     \label{app:fig:applis:weasels:BM_OU_OU_estimates}
\end{figure}

\FloatBarrier
\newpage


\subsection{Virulence Heritability in Human Immunodeficiency Viruses (HIV)}\label{app:appli:sec:hiv}



\begin{figure}
\centering
\begin{knitrout}
\definecolor{shadecolor}{rgb}{0.969, 0.969, 0.969}
\input{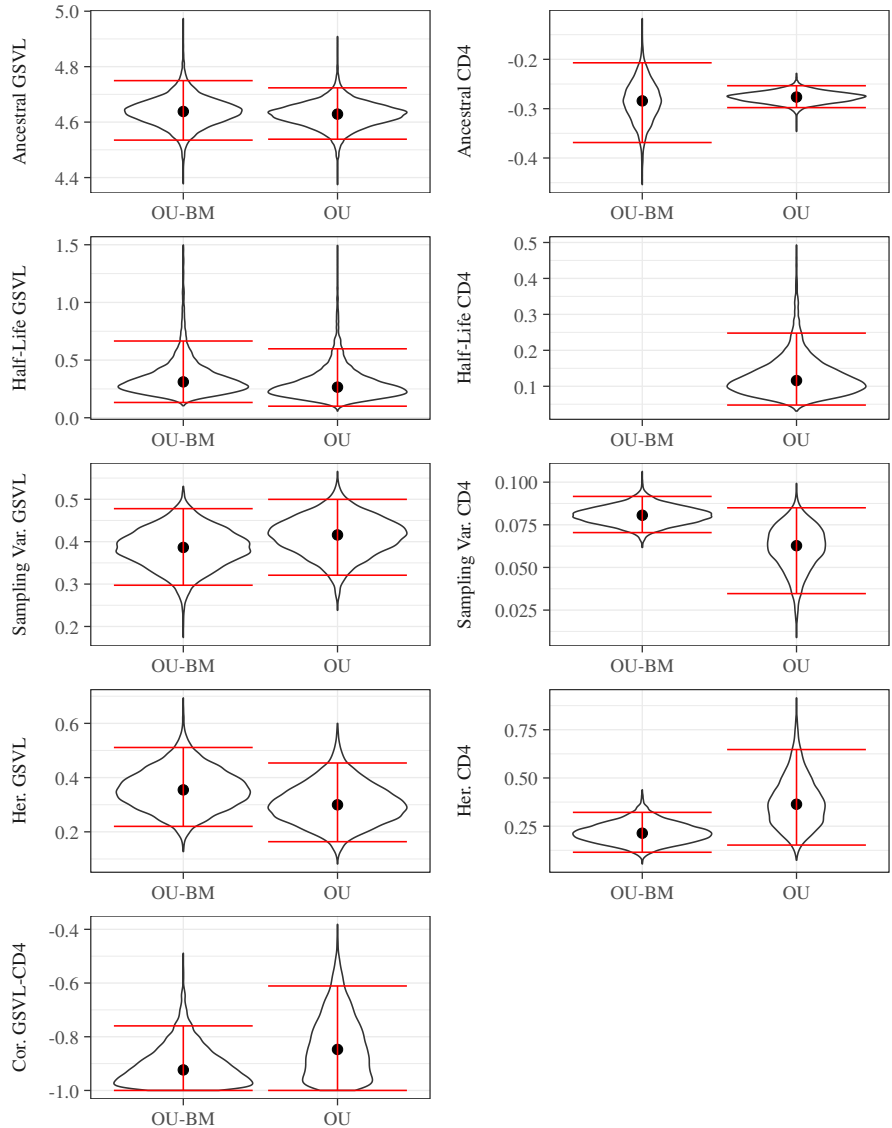}

\end{knitrout}
     \caption{Comparison between the OU and OU-BM models with scaled sampling
              variance.
              Under the OU-BM model, the GSVL evolves as an OU, while the
              CD4 evolves as a BM.
              Under the OU model, both evolve as an OU.
              HPD intervals overlap on both models for all parameters.
              Using the OU-BM model instead of the OU leads to a slightly higher
              estimate for the heritability of GSVL, and
              a lower estimate for the heritability of CD4.
              Consistently, the sampling variance of CD4 under the OU-BM model
              is estimated to be higher.
              This behavior is expected, as the OU allows for a higher
              disparity between trait measurements at the tips, reducing
              the need for the residual sampling variance.
              Black points are the median, and red bars the 95\% HPDI.
              For the half-life, outliers were omitted from the plot
              (representing, respectively,
              $0.75\%$ and
              $0.61\%$
              of the sampled data points.}
     \label{app:fig:applis:hiv:OU_OUBM_estimates_plot}
\end{figure}


\FloatBarrier
\paragraph*{Pagel's $\lambda$ Analysis}
Pagel's $\lambda$ \citep{Pagel1999} has been proposed as an estimator of
phylogenetic heritability \citep{Alizon2010,Vrancken2015}.
Given a \BMa model of evolution on a tree (with no observation model layer),
the $\lambda$ model is obtained by
scaling all the internal branch lengths by a factor $\lambda$, and adjusting the
length of all external branches so that all the tips remain at the same height
(i.e.
$\branchLength_\iNode(\lambda) = \lambda \branchLength_\iNode$ for all
internal node $\iNode$, and
$\branchLength_\iObs(\lambda) = \branchLength_\iObs + (1-\lambda) \nodeHeight_{\pa(\iObs)}
        = \lambda \branchLength_\iObs + (1-\lambda) \nodeHeight_\iObs)$
for all tips $\iObs$).
For a univariate trait, this model has been shown to be equivalent to a \BMa
with a scaled error observation model, with the $\lambda$
parameter equal to the heritability \citep{Leventhal2016}.
%
For a multivariate model, the same $\lambda$ parameter applies to all
the traits through the rescaling of the underlying tree. The model is then equivalent
to a \BMa with an observation error that is equal to the variance parameter of
the process, up to the tree re-scaling, and a unique scaling parameter
$\mu = (1 + \lambda)^{-1}$. This model enforces that all the traits share the same
heritability $\lambda$.

\paragraph*{Pagel's $\lambda$ Results}
The independent Pagel's $\lambda$ model is associated with the lowest support, with the
second weakest model (multivariate \BM with independent, non scaled errors)
having a relative log Bayes factor
of
$3$.
In contrast, the multivariate Pagel's $\lambda$ model yields the strongest support,
with a log Bayes factor
of
$1$
compared to the second best fitted model, that has
$3$ additional
parameters.
Under this model, the common heritability is estimated to be
$0.13$ (HPDI $95\%$: $[0.05, 0.22]$).

\paragraph*{Note on the Phylogenetic Correlation Parameter}
Figure~\ref{app:fig:applis:hiv:OU_OUBM_estimates_plot} (last panel) shows that
the estimated phylogenetic correlation variable between the two traits is very high
in absolute value,
with, for the \OUBMa model, a median of
$-0.92$ ($95\%$ HPDI $[-1, -0.76]$).
This result might look surprising at first sight, and in opposition with the empirical
(non-phylogenetic) correlation of $-0.23$ observed between the variables.
This is however a mechanical result of the model, as the tip variance is the
sum of the phylogenetic variance and the observation variance
(see Definition~\ref{def:general_model}).
When the later forced to be diagonal, any observed correlation must be phylogenetic.
But, when the heritability is low, the observation variance has a stronger relative impact
on the total variance than the phylogenetic variance, so that the phylogenetic
correlation must be high in order to have a small impact on the total observed variance.
This effect partly goes away when the observation error is allowed to be correlated,
as the phylogenetic variance is then estimated to
$-0.71$ ($95\%$ HPDI $[-1, -0.22]$)
for the same \OUBMa model.
In both cases however, the mean total correlation at the tips given by the model
is stable, of, respectively,
$-0.27$ and $-0.28$
for the independent and correlated observation models,
which is consistent with the empirical correlation.

\FloatBarrier
\newpage


\subsection{Exploration of Model Selection in an Heritability Estimation Context}\label{app:sec:simus}


\FloatBarrier
\begin{figure}
\centering
\begin{knitrout}
\definecolor{shadecolor}{rgb}{0.969, 0.969, 0.969}
\input{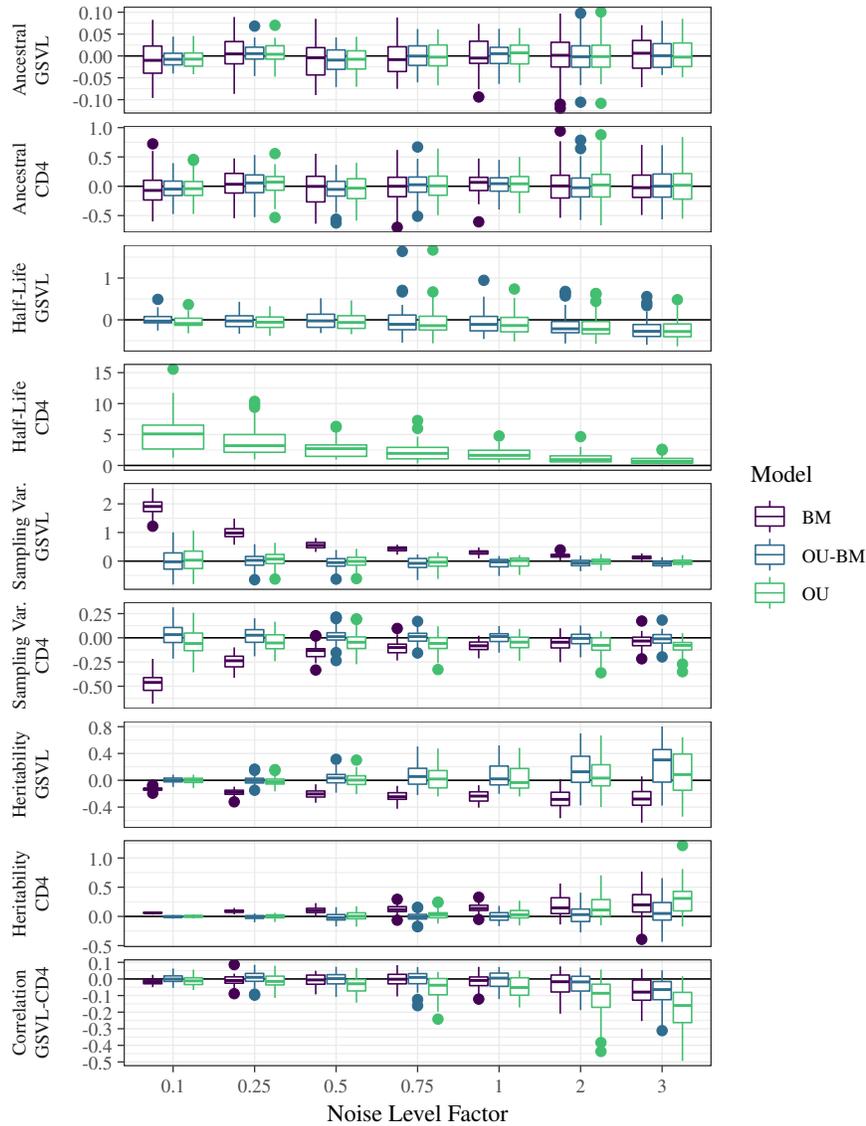}

\end{knitrout}
     \caption{Boxplot of the normalized estimates of the parameters of the process when a
              BM, an OU-BM (true model) or an OU is used for the inference and
              various levels of noise.
              The normalized estimates should converge to 0 (vertical line).
              Aside from some outliers, the ancestral state (optimal value)
              is well estimated, for any noise level and irrespective of the model
              used for the inference.
              The phylogenetic half-life tend to be under estimated when the noise
              becomes higher. That is consistent with the fact that a higher noise
              means less phylogenetically related measures, which can be obtained
              by higher selection strengths.
              For the CD4, the true model is a BM, so that the true half-life
              is formally equal to infinity. We can see that when an OU is inferred
              on this dimension, it can get a very short time-scale for high
              level of noise, indicating falsely a strong strength of selection.
              When the noise gets higher, its parameter, the sampling variance, gets
              better estimated.
              The heritability tends to be over-estimated with high levels of noise.}
     \label{app:fig:applis:simus:all_estimates_plot}
\end{figure}

\begin{figure}
\centering
\begin{knitrout}
\definecolor{shadecolor}{rgb}{0.969, 0.969, 0.969}
\input{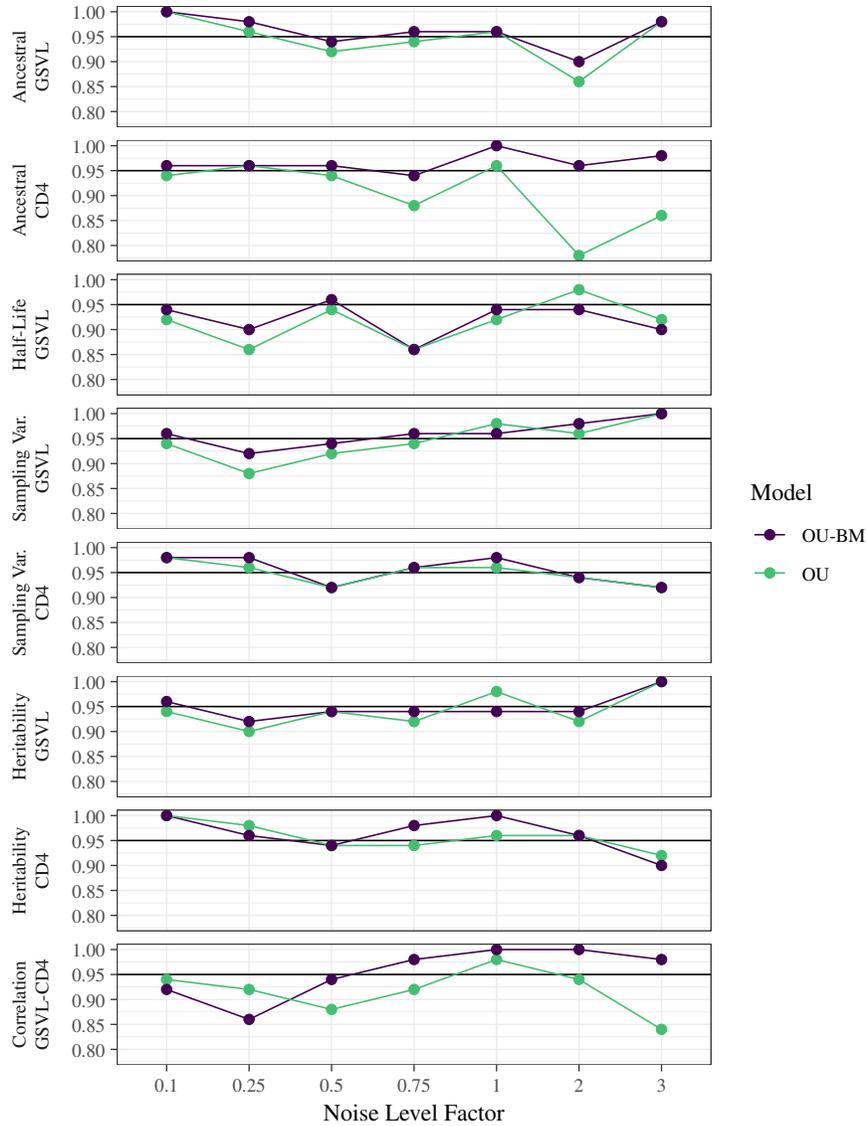}

\end{knitrout}
     \caption{Coverage of the HPD intervals for the parameters
              estimated under the true model (OU-BM) or the OU model.
              Black line is the target $0.95$.}
     \label{app:fig:applis:simus:all_coverage_plot}
\end{figure}

\FloatBarrier
\newpage


\subsection{Efficiency Gains of the HMC}\label{app:timing}
In this section, we compared the relative efficiency of a Random Walk (RW)
Metropolis-Hasting sampler, and a Hamiltonian Monte Carlo (HMC) sampler
to sample the posterior distribution.

\subsubsection{Datasets}
We use the two datasets studied in the previous section, that are distinct
in size and shape of the underlying tree, and in the number of traits considered.
For the \weasels dataset (see Section~\ref{sec:weasels}),
we used a fixed tree (obtained of through the total evidence approach with an OU model),
trimmed from all fossils so that it is ultrametric,
with 77 taxa and 3 continuous traits.
For the HIV dataset (see Section~\ref{sec:hiv}),
we used the same fixed tree as in the main text,
that is sampled in time, with 1171 tips and two continuous traits.

\subsubsection{Models}
On each datasets, we fit two models:
the BM with no residual error
and the diagonal OU with correlated residual errors.
Note that the second model has twice as many parameters to estimate
than the first one.

\subsubsection{Set-Up}
We used a set-up similar to our previous analyses for comparisons.
For the RW Metropolis Hastings, the chain was run for one million iterations,
sampled every one thousand steps.
For the HMC method, the chain was run for five thousands iterations,
sampled at every step,
with one hundred leap-frog steps of size 0.01.
Each analysis was reproduced 10 times.
The efficiency of the two samplers was assessed through the
average Efficient Sample Size for each parameter,
normalized by the time in hours (ESS per minute).
The \code{xml} files used for the analyses
as well as the \code{R} scripts for the formatting of the results
are available in the companion GitHub repository,
see Section~\ref{sec:data_scripts}.

\subsubsection{Results}
The results are presented in Table~\ref{table:timing}.


\begin{table}[!ht]
\begin{center}

\begin{tabular}{>{}l|rr>{}r|rr>{}r|rr>{}r|rr>{}r|}
\toprule
\multicolumn{1}{c}{ } & \multicolumn{6}{c}{Musteloidea} & \multicolumn{6}{c}{HIV} \\
\cmidrule(l{3pt}r{3pt}){2-7} \cmidrule(l{3pt}r{3pt}){8-13}
\multicolumn{1}{c}{ } & \multicolumn{3}{c}{BM no Error} & \multicolumn{3}{c}{OU with Error} & \multicolumn{3}{c}{BM no Error} & \multicolumn{3}{c}{OU with Error} \\
\cmidrule(l{3pt}r{3pt}){2-4} \cmidrule(l{3pt}r{3pt}){5-7} \cmidrule(l{3pt}r{3pt}){8-10} \cmidrule(l{3pt}r{3pt}){11-13}
  & RW & HMC & \textbf{Sp. Up} & RW & HMC & \textbf{Sp. Up} & RW & HMC & \textbf{Sp. Up} & RW & HMC & \textbf{Sp. Up}\\
\midrule
Median & 94.8 & 449.2 & \textbf{4.7$\times$} & 36.2 & 75.1 & \textbf{2.1$\times$} & 10.5 & 662.9 & \textbf{63.1$\times$} & 2.8 & 16.9 & \textbf{6.0$\times$}\\
Min & 88.8 & 332.9 & \textbf{3.8$\times$} & 23.5 & 24.1 & \textbf{1.0$\times$} & 10.4 & 66.1 & \textbf{6.4$\times$} & 0.6 & 8.2 & \textbf{13.3$\times$}\\
\midrule
Var 1 & 93.7 & 334.7 & \textbf{3.6$\times$} & 38.5 & 86.0 & \textbf{2.2$\times$} & 10.9 & 66.1 & \textbf{6.1$\times$} & 2.1 & 22.5 & \textbf{10.7$\times$}\\
Var 2 & 96.3 & 332.9 & \textbf{3.5$\times$} & 28.4 & 56.6 & \textbf{2.0$\times$} & 10.5 & 662.9 & \textbf{63.1$\times$} & 0.6 & 8.3 & \textbf{13.4$\times$}\\
Var 3 & 96.9 & 342.6 & \textbf{3.5$\times$} & 23.5 & 24.1 & \textbf{1.0$\times$} & - & - & \textbf{-} & - & - & \textbf{-}\\
\hline
Corr 12 & 88.8 & 468.1 & \textbf{5.3$\times$} & 35.5 & 50.3 & \textbf{1.4$\times$} & 10.6 & 662.9 & \textbf{62.4$\times$} & 5.5 & 26.4 & \textbf{4.8$\times$}\\
Corr 13 & 90.8 & 434.8 & \textbf{4.8$\times$} & 36.7 & 78.4 & \textbf{2.1$\times$} & - & - & \textbf{-} & - & - & \textbf{-}\\
Corr 23 & 94.6 & 449.2 & \textbf{4.7$\times$} & 31.6 & 51.6 & \textbf{1.6$\times$} & - & - & \textbf{-} & - & - & \textbf{-}\\
\hline
Mean 1 & 96.1 & 624.5 & \textbf{6.5$\times$} & 39.6 & 172.4 & \textbf{4.4$\times$} & 10.4 & 289.0 & \textbf{27.8$\times$} & 6.5 & 17.9 & \textbf{2.8$\times$}\\
Mean 2 & 96.8 & 544.0 & \textbf{5.6$\times$} & 39.8 & 179.7 & \textbf{4.5$\times$} & 10.4 & 662.9 & \textbf{63.7$\times$} & 6.4 & 18.9 & \textbf{3.0$\times$}\\
Mean 3 & 94.8 & 598.5 & \textbf{6.3$\times$} & 37.7 & 195.8 & \textbf{5.2$\times$} & - & - & \textbf{-} & - & - & \textbf{-}\\
\hline
Sel S 1 & - & - & \textbf{-} & 37.0 & 59.5 & \textbf{1.6$\times$} & - & - & \textbf{-} & 2.6 & 15.9 & \textbf{6.1$\times$}\\
Sel S 2 & - & - & \textbf{-} & 33.1 & 51.2 & \textbf{1.5$\times$} & - & - & \textbf{-} & 1.2 & 9.1 & \textbf{7.4$\times$}\\
Sel S 3 & - & - & \textbf{-} & 30.9 & 39.4 & \textbf{1.3$\times$} & - & - & \textbf{-} & - & - & \textbf{-}\\
\hline
Err V 1 & - & - & \textbf{-} & 38.9 & 132.3 & \textbf{3.4$\times$} & - & - & \textbf{-} & 3.0 & 13.1 & \textbf{4.4$\times$}\\
Err V 2 & - & - & \textbf{-} & 34.4 & 116.4 & \textbf{3.4$\times$} & - & - & \textbf{-} & 0.8 & 8.2 & \textbf{10.4$\times$}\\
Err V 3 & - & - & \textbf{-} & 23.6 & 71.8 & \textbf{3.0$\times$} & - & - & \textbf{-} & - & - & \textbf{-}\\
\hline
Err C 12 & - & - & \textbf{-} & 38.0 & 151.8 & \textbf{4.0$\times$} & - & - & \textbf{-} & 5.1 & 24.3 & \textbf{4.8$\times$}\\
Err C 13 & - & - & \textbf{-} & 36.6 & 105.5 & \textbf{2.9$\times$} & - & - & \textbf{-} & - & - & \textbf{-}\\
Err C 23 & - & - & \textbf{-} & 35.8 & 63.9 & \textbf{1.8$\times$} & - & - & \textbf{-} & - & - & \textbf{-}\\
\bottomrule
\end{tabular}

\caption{
Average Effective Sample Size (ESS) per minute
for the Random Walk (RW) and the Hamiltonian Monte Carlo (HMC) samplers
and associated relative speed-up (Sp. Up) of the HMC.
The two datasets are the one presented in Sections~\ref{sec:weasels} and~\ref{sec:hiv}.
The two models are the BM with no residual errors and the OU with correlated residual errors.
For each analysis, we first present the median and the minimum of the average ESS per minute among all parameters.
All the parameters of each models are then detailed
(Var: diagonal variance;
Corr: off-diagonal covariance;
Mean: root or optimum value;
Sel S: diagonal selection strength;
Err V: diagonal variance of the residual error;
Err C: off-diagonal covariance of the residual error).
}
\label{table:timing}
\end{center}
\end{table}

\FloatBarrier


\bibliographystyle{imsart-nameyear}
\bibliography{library.bib}

\end{document}